\documentclass[12pt]{iopart}

\usepackage{graphicx}
\def\P3{{\cal P}_t}
\def\J3{{\cal J}}
\def\T3{{\cal T}}

\def\v#1{{\bf#1}}
\def\bra{\langle}
\def\ket{\rangle}

\def\egu{\, =\, }
\def\plus{\, +\,}
\def\minus{\, -\,}
\def\cap{\noindent}
\def\hstm{ {{\hbar^2}\over {2 m}} }

\def\tpc{(2\pi)^3}

\def\kk{{\cal K}}

\def\barr#1{\overline{#1}}
\newcommand{\Vef}{V_{\mbox{\scriptsize eff}}}
\def\beq{\begin{equation}}
\def\eeq{\end{equation}}
\newcommand{\be}{\begin{eqnarray}}
\newcommand{\ee}{\end{eqnarray}}
\def\bar{\begin{array}[b]}
\def\barc{\begin{array}}
\def\bart{\begin{array}[t]}
\def\ear{\end{array}}
\def\le#1{\label{eq:#1}}
\def\re#1{\ref{eq:#1}}

\def\creas{{}\!^\dagger}
\begin{document}

\title{Properties of the nuclear medium}

\author{M. Baldo and G.F. Burgio}

\address{Instituto Nazionale di Fisica  Nucleare,

Sez. di Catania,

Via S. Sofia 64

95123 Catania, I} \ead{marcello.baldo@ct.infn.it, fiorella.burgio@ct.infn.it}
%

\begin{abstract}
We review our knowledge on the properties of the nuclear medium that have been studied, along many years, on the
basis of many-body theory, laboratory experiments and astrophysical observations. Throughout the presentation
particular emphasis is put on the possible relationship and links between the nuclear medium and the structure of
nuclei, including the limitations of such an approach. First we consider the realm of phenomenological laboratory
data and astrophysical observations and the hints they can give on the characteristics that the nuclear medium
should possess. The analysis is based on phenomenological models, that however have a strong basis on physical
intuition and an impressive success. More microscopic models are also considered, and it is shown that they are
able to give invaluable information on the nuclear medium, in particular on its Equation of State. The interplay
between laboratory experiments and astrophysical observations are particularly stressed, and it is shown how their
complementarity enriches enormously our insights into the structure of the nuclear medium. We then introduce the
nucleon-nucleon interaction and the microscopic many-body theory of nuclear matter, with a critical discussion
about the different approaches and their results. The Landau Fermi Liquid theory is introduced and briefly
discussed, and it is shown how fruitful it can be in discussing the macroscopic and low energy properties of the
nuclear medium. As illustrative example, we discuss neutron matter at very low density, and it is shown how it can
be treated within the many-body theory. The general bulk properties of the nuclear medium are reviewed to indicate
at which stage of our knowledge we stand, taking into account the most recent developments both in theory and
experiments. A section is dedicated to the pairing problem. The connection with nuclear structure is then
discussed, on the basis of the Energy Density Functional method. The possibility to link the physics of exotic
nuclei and the astrophysics of neutron stars is particularly stressed. Finally we discuss the thermal properties
of the nuclear medium, in particular the liquid-gas phase transition and its connection with the phenomenology on
heavy ion reactions and the cooling evolution of neutron stars. The presentation has been taken for
non-specialists and possibly for non-nuclear physicists.

\end{abstract}

\maketitle

\section{Introduction}

Nuclear Physics has so many facets that it looks impossible to find a common theoretical picture that is able to
unify under a common view, at least to a certain extent, the whole realm of phenomena where nucleonic systems play
a role. Indeed, the structure of nuclei, their excitations, nuclear collisions, the structure of Neutron Stars,
Supernovae explosion, very many astrophysical phenomena and processes, are all directly connected to that area of
Physics that can be called ''Nuclear Physics". The possible unification can come from the fundamental theory of
strong and weak interactions, Quantum Chromo Dynamics (QCD), and the so called Standard Model. However, besides
the difficulty to solve QCD for multi-baryonic systems with the necessary accuracy, this would be hardly useful
for the understanding on simple physical basis of the rich structure that nuclear systems display in different
contexts. \par From a semi-classical or macroscopic point of view, all nuclear systems can be considered as pieces
of a quite particular matter, the nuclear medium. The hypothetical uncharged infinite and homogeneous system
formed by the nuclear medium is usually called Nuclear Matter. Actually, as we will discuss, in first
approximation, Supernovae and Neutron Stars contain macroscopic portions of nuclear matter. From this point of
view, nuclei are considered as droplets of nuclear matter, and indeed this is the basis of the Liquid Drop Model
of nuclei. The macroscopic view cannot of course exhaust all the numerous aspects of nuclear structure, where
microscopic many-body effects are essential. It is however physically meaningful to ask for the properties of the
nuclear medium, since this is a state of matter of fundamental relevance. \par In this brief review paper we will
present the status of our knowledge on the nuclear medium as can be extracted phenomenologically and established
theoretically. On the other hand, we will discuss, on the basis of the works performed in the last few years, the
possibility of using the properties of the nuclear medium, noticeably its Equation of State (EoS), to guide the
nuclear structure theory of normal and exotic nuclei. Along the same lines it can be of great physical insight to
try to establish, to the extent that this is possible, a link between the macroscopic view and the general
properties of finite nuclei.
\par The style of the review is intended for non-specialists. We introduce each subject by reporting standard
results, leaving formal arguments to textbooks or original papers, before going to more advanced developments and
the discussion about on-going research works. The presentation is of course guided by the personal views of the
authors as well as by their limitations.

\section{The free Fermi gas of nucleons}
Before going to the microscopic many-body theory of nuclear matter, we remind the elementary properties of a free
Fermi gas of nucleons. This will serve as starting point when the nuclear interaction will be introduced and at
the same time as reference for comparison with the realistic treatments and results.

\subsection{The Equation of State \label{EoS}}

If we assume that no interaction takes place between $N$ nucleons inside a box of large volume, we have the
simplest model of nuclear matter, the free fermion gas. We remind here some elementary results, that will be
useful in the sequel. The total energy $E$ of the system is the sum over the single particle energies \beq E \egu
\sum_k {{\hbar^2\v{k}^2}\over {2 m}} \egu
 g\sum_{\v{k}} {{\hbar^2\v{k}^2}\over {2 m}} \egu
 {{g V \hbar^2}\over {2 m}} \int_{|\v{k}| < k_F} {{\v{k}^2 d^3k}\over
 (2\pi)^3} \ ,
\le{en} \eeq \cap and the energy per particle $e$ is given by \beq
 e \egu {E\over N} \egu {{g \hbar^2}\over {2 m \rho}}
 \int_{|\v{k}| < k_F} {{\v{k}^2 d^3k}\over (2\pi)^3}
 \egu \left( {{3\hbar^2 k_F^2}\over 10 m} \right) \egu {3\over 5} E_F
 \ .
\le{enp} \eeq \cap In equation (\re{enp}) we have used equation (\re{en}) and introduced the Fermi energy $E_F =
\hbar^2 k_F^2 / 2m$, the energy of the highest occupied level. From Eqs. (\re{enp}) and (\re{dens}), one gets \beq
 e \egu {3\over 5} {\hbar^2\over {2m}} \left( {{3\pi^2}\over 2}\right)
^{2\over 3} \rho^{2\over 3} \le{eos1} \eeq \cap which relates the energy per particle $e$ to the density $\rho$,
and therefore it is the EOS (the simplest one) for a free symmetric nucleon gas at zero temperature. If one
measures the energy in MeV, the length in femtometers $fm$ (otherwise also called ``fermi''), and adopts for the
nucleon mass $mc^2 = 938.9$ MeV, an average value between neutron and proton masses, for simplicity, then
$\hbar^2 / 2m = $ 20.74, and \beq
 e \egu 75.03 \,\, \rho^{2\over 3} \ \ {\rm MeV} \ .
\le{eosr} \eeq \cap This well-known result indicates that the energy of a free fermion gas increases monotonically
with the density. If nuclear matter must be stable in mechanical equilibrium at a density $\rho = \rho_0 \approx
0.16 fm^{-3}$, the so-called saturation density, a net attractive potential energy must be present around this
density. This attraction, coming from the nucleon--nucleon interaction, must produce a minimum in the EOS, namely
in the curve $e = e(\rho)$, at $\rho = \rho_0$. This requirement originates from the phenomenological observation
that the central density of medium and heavy nuclei (as extracted from e.g. electron scattering data) is pretty
constant along the nuclear mass table and close to the above mentioned value of $\rho_0$. This is interpreted as
being the mechanical equilibrium density of nuclear matter and it is the starting point for the development of the
empirical mass formula in its different versions. The latter is discussed at the beginning of the next section.
Before doing that, some considerations on the free gas model and some of its applications will be discussed.
\subsection{The incompressibility}
Another way of presenting the free gas EOS (at zero temperature) is to consider the pressure
\beq
\bar{rl}
 p\!\!\! &\egu -\left({dE\over dV}\right)_N \egu {\rho^2 \over N}
\left({dE\over d\rho}\right)_N \egu \rho^2\left({de\over d\rho}\right)
\egu\\
   &             \\
 &\egu {2\over 5} \hstm \left({{3\pi^2}\over 2}\right)^{{2\over 3}}
 \rho^{{5\over 3}}\, \approx \, 50.02 \,\, \rho^{{5\over 3}} \ \
 {\rm MeV} fm^{-3} \  ,
\ear
\le{pre}
\eeq
\cap which can be considered the ``Pauli pressure'', namely the pressure due to the exclusion principle, a typical
quantal effect. From the pressure, the incompressibility $K_0$ can be derived according to the usual definition
\beq \bar{rl} K_0\!\!\! &\egu -V\left({dp\over dV}\right) \egu
\rho\left({dp\over d\rho}\right) \egu \\
      &             \\
    &\egu {2\over 3} \hstm \left({{3\pi^2}\over 2}\right)^{{2\over 3}}
    \rho^{{5\over 3}} \, \approx 83.36 \,\, \rho^{{5\over 3}} \ \ .
\ear \le{com}
\eeq
\cap The definition of equation (\ref{eq:com}) is in agreement with the usual one adopted in textbooks on basic
mechanics and thermodynamics. For practical reasons, it is more customary among nuclear physicists to use the
alternative definition \beq K \egu k_F^2\left({d^2e\over dk_F^2}\right)  \ \ , \le{comn} \eeq \cap which has the
dimension of an energy. The following relationship can be easily checked \beq K_0 \egu {4\over 3}p \, +\, {1\over
9}\rho K  \ \ . \le{comr} \eeq \cap Equation (\re{comr}) is not restricted to the free gas model, but it is valid
in general. At saturation $p = 0$, and the two definitions have a very simple connection. If one drops the first
term on the right hand side of equation (\re{comr}) and adopts for $K_0$ the free gas value given by equation
(\re{com}), one gets \beq K \egu 6 E_F \, \approx\, 221 \ \ {\rm MeV}  \ , \le{comf} \eeq \cap where the numerical
value is taken at $\rho \approx \rho_0$. This value is close to the values obtained in several phenomenological
analysis of the data on the monopole frequency in heavy nuclei \cite{monop}. It is appreciably lower than the
value of $ 240 $ MeV obtained in reference \cite{Swi} on the basis of a Skyrme force fit to the properties of a
wide set of medium-heavy nuclei. This approximate agreement must be considered essentially fortuitous. In fact,
the monopole frequency is determined by the mechanical incompressibility $ K_0$, but for the free Fermi gas the
pressure term $4/3 p$ of equation (\re{comr}) is quite large at $\rho = \rho_0$. Therefore, the procedure we
followed to extract $K$ is clearly inconsistent. The agreement is the result of some ``compensation of errors''.
Of course, one can always {\it define} the incompressibility as $\,\, K' \egu 9 K_0/\rho\,\, $ instead of equation
(\re{comn}) for all densities, in which case for a free Fermi gas indeed $K' \egu 6E_F$. Anyhow, the connection
between monopole frequency and incompressibility is less obvious than at first sight \cite{Blaizot}.
\subsection{Momentum distribution}
The ground state of the free fermion gas is characterized by the filling of the lowest single particle levels,
i.e. the occupation number of the states $k$ is one below the Fermi momentum $k_F$ and zero above, as indicated in
figure (1a).
\begin{figure} [ht]
 \begin{center}
\vskip 0.5 cm
\includegraphics[bb= 140 0 300 790,angle=90,scale=0.4]{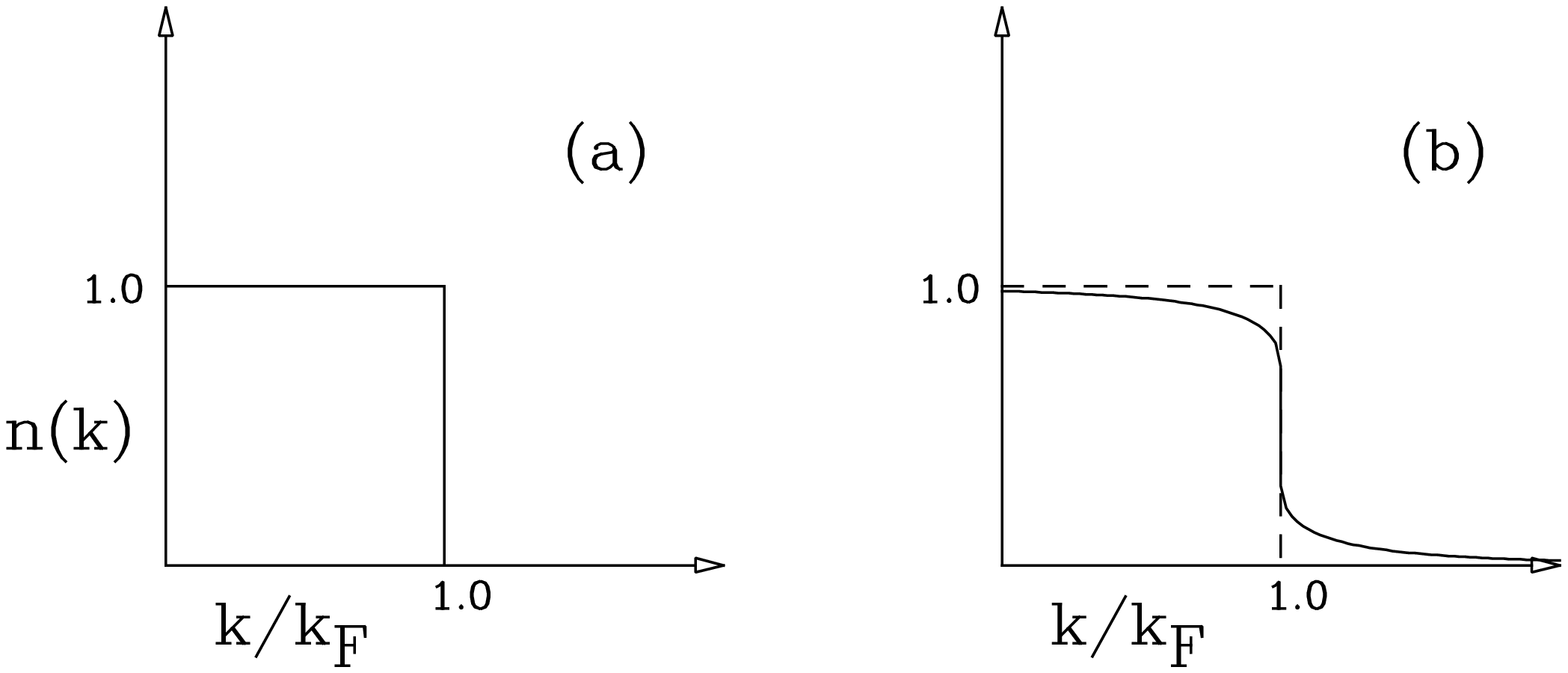}
\end{center}
\vspace{0.3 cm} \caption{ Schematic representation of the momentum distributions
   in a free fermion gas (a) and in an interacting fermion gas (b).}
    \label{Fig1}
\end{figure}
This picture is expected to be modified by the nucleon--nucleon interaction, as shown in figure (1b). Here the
discontinuity at the Fermi energy is assumed to persist despite the nucleon--nucleon correlations. The Fermi
liquids that have this property are called ``normal'' Fermi liquid. The deviation of the discontinuity from one is
a measure of the strength of the correlations. The persistency of the discontinuity at $k_F$ is the basis of the
Landau theory of Fermi liquid and of the concept of quasi-particle \cite{Mig}, to be discussed in section
\ref{landau}. If nuclear matter is superfluid, as it appears to be in a range of density, the discontinuity
disappears. Apart from the possible onset of superfluidity, which affects only weakly the gross properties of the
EOS, nuclear matter appears to be a normal Fermi liquid. Superfluidity changes of course dramatically the
transport properties of nuclear matter.
\subsection{The symmetry energy}
 If the proton number $N_p$ is different from the neutron number $N_n$, with $\, N = N_n + N_p\, $, then the
neutron and proton Fermi momenta are different, since the neutron and proton densities are different. Accordingly,
the EOS of equation (\re{eos1}) has to be generalized. Defining \beq
   \beta \egu {{N_n - N_p}\over {N_n + N_p}} \egu
    {{\rho_n - \rho_p}\over \rho}
\le{as} \eeq \cap as the ``asymmetry'' parameter, one easily gets \beq \bar{rl}
 E\!\!\! &\egu  E_n \, +\, E_p \egu N_p {3\over 5} E_F^{(p)} \plus
 N_n {3\over 5}E_F^{(n)} \\
   &              \\
 e\!\!\! &\egu  {E\over N} \egu {3\over 10} \hstm
 \left( {{3\pi^2}\over 2}\right)
^{{2\over 3}} \rho^{{2\over 3}} \left[ (1 + \beta)^{5\over 3}
\plus (1 - \beta)^{5\over 3} \right] \\
   &             \\
 \phantom{e}& \, \approx\, e(\beta = 0) \plus a_{sy} \beta^2 \plus
 \cdots\cdots \\
   &             \\
a_{sy}\!\!\! &\egu {1\over 3} E_F  \ \ \ \ . \ear \le{enas} \eeq \cap Thus, for a fixed value of the total
density $\rho$, the energy per particle $e$ has a minimum at $\beta = 0$. The coefficient $a_{sy}$ is called the
symmetry energy. At $\rho \approx \rho_0$, one finds $a_{sy} \approx 12$ MeV. From the systematics on the
asymmetry dependence of the binding energy of medium--heavy nuclei, $a_{sy}$ turns out to be more than twice
larger than this value. Here the interaction must play a major role. The density dependence of $a_{sy}$ is one
of the most relevant issue in nuclear astrophysics, but also in nuclear structure.
\subsection{The single particle density of states}
 For many physical phenomena the single particle density of states at the Fermi energy is a relevant quantity.
For the free gas model one can readily get an explicit expression \beq \bar{rl} D(E_F)\!\!\! &\egu \sum_k
\delta\left(E_F - {{\hbar^2 \v{k}^2}\over 2m} \right) \egu {V \cdot g\over (2\pi)^3}\int d^3 k
\delta\left(E_F - {{\hbar^2 \v{k}^2}\over 2m} \right)\nonumber \egu \\
      &               \\
 &\egu {N\over \rho} {g\over \tpc} 4\pi {m\over \hbar^2} k_F
 \egu {3 N\over 2 E_F} \, \approx\, {N\over 23} \ \ {\rm MeV}^{-1}  \ \ ,
\ear \le{ds} \eeq \cap where the last equality holds at $\rho \approx \rho_0$. This elementary result is expected
to be modified by the presence of the interaction \cite{BM}. The effect of the nucleon--nucleon correlations on
$D(E_F)$ can be introduced by substituting the free nucleon mass with the so called nucleon ``effective mass",
which will be shortly discussed in the section on Landau theory. Another related useful quantity is the single
particle level density per unit volume, that for symmetric matter can be written
 \beq
d(E_F) \egu D(E_F)/V \egu {2m\over \pi^2\hbar^2} k_F \ \ \le{dsv} \eeq \cap It depends only on the nuclear
matter density.

\subsection{Other microscopic physical quantities}

In order to characterize the properties of the nuclear medium other quantities are necessary. First of all the
nuclear surface properties are characterized by the values of surface thickness and the surface tension. Bulk and
shear viscosity are essential to describe the macroscopic dynamics of Neutron Stars. They are dominated by the
nucleon-nucleon interaction, and therefore they will be discussed after the correlations among nucleons will be
introduced and discussed. In finite nuclei viscosity must to be treated in a different scheme than in nuclear
matter, since the presence of the nuclear surface plays a major role. This issue will be also discussed in the
section on viscosity.

\section{Basic phenomenology \label{pheno}}
\subsection{Mass formula and saturation \label{mass}}

Some of the basic phenomenological data on the nuclear medium come from the semi-empirical mass formula
\cite{We,BB}. The aim of the mass formula is to express the total binding energy $B(A,Z)$ of a nucleus as a
smooth function of the mass number $A$ and the atomic number $Z$. Several versions of the formula exist. In any
case, the physical basis is the Liquid Drop Model or the so-called Droplet Model. In these models, the nucleus
is described as a drop of a quantal liquid, the nuclear medium, whose properties are derived as for a classical
liquid, with the addition of some quantal corrections, typical of the nuclear systems. A set of parameters are
introduced, some macroscopic in character, some other more connected to a Fermi liquid behavior.
The refined versions of the purely phenomenological mass formula contain several terms and can be written
\beq
%
%
\hskip -2cm   B(A,Z) \egu a_V A \plus a_S A^{2\over 3} \plus (a_I + a_{IS}/A^{1\over 3})
 ({N - Z\over A})^2 \plus a_C {Z^2 \over A^{1\over 3}}
 \minus \delta_P  \plus E_D
%
\label{eq:massf} \eeq
 \cap which, in the written order, contains the bulk contribution (parameter $a_V$), the surface
correction ($a_S$), the bulk and surface symmetry energies ($a_I$ and $a_{IS}$ respectively), the coulomb energy
($a_C$), the pairing energy ($\delta_P$),  to be discussed in detail in section \ref{super},
 and the deformation energy ($E_D$). The overall trend of the empirical
binding energy of nuclei and the way it can be reproduced by this simple formula, by adjusting the set of
parameters $a$, are discussed in basic books \cite{RS}, where the meaning and possible forms of the different
terms are discussed in more detail. The values of the parameters depend slightly on the particular form used for
$\delta_P$ and $E_D$ \cite{MS66,My69,MS69}.
The value of the bulk energy $ a_V$ in all cases is very close to $- 16 $ MeV. This formula provides an excellent
fit to the smooth part of the binding energy of nuclei throughout the nuclear mass table with few parameters. This
fact supports the interpretation of each term as schematically indicated above. A partial justification of the
mass formula can be obtained within the semi-classical scheme of approximation. In fact, it is possible to show
\cite{RS} that the smooth function $B(A,Z)$ can be considered the first term of the expansion in $\hbar$ of a mean
field estimate of the nuclear binding energy. The deviations, which are actually in percentage very small, are
therefore interpreted as ``shell corrections'' \cite{BM}, i.e. corrections coming from the quantal effects related
to the finite size of nuclei. Systematic methods to estimate these effects have been devised by many authors, in
particular by Strutinsky \cite{Strut}. They will be discussed in the section \ref{TF}.
\par
The very fact that in the fitting procedure a constant term $a_V$ can be well identified as one of the relevant
term indicates that this term can be indeed interpreted most naturally as the bulk part of the binding energy,
namely the energy per particle of the infinite symmetric nuclear matter. This can be also seen if one extrapolates
the formula for $A\rightarrow \infty$, provided $N = Z$ and the Coulomb energy is neglected. Then, in this case,
only the first term survives. Similarly the coefficient $a_I$ can be identified with the nuclear matter symmetry
energy per particle. However, it has been argued recently \cite{Dani1,Dani2} that these parameters could reach the
asymptotic values only at exceedingly large value of the mass number $A$, and therefore along the nuclear mass
table they still contain a smooth dependence on the mass and atomic number.
\par
The Droplet Model \cite{Nix1,Nix2} includes additional contribution with respect to the Liquid Drop Model, in
particular a curvature term and a term taking into account the possibility of a slight compression of the nuclear
medium in the nucleus. \par
The appealing physical feature of these models is the direct relationship between each parameter and a definite
property of the nuclear medium. In principle, the phenomenological analysis based on these models can provide
basic physical quantities which characterize the nuclear medium, both in its homogeneous macroscopic phase and
in finite nuclei.
\par
One has to mention a similar approach, the so called microscopic-macroscopic models \cite{Swiate}. In this case
some hints are taken from a more microscopic treatment of the binding energy, like the Thomas-Fermi, see section
\ref{TF}.
\par
Of course the mass formula contains information only at the saturation density $\rho_0$, and therefore the
knowledge of the complete EOS goes well beyond the content of equation (\re{massf}). As mentioned briefly in the
introduction, information on the EOS at $\rho \neq \rho_0$, finite temperature and large asymmetry are expected to
come from heavy ion collision experiments and astrophysical observations.
\par
Finally, one has to mention that the notion itself of saturation is coming also from the observation that the
central density of medium-heavy nuclei is pretty constant throughout the nuclear chart. This fundamental
phenomenological result has been obtained mainly from elastic electron scattering, which provides the whole charge
distribution in nuclei. The total density is then obtained by assuming that the neutron density scales as $N/Z$
with respect to the proton density. For a recent analysis see references \cite{elect1,elect2}. The value of the
density is around $0.16$ fm$^{-3}$ and it is interpreted as the density at which symmetric nuclear matter displays
its minimal energy (saturation point). Till now asymmetry effects are within the overall phenomenological
uncertainty on the saturation point.

\subsection{Giant Resonances in nuclei \label{giant}}

If nuclei are viewed as droplets of nuclear matter, it is natural to consider the possibility of their
excitations. The quantization of these modes correspond to collective excitations of the nucleus as a whole. This
is the physical basis of the Bohr-Mottelson model \cite{BM} for the nuclear modes of excitation. These collective
modes have found a clear and extensive experimental evidence \cite{BM,RS}. The vibrational excitations are
classified according to the multipolarity of the surface oscillations and their isospin character, i.e. if
neutrons and protons oscillate with the same or opposite phase (for simplicity we neglect spin flip). They are
universally called "Giant Resonances", since they usually carry a large fraction of the total strength of the
corresponding spectral function.\par The simplest oscillation is the isoscalar monopole vibration, corresponding
to a compressional mode of the nucleus. The question that arises naturally is then if it is possible to extract
from the study of the monopole excitation, in particular from its energy, the compression modulus of nuclear
matter at saturation. This possibility has been explored extensively since many years \cite{Blaizot,RS}. From a
purely macroscopic point of view there are essentially two difficulties along this line : a) To calculate the
excitation energy, as for an harmonic oscillator, not only the incompressibility is needed, giving the restoring
force, but also the dynamical mass that should be used, and b) the surface tension of the nucleus should play some
role, but there is not any obvious relationship between surface tension and incompressibility. As for point a),
microscopically there is a general method to estimate the collective mass of an excitation, the so-called
"crancking mass" \cite{BM,RS}. At macroscopic level it can be estimated assuming a particular velocity flow, in
particular the scaling hypothesis implies a radial velocity proportional to the radius, in which case the inertial
parameter has an analytic expression and it is proportional to the radius square of the nucleus
\cite{Blaizot,Xavier}. It is difficult to handle point b) with a satisfactory accuracy at macroscopic level, and
it is necessary to introduce some microscopic elements in the theory. The most successful semi-microscopic method
is the Skyrme functional method, to be discussed with some detail in section \ref{EDF}. According to this well
known method, an effective nucleon-nucleon interaction is introduced and the energy of the nucleus is assumed to
be equal to the Hartree or Hartree-Fock energy calculated with such a force, i.e. minimizing the mean field energy
functional calculated with the force. The effective force is semi-phenomenological in character and therefore it
contains few parameters. With the same force it is possible to calculate also the nuclear matter EoS, and to tune
the parameters in order to obtain the correct saturation point and a given incompressibility modulus. In this way
it is possible to check if a correlation exist between the energy of the isoscalar monopole vibration and the
nuclear matter incompressibility. All the parameters are fitted in any case to reproduce the binding energy of a
large set of nuclei, as well as other phenomenological data. For an extensive application of this method, see e.g.
reference \cite{Chamel1,Chamel2}. One finds indeed that a correlation exists between incompressibility and
position of the monopole Giant Resonance, so that, in principle it is possible to extract from the experimental
data the value of the incompressibility in nuclear matter. For the method to be reliable, the result should be
essentially independent of the particular Skyrme force used in the calculations. Unfortunately this is not the
case. It was shown more recently \cite{Colo} that the relationship between the centroid of the monopole excitation
and the value of the incompressibility is not unique, but it depends also on other details of the force, mainly it
is correlated also to the density dependence of the energy density and symmetry energy of the force
\cite{Colo,Sharma}. This also explains, at least partially, the reason why the incompressibility extracted from
relativistic mean field functional tends to be systematically higher than the one extracted for non-relativistic
Skyrme functional. At present, the constraints on the value of the nuclear matter incompressibility from the
monopole excitation are not so tight. It is fair to say that it can be approximately constrained between 210 and
250 MeV. More refined value can be expected to come out in the near future from additional analysis of
phenomenological data.\par The prototype of Giant Resonance is surely the dipole mode, where neutron and proton
oscillates against each other. The restoring force in this case is the symmetry energy. Both volume and surface
contribution can be present. In fact, recent analysis \cite{Trippa,Xavier2} on the correlation between dipole
resonance energy and symmetry energy indicates that such correlation can be obtained if the symmetry energy is
taken at $0.1$ fm $^{-3}$, about $2/3$ the saturation density. In any case it is difficult to get a strong
constraint on the nuclear matter symmetry energy at saturation from the Giant Dipole Resonance. \par The isoscalar
quadrupole mode is more connected with the surface tension in nuclei, since, in first approximation, the mode
occurs at constant volume. However, this correlation has not been explored, probably because in this case it is
more difficult to estimate the collective mass term.
\par The other Giant Resonances do not involve only one or few characteristics of the effective forces, and
therefore they can hardly be used to study definite physical properties of the nuclear medium.
\par
Finally it has to be mentioned the study of the Giant Resonance damping, which is measured by their width. From a
macroscopic point of view such a damping should be connected to some sort of viscosity of the nuclear medium.
Unfortunately the physical situation is much more complicated. First of all, one should take into account that we
are dealing with a quantal liquid, as discussed in section \ref{landau}. Therefore Giant Resonances actually
should be considered as zero sound modes, and, in principle, no hydrodynamical picture should be adopted.
Furthermore, the presence of the nuclear surface introduces a different type of damping, the so called one-body
dissipation \cite{Bloc1,Bloc2,Jar,Bloc3}. The applicability of such damping mechanism requires the onset of a
certain degree of chaos in the single particle dynamics \cite{chaos}, and therefore it seems not suited to Giant
Resonance of low multipolarity. Probably the octupole vibration can be partly affected by such a type of
dissipation. Finally the damping can be produced by the emission of nucleons, the so called decay damping. At
least for all these reasons the extraction from the width of any sort of viscosity is strongly hampered, and the
study of the Giant Resonance width must rely completely on nuclear structure analysis \cite{PierF,Dussel}. Shear
and bulk viscosity in nuclear matter, as present in Neutron Stars, must be predicted only on purely theoretical
basis when microscopic models of astrophysical phenomena are developed.

\subsection{Heavy ions \label{ions}}

In a period that includes at least the last two decades intensive studies of heavy ion reactions at energies
ranging from few tens to several hundreds MeV per nucleon (hereafter indicated as MeV/A) have been performed in
different laboratories throughout the world. One of the main goal, probably the principal one, has been the
extraction from the data on suitable observable quantities the gross properties of the nuclear Equation of State.
An enormous literature exists on the subject, and therefore we will focus on few items that, according to our
personal view, are connected with established and insightful results.

\subsubsection{Flows and differential flows}
\par
It can be expected that in heavy ion collisions at large enough energy nuclear matter is compressed and that, at
the same time, the two partners of the collisions produce flows of matter. In principle the dynamics of the
collisions should be connected with the nuclear medium EoS and its viscosity. \par However at low enough energy
the cross section is dominated by deep inelastic processes, where target and projectile keep their identity during
the collision, stick  together for a while and separate again. This reaction mechanism persists up to about 10
MeV/A. At increasing energy the so called ''multifragmentation" regime is encountered, where after the collision
numerous nucleons and fragments of different sizes are emitted. Usually, at non-central collisions, one
distinguishes target-like and project-like fragments, the so called spectators, and the participant region, where
matter is partly stopped and tends to form a partly equilibrated zone. In semi-classical simulations of heavy ion
collisions two main ingredients are introduced, the single particle mean field $U$ and the in-medium NN scattering
cross section ${d\sigma \over \Omega}$. A Boltzmann-like kinetic equation is assumed for the nucleons
\beq
 {\partial f \over \partial t} \, +\, {\vec \nabla}_p\epsilon \cdot {\vec \nabla}_r f \, -\,
                                  {\vec \nabla}_r\epsilon \cdot {\vec \nabla}_p f \, =\,
                                  I\Big\{{d\sigma \over \Omega}\Big\}
\label{eq:BUU} \eeq
\noindent where $n = n(\mathbf{r},\mathbf{p},t)$ is the single particle density distribution in phase space,
$\epsilon = p^2/2M \, +\, U(\mathbf{r},\mathbf{p},t)$ the local single particle energy and $I$ the two-body
collision integral that describes the loss and gain of particles, at a given phase space point, due to scattering
of nucleons in the medium. The single particle potential $U$ can be written
\beq
U(\mathbf{r},\mathbf{p},t) \, =\, \int d^3r'd^3p'\, v_{eff}(\mathbf{r},\mathbf{p};\mathbf{r}',\mathbf{p}')
                                     n(\mathbf{r}',\mathbf{p}',t)
\label{eq:Upot} \eeq
\noindent where $v_{eff}$ is the effective NN interaction in the medium. In general $U$ is identified with the
single particle potential in nuclear matter at the local density, e.g. the Brueckner potential. \par In practice
it is not possible to get directly from the data indications on the EoS and the scattering cross section in the
medium. Even if $4\pi$ detectors, with which is possible an (almost) complete reconstruction of the collision
dynamics, have been developed, the interpretation of the data is not unique. The usual procedure is to assume a
set of possibility for the potential $U$ and, more rarely, for $\, d\sigma / \Omega\,$, and to find which ones in
the considered sets fit better the data. Once $U$ is chosen, the EoS can be calculated, since the single particle
potential fixes the interaction energy per particle. Analogously, the scattering cross section determines the
transport properties of the nuclear medium. However, very often the NN scattering processes have only the effect
of driving the system toward equilibrium, or quasi-equilibrium, at least in the participant zone, so that the
information on the cross section is often too indirect to be accessible. In any case it is true that the results
of the simulations depend in general on both quantities. \par
One of the quantity that is more often analyzed is the so called transverse momentum, also in its differential
form. If the reaction plane is the $(x,y)$ plane and  the initial direction of the two colliding nuclei is along
the $y$ axis, one calculates the average momentum $p^x$ along the $x$-axis of the nucleons as a function of
their velocity $y$ (or ''rapidity") along the $y$-axis
\beq
F(y) \, =\, <p^x>_y \ \ \ \ \ ; \ \ \ \ \ F'(y) \, =\, d <p^x>_y /dy
\label{eq:flow} \eeq
At high enough energy the flow is strongly affected by the matter compression during the collision and dominated
by the corresponding pressure. Then the initial flow undergoes a strong repulsion from the interaction zone, which
means that $F(y)$ turns from negative to positive values as $y$ changes sign, with a well defined slope. In fact
the negative and positive values of $y$ label the target-like and projectile-like fragments or nucleons, at least
for $y$ around zero (''mid-rapidity" region). It was hoped that the slope $F'(0)$ could characterize sharply the
nuclear matter EoS. Unfortunately in the simulations only a weak dependence on the EoS stiffness is observed,
somehow obscured by the numerical uncertainty of the simulations themselves \cite{Aichelin}. This is probably due
to the competing effect of the NN collisions incorporated in the collision integral $I$. For the same reason, it
is difficult to extract any solid information on the in-medium cross section. To complicate further the situation,
different simulation methods, like the Quantum Molecular Dynamics \cite{Ono,Hart,Maru,AichPR}, give slightly
different results in the same physical conditions, which increases the uncertainty on the EoS that can be
extracted.\par
\begin{figure} [ht]
 \begin{center}
\vskip 0.7 cm
\includegraphics[bb= 140 0 300 790,angle=270,scale=0.35]{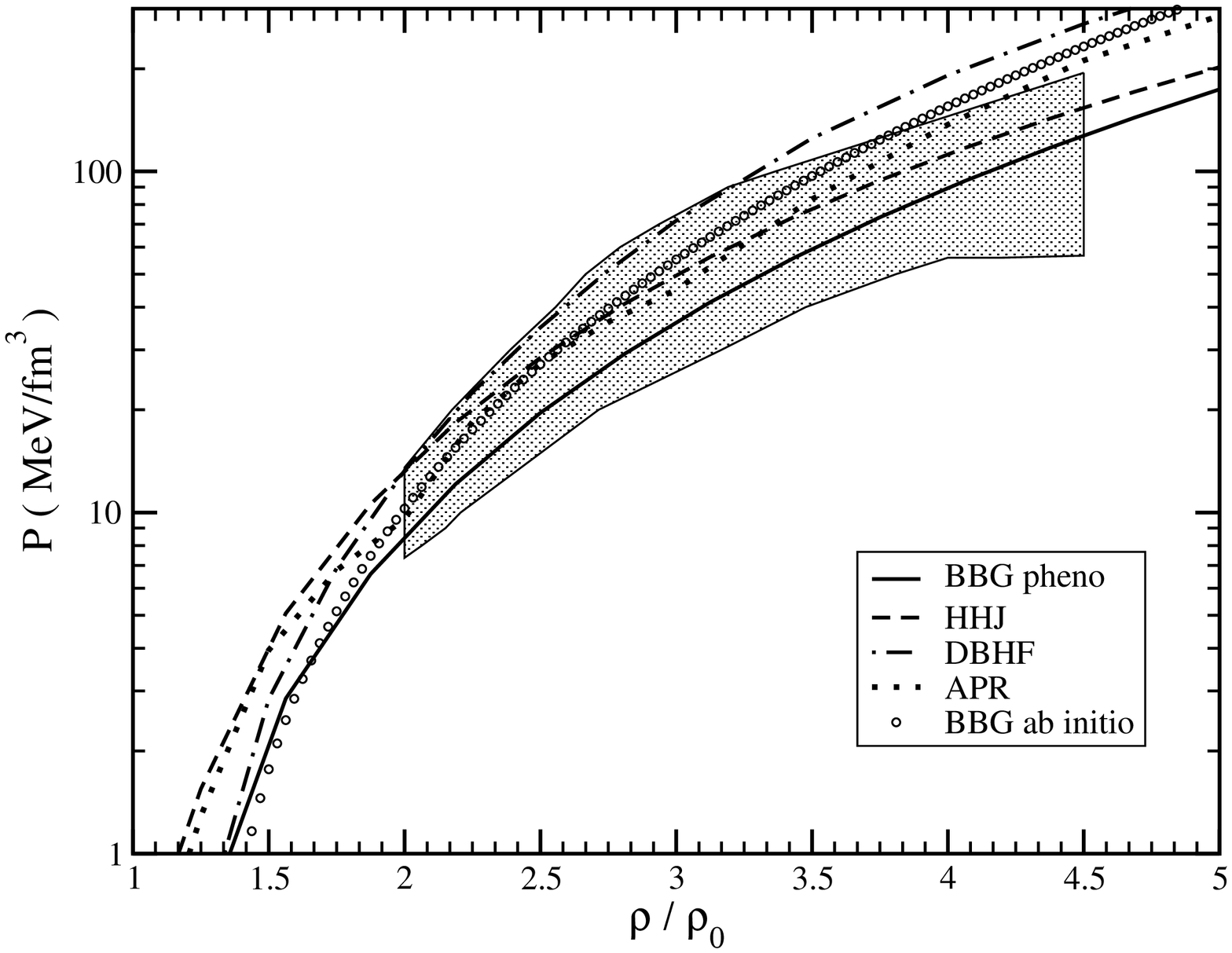}
\end{center}
\vspace{2.8 cm} \caption{Different EoS in comparison with the phenomenological constraint
    extracted by Danielewicz et al. (shaded area), where $\rho_0 = 0.16 $
    fm$^{-3}$.
    Full line: EoS from
    the BBG method with phenomenological TBF \cite{fiorella}. Dashed line : modified
    variational EoS of Heiselberg and Hjorth--Jensen \cite{martino}. Dotted line : variational
    EoS of Akmal et al. \cite{AP}. Open circle : EoS from the BBG method with ``ab initio"
    TBF \cite{fiorella}. Dash--dotted line : EoS from Dirac--Brueckner method
    (van Dalen et al. \cite{dalen}).   }
    \label{fig:DL}
\end{figure}

 Despite all these difficulties, some gross constraints on the nuclear EoS can be extracted. In
reference \cite{DL} this effort was summarized by plotting the region where any reasonable EoS should pass through
in the pressure vs. density plane. The pressure was taken at the center of the interaction zone at the moment of
maximal density during the collision, using all the simulations with different EoS compatible with different data
and including the uncertainties. The plot is reproduced in figure (\ref{fig:DL}), as taken from reference
\cite{jpg}, in comparison with some microscopic calculations to be discussed in section \ref{micro}. In particular
the EOS labeled BBG is the one calculated win the Brueckner approximation including three-body forces, as
presented in detail in section \ref{micro}, that looks in agreement with the data in the full density range. It
has to be stressed that the EoS employed in such a constructions have incompressibility that ranges from $K = 167$
MeV to $K = 380$ MeV. Only the densities above twice the saturation density were included, since this guarantees
that quasi-equilibrium in the participant zone was reached. This shows that only the EoS at high density can be
studied in heavy ion collisions. The values of the incompressibility do not characterize completely the EoS, since
it is actually density dependent, but in any case the analysis indicates the broad constraints on the EoS that can
be obtained from heavy ion collisions. Despite they are somehow a little loose, they are able to exclude some of
the phenomenological EoS \cite{David}.

\subsubsection{K-meson production}

Strange particles production in heavy ion collisions can probe the central part of the participant zone. In fact
near threshold strange particles are mainly produced in the high density region and, once produced, they interact
weakly with the matter. This is due to strangeness conservation in reactions produced by strong force, which
implies that strange particles are always produced in pairs and they cannot be directly re-absorbed by nucleons.
However this does not necessarily mean that the interaction can be completely neglected. In particular, let us
consider the lightest strange particle, the $K$-meson (kaon). In this case one has to distinguish between $K^-$
and $K^+$, since the negative kaon forms resonances with nucleons even at low energy and therefore their
interaction with the nuclear medium cannot be neglected. Unfortunately the interaction potential felt by $K^-$ in
the medium is not so well known theoretically, and this introduces uncertainty in the analysis of the experimental
data. The situation for $K^+$ is different, since no resonance with nucleons is present and the interaction can be
treated almost perturbatively, and actually the uncertainty is much reduced. The main mechanism of $K^+$
production is through the excitations of a nucleon to a $\Delta$, that in turn decays in a $\Lambda$ and a $K^+$
\beq
  NN \, \longrightarrow\, N \Delta \, \longrightarrow\, N \Lambda K^+
\eeq
\noindent It is then clear that the simulations must include nucleon excitations and must be relativistic. The
uncertainty is mainly due to the not so well known potential of the $\Delta$ in the nuclear medium. Fortunately
this does not affect too much the final results, which look to be under control also numerically and almost
independent on the simulation method. An excellent and extensive review of the subject, both at experimental and
theoretical level, can be found in \cite{Fuchs,Aichelin}. Here we restrict to some of the conclusions that can
be drawn from this line of research, that was developed along several years.\par
\begin{figure}  [h]
 \begin{center}
\vskip -6.8 cm
\includegraphics[bb= 280 0 300 790,angle=0,scale=0.5]{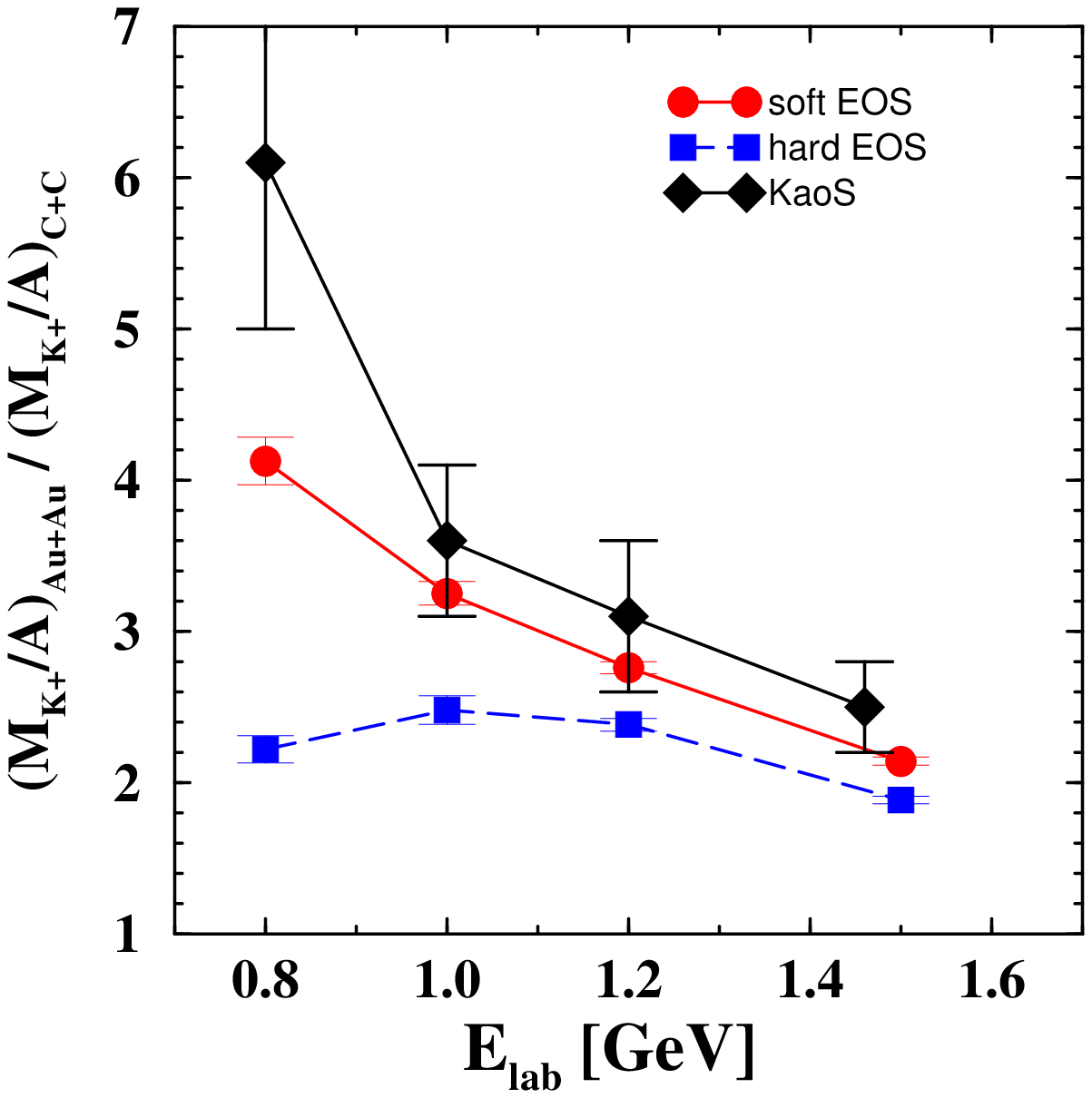}
\end{center}
\vspace{-1.7 cm} \caption{Excitation function of the $K^+$ multiplicity ratio between inclusive Au + Au over C+C
reactions. The simulations are performed with hard/soft nuclear EoS and compared with the data from the KaoS
collaboration \cite{Kaosdata} }
    \label{fig:kplus}
\end{figure}

The optimal energy for this type of investigation is close or even below two-body threshold, since then the only
way to produce the kaons is by compression of the matter. Since at threshold the production rate increases
steeply, there is a strong sensitivity to the value of the maximum density reached during the collision, and this
is an ideal situation for studying the EoS and its incompressibility.
 The comparison of the simulations with the experimental data on $K^+$ production, noticeably the ones from the
KaoS \cite{KaoS} and FOPI \cite{FOPI} collaborations, points in the direction of a soft EoS. More precisely, the
interval of compatible incompressibility is narrower than the ones obtained from the analysis of flows
\beq
 180 \ \  \leq \ \  K \ \  \leq \ \ \ {\rm 250 \ \ MeV}
\eeq
\noindent These values are compatible with the ones obtained from the monopole oscillations, see Section
\ref{pheno}. However it has to be kept in mind that, in the simulations, kaon production occurs at density about
2-3 times larger than saturation, $\rho \geq 2-3\rho_0$, and therefore the two sets cannot be fairly compared. In
any case, a stiff EoS above saturation seems to be excluded from this analysis, as it is apparent from figure
(\ref{fig:kplus}), taken from reference \cite{Fuchs}.
\subsection{Astrophysics \label{astro} }
The nuclear medium is directly and massively involved in core-collapse supernova explosions, where nuclear matter
is compressed at supra-saturation density and trigger the shock wave that is the main agent of the outflow. The
properties of nuclear matter determine completely the structure of Neutron Stars and the phenomena that occur in
their interior or at the surface. Indirectly, the peculiarities of the nuclear medium are relevant for many other
processes, like nucleosynthesis.

\subsubsection{Supernovae}
One of the major puzzle in Astrophysics is the mechanism that drives the explosion of core-collapse Supernovae. In
the complex simulations of the after-bounce stage of the supernovae the shock wave is stalling due to the energy
loss, mainly produced by the disintegration of nuclei in the envelope. It is common wisdom that only the revival
of the shock by the blast of neutrinos, initially trapped, can produce the final explosion \cite{Wilson}. For a
long time it was believed that a direct link should exist between the possibility of explosion and the value of
the incompressibility in nuclear matter. Many years of effort on supernova event simulations have disproved this
believing and the difficulty of getting a real explosion in the computer has indicated that only a detailed
treatment of the different aspects of the phenomenon could lead to a definite solution of the puzzle
\cite{Janka2007}. It seems now \cite{Burrows,Janka2010} that the key ingredient is the 3-dimensional character of
the process, that makes more tiny the structure of the turbulent flow of matter and renders more efficient (with
respect to a 2-dimensional situation) the energy deposition of neutrinos on the matter close to the shock. The
possibility of the explosion is therefore not determined by the details of the nuclear EoS. However many
quantitative aspects of the explosion do depend on the EoS, like the total energy release, neutrino luminosity and
the timing of the whole phenomenon as determined by the neutrino mean free path (for which nucleon correlations in
nuclear matter is essential). Unfortunately we are not yet at a stage where detailed quantitative predictions and
comparisons can be made in relation to phenomenology because of the enormous complexity of the supernova
explosion.
\par
It has to be kept in mind that the nuclear medium present in supernovae, as well as in Neutron Stars, is very
asymmetric and in different physical conditions than in heavy ion collisions. In particular the time scale is
different by several order of magnitude, so that  many processes that the nuclear medium can undergo in
supernovae cannot occur in heavy ion reactions because the time is too short.
\subsubsection{Neutron Stars}
\begin{figure}
 \begin{center}
\vskip -11.6 cm
\includegraphics[bb= 200 0 360 790,angle=0,scale=0.8]{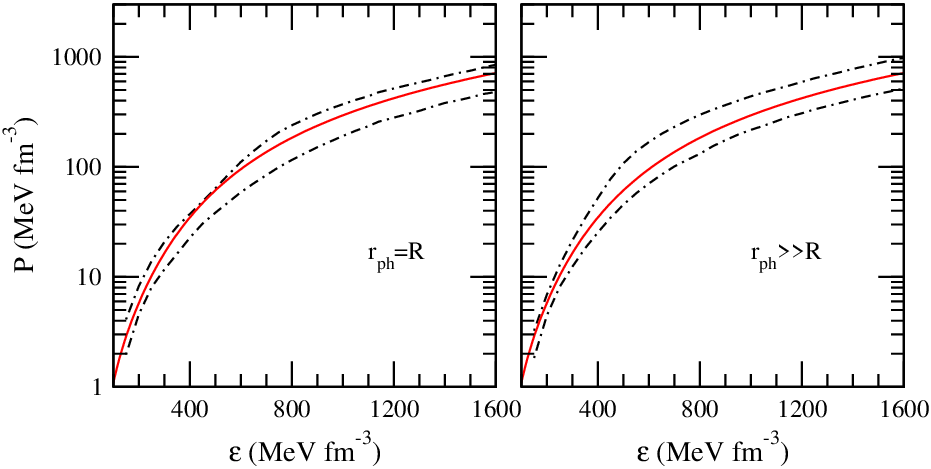}
\end{center}
\vspace{-2.5 cm} \caption{Comparison of the phenomenological allowed region (within the dotted-dashed lines) for
the Neutron Star matter EoS with the corresponding microscopic EoS from the BHF method (full line).
Phenomenological data are from reference \cite{baye}.}
    \label{fig:baye}
\end{figure}

The compact remnant of a supernovae explosion, if the collapse does not end in a black hole, is a Neutron Star,
(NS) an extremely dense object, that in the standard view is composed of nuclear matter, electron and muons.
Leptons are necessarily present because initially the star is of course neutral. The structure of the NS is
determined by the properties of the nuclear medium in an extremely wide range of density, from few times the
saturation one at the center, down to values several order of magnitude smaller close to the surface. An old
enough NS is virtually at zero temperature and its exterior part is actually formed by a Coulomb crystal of nuclei
\cite{Baym,Dima}. Below this outer crust, an inner crust is present, where nuclei are surrounded by a gas of
dripping neutrons. In this region the EoS of pure neutron matter at low density is relevant. At the same time
nuclei are very exotic, challenging our knowledge of the nuclear medium at large asymmetry. Below the crust
asymmetric homogeneous nuclear matter fills the whole space. Again the asymmetry is very large, and a direct link
with nuclear structure and heavy ion reactions in terrestrial laboratories is not possible. The challenge for
microscopic many-body theory is just to establish this link and try to test it by  comparison with phenomenology.
In this section we discuss the item of the NS maximum mass and few related issues, leaving other issues to
sections 6-9. A NS is bound by gravity, and it is kept in hydrostatic equilibrium only by the pressure produced by
the compressed nuclear matter. It is then apparent that the nuclear matter EoS is the main medium property that is
relevant in this case, as can be seen in the celebrated Tolman-Oppenheimer-Volkoff \cite{OV,Tol} equations, valid
for spherically symmetric NS
\be
{d P\over d r} \, &=\,& - G\, {\varepsilon\, m \over r^2} \left( 1 \, +\, {P \over \varepsilon} \right)
                                            \left( 1 \, +\, {4\pi P r^3\over m } \right)
                                            \Bigg( 1 \, -\, {2 G m \over r } \Bigg)^{-1} \\ \nonumber
\  &\ \ &          \\ \nonumber
{d m\over d r} \, &=\,& 4\pi r^2 \varepsilon \\
\label{eq:OV} \ee
\noindent where $G$ is the gravitational constant, $P$ the pressure, $\varepsilon$ the energy density, and $r$ the
(relativistic) radius coordinate. To close the equations we need the relation between pressure and density, $P \,
=\, P(\varepsilon)$, i.e. just the EoS. In the Newtonian limit the energy density is just the mass density and in
each parenthesis the second term is neglected, and we get the equations of hydrostatic equilibrium in
non-relativistic mechanics. The use of General Relativity (GR) is demanded by the strong gravitational field.
Integrating these equations one gets the mass and radius of the star for each central density. Typical values are
1-2 solar masses ($M_\odot$) and about 10 Km, respectively. This indicates the extremely high density of the
object. It turns out that the mass of the NS has a maximum value as a function of radius (or central density),
above which the star is unstable against collapse to a black hole. The value of the maximum mass depends on the
nuclear EoS, so that the observation of a mass higher than the maximum one allowed by a given EoS simply rules out
that EoS. Up to now the best microscopic EoS are compatible with the largest observed masses, that are close to
$1.7$ solar mass \cite{Yako}. It would be of course desirable to have some phenomenological data also on the
radius of NS. Unfortunately this is quite difficult, but some tentative analysis look promising \cite{Baym-O}. In
particular a recent analysis of the data on six NS based on Bayesian statistical framework \cite{baye} has led to
a tentative constraint on the nuclear EOS. Depending on the hypothesis made on the structure of the NS, the
results are slightly different. The overall allowed region where the EOS should lie is reported in figure
(\ref{fig:baye}), where the theoretical EoS from the BHF calculations, to be discussed in the next section, is
also reported.
The theoretical EoS appears to be compatible with the extracted observational constraints. It turns out that other
microscopic EoS do not show the same agreement, in particular the EoS of reference \cite{AP} looks too repulsive
at high density \cite{baye}. These boundaries obtained from astrophysical data are complementary to the ones
obtained from heavy ion reactions, see figure (\ref{fig:DL}) in the previous sub-section. In fact, in heavy ion
collisions the tested matter is essentially symmetric, while in NS the matter is highly asymmetric. Considered
together, the two types of constraints probe the density dependence of the symmetry energy.
\par
Unfortunately the theoretical situation for the EoS in NS and for the maximum mass is actually much more
complicated. In fact, in NS weak processes have time to develop and, if energetically convenient, they can produce
strange particles like hyperons, and then change the composition of the nuclear medium. This is clearly at
variance with what can happen in heavy ion reactions, where the collision time is short and the multiplicity of
strange particles is so small that a bulk strange matter cannot be formed. On the contrary in NS, at least above a
certain density, the difference of the neutron and proton chemical potentials is so high to overcome the mass
difference between hyperons and neutrons. This is indeed the case, according to microscopic calculations
\cite{Hans}. Above 2-3 times saturation density $\Sigma^{-}$ or $\Lambda$ hyperons appear. This soften so much the
EoS that the maximum mass becomes smaller than the most established NS mass \cite{Taylor,Yako}. This result seems
to be quite robust and not dependent on the not so well known hyperon-nucleon or hyperon-hyperon interaction
\cite{Bomb}. The only way out seems to be, up to now, a possible phase transition to quark matter. Indeed,
calculations on the basis of simple models (\cite{q1}-\cite{q5}) can result in a maximum mass that is (marginally)
compatible with the observed largest mass. Of course it could also be that the quark matter EoS is stiffer than
assumed in simple models \cite{Alford,Alford_N}, but in any case it seems that we are close to test our knowledge
on the QCD deconfined phase at high density. All that makes clear that the NS physics connected with the central
high density core is quite different from the ones in heavy ions, where in ultra-relativistic collisions at LHC
the deconfined QCD phase is tested at zero baryon density and high temperature. However it is a basic challenge to
the theory to be able to connect the transition to quark matter in this two extreme different physical situations.
Advances both in phenomenological observations and theoretical methods are needed.\par The maximum mass problem
clearly can lead far from the physics of "nuclear medium", at least as it is considered in traditional nuclear
physics. However the distinction between traditional nuclear physics and QCD physics is partly artificial, and
they should considered as the two complementary aspects of the same physical realm.\par Finally one has to observe
that an observation of a maximum mass of 2 solar masses or higher would be a real breakthrough of our knowledge on
high density nuclear medium, since it would question the simple models of quark matter. Recent observations
\cite{2solmass} on the pulsar of the binary system PSR J1614-2230 seem to indicate such a possibility, and as
anticipated in reference \cite{q1}, it would imply the necessity of a repulsive interaction in quark matter
\cite{Alford2010}.

\section{From the Nuclear Interaction to the Correlated Nuclear Medium \label{micro}}
The properties of the nuclear medium are determined or strongly affected microscopically by the features of the
nucleon-nucleon (NN) interaction. In particular, one of the main characteristics of the NN interaction is the
presence of a hard repulsive core, whose relevance can be hardly overlooked. Furthermore, any realistic NN
potential must include a complex structure of operators involving spin, isospin and orbital angular momentum. This
non-trivial structure is one of the main reasons that renders the microscopic many-body theory of nuclear matter
and nuclei so hard to be handled. Another characteristics of the NN interaction is the presence of a quasi-bound
state ($^1S_0$ channel) and a bound state ($^3S_1 - ^3D_1$ channel) in the s-wave. This peculiar feature is
probably unique in nature and strongly affects the structure of low density nuclear matter. At increasing density
the effect of the bound and quasi-bound states tends to be reduced and this indicates that  many properties of the
nuclear medium should change strongly with density. Furthermore at very low density we know that nuclear matter at
not too high temperature must form clusters, i.e. light nuclei, and this has a decisive role for the Neutron Star
or proto-neutron star crust, as well as for heavy ion collision processes.
In order to illustrate these fundamental features of the medium as a many Fermion system, in this section the NN
interaction is introduced on the basis of the meson-nucleon model of the strong interaction in the baryon sector,
and the many-body theory of nuclear matter is then schematically developed following the most established methods.
Each microscopic many-body theory has a particular scheme to treat the hard repulsive core, that, implicitly or
explicitly, introduces an effective NN interaction, more manageable than the original NN interaction.
\subsection{Sketch of the nucleon-nucleon interaction}
\begin{figure}
\vspace{-7 cm}
\begin{center}

\includegraphics[bb= 140 0 350 790,angle=0,scale=1.0]{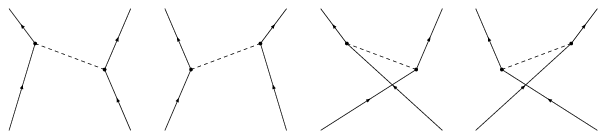}





\end{center}
\vspace{-18 cm}

\caption{The four possible meson exchange processes. Dashed lines indicate mesons, full lines nucleons.}
\label{fig:OB}
\end{figure}

The nucleon-nucleon interaction was intensively studied when Nuclear Physics started developing. Along the years
the phenomenological analysis has been more and more refined. Presently the phase shifts in different two-body
channels are known with high precision up to an energy of about 300 MeV in the laboratory, even if discrepancies
between the results of different groups still persist \cite{Machjpg}. For future reference in the paper we remind
very briefly the connection between the two-body interaction and the cross section, the quantity which is actually
measured. If one assumes that the interaction can be described by a static non-relativistic potential $v$, the
scattering process at the energy $E$ can be described by the $T$-matrix, that here we take with the standing wave
boundary conditions (often called $R$-matrix). It can be calculated solving the integral equation
\beq
 T(E)  \, =\, v \, +\, v {P \over {E \, -\, H_0 } } T(E)
\label{eq:Rmat}\eeq
\noindent where $H_0$ is the free kinetic energy hamiltonian and $P$ indicates the principal value for the
integral, which fixes the stationary wave boundary conditions. For a central potential, the phase shift in a
given channel $l$ is given by
\be
\tan \delta_l \, =\, -q \mu (q | T(E) |q)
\ee
\noindent where $q$ is the relative momentum and $\mu$ the reduced mass. The differential cross section is given
by
\be
 {d\sigma \over {d\Omega} } \, =\, {1 \over {q^2} } |\Sigma_l (2 l + 1) (e^{2i\delta_l } - 1) P_l (\theta) |^2
\ee
\noindent The nucleon-nucleon interaction, as we will see, contains not only a central interaction part but also
more complex operators, and the summation is extended to single and coupled channels $\alpha$, characterized by
the total angular momentum $J$, total spin $S$, total isospin $T$ and the orbital angular momenta $l,l'$ ($l = l'$
for single channel). Fitting the data on the cross sections at different energy, the phase shifts $\delta_\alpha$
for each channel can be extracted. For the NN interaction a particular form is assumed, as suggested by the
meson-nucleon theory of strong interaction, that contains several parameters that are fitted to reproduce the
phase shifts. In this way one can fix the nucleon-nucleon potential, which however partly remains model-dependent.
For details, see e.g. reference \cite{Machleidt2000}. Here we  sketch the main ideas of the meson theory of the
nucleon-nucleon interaction. For simplicity we use a non-relativistic treatment, as a first schematic introduction
to the theory of nuclear forces. Of course, the correct framework is the relativistic field theory of the
meson-nucleon system.
%
Let us consider the simplest possible case, the coupling of nucleons with a spinless neutral meson. If we
indicate by $b^\dagger{ }_q$ ($b_q$) the creation (annihilation) operator of a meson with momentum $q$, the
simplest coupling term is the scalar one
\begin{eqnarray}
 H_c& \, = \, G_s\int d^3x \psi^\dagger(x)\psi(x)\phi(x)
 \ \ \ \ \ \ \ \ \ \ \ \ \ \ \ \ \ \ \ \ & \ \nonumber\\
 & \ & \ \nonumber\\
\phantom{H_c}& = \,\,\,\, G_s{1\over \sqrt{V}}\sum_{k q} \sqrt{{\hbar\over 2\omega(q)}}
     (a^\dagger{}_k a_{k+q} b^\dagger{}_q + a^\dagger{}_k a_{k-q} b_q )
 & \ ,
\label{eq:mes}\end{eqnarray}
\cap which describes the processes of emission and absorption of a meson. The two processes are included together,
with the same weight, as required by the hermiticity of the interaction $H_c$ and of the scalar field $\phi(x)$.
The momentum conservation has been explicitly worked out and the constant $G_{s}$ is the meson-nucleon coupling
constant. In equation (\ref{eq:mes}) $\omega(q)$ is the meson energy and the factor in front $1 / \sqrt{2\omega}$
comes from the usual quantization of the boson (meson) field in a set of harmonic oscillators \cite{Mandl}. The
creation and annihilation meson operators satisfies the usual boson commutation relations
\beq
 [b_{k'} , b^\dagger{}_k]_{-} \egu - [b^\dagger{}_k , b_{k'}]_{-}
 \, \equiv \, b_{k'} b^\dagger{}_k -  b^\dagger{}_k b_{k'} \egu
 \delta_K (k - k')  \ \ \ .
\eeq
\cap In equation (\ref{eq:mes}) the product $a^\dagger a $ includes a scalar product in the spin component,
$a^\dagger a \, \equiv \, \sum_\sigma a^\dagger{}_\sigma a_\sigma$. \cap The total Hamiltonian will include,
besides the free nucleon Hamiltonian, also the free meson part
\beq
  h_0 \egu \sum_q \omega(q) b^\dagger{ }_q b_{-q}
\eeq
\cap and we can take the relativistic expression $\omega(q) = \sqrt{(mc^2)^2 + q^2c^2}$ since the meson mass $m$
is usually much smaller than the nucleon mass, and therefore its kinematics is surely relativistic. We will
indicate by $H_0$ the non-interacting part of the Hamiltonian. The inclusion of the coupling term of equation
(\ref{eq:mes}) in a non-relativistic framework is somehow problematic. In fact, since it involves the creation or
annihilation of a particle, the center of mass energy in such a process cannot be conserved and therefore Galilei
invariance is manifestly broken. The breaking is proportional to the ratio between the meson and the nucleon
masses, and therefore it vanishes in the limit of infinitely heavy nucleons. This is indeed the limit in which the
concept of a static nucleon-nucleon potential has a meaning. Perturbation theory in $H_c$ of different physical
quantities can easily be developed and the different terms can be represented by diagrams. For our purposes only
the lowest order has to be considered. In the framework of the meson-nucleon theory, the effective nucleon-nucleon
interaction can be identified with the irreducible part of the two-nucleon scattering matrix $T^{(2)}$. By
``irreducible" here we mean the set of (connected) diagrams which cannot be separated into two distinct parts by
cutting two nucleon lines at any given level along the diagram. The general perturbation theory for $T^{(2)}$ can
be obtained from the usual expansion for the scattering matrix
\begin{eqnarray}
 T^{(2)}(E) &\!\! = H_c + H_c {1\over E - H_0} T^{(2)} \ \ \ \ \ \ \
\ \ \ \ \ \ \ \ \ \ \ \ \ \ \ \ &\ \nonumber\\
\phantom{T^{(2)}(E)} &= H_c + H_c {1\over E - H_0}H_c + H_c ({1\over E - H_0}H_c)^2 \ldots & \
\end{eqnarray}
\cap In this expansion we have to select the processes which indeed correspond to the scattering of two nucleons
and are irreducible. The lowest order which can contribute is the second order, since to first order the
coupling term $H_c$ can describe only emission or absorption of a meson. Let us denote by $\vert k k' \ket$ the
free (antisymmetrized) state $a^\dagger_k a^\dagger{ }_{k'} \vert 0 \ket$ of two nucleons with momenta $k$ and
$k'$. The amplitude for the scattering from the state $\vert k_0 k'{ }_0 \ket$ to the state $\vert k_1 k_1{ }'
\ket$ can be extracted from the second order term of $T^{(2)}$
\beq
\bra k_1 k_1{ }' \vert T^{(2)} \vert k_0 k_0{ }' \ket \approx \bra k_1 k_1{ }' \vert H_c {1\over E - H_0}H_c \vert
k k' \ket \ \ \ .
\eeq
\cap If we insert the expression of equation (\ref{eq:mes}) for $H_c$, since by definition $H_0$ is diagonal in
the free state representation, we can use Wick' s theorem for the vacuum state in a straightforward way. For the
meson operators this is trivial (they commute with the nucleon operators). The four contractions which can
contribute can be depicted as in figure (\ref{fig:OB})
\cap which give four distinct contributions. The corresponding analytical expressions for the the two-body
scattering matrix is given by
\begin{eqnarray}
 \bra k_1 k_1{ }' \vert T^{(2)} \vert k_0 k_0{ }' \ket = \sum_q {\hbar\over V} {G_{s}^2\over 2\omega_q}
\delta_K (k_0 + k_0{ }' - k_1 - k_1{ }')
 \times  \ \ \ \ \ \ \ \ \ \ \ \ \ \ \ \ \ \ & \ & \ \nonumber\\
 \{ ( \delta_K (q - k_0 + k_1)
 - \delta_K (q - k_0 - k_1{ }') )
 {1\over E_0 - E + E_{k_0 - q} - E_{k_0} + \omega_q} + & \ & \ \nonumber\\
                      ( \delta_K (q - k_1{ }' + k_0)
 - \delta_K (q - k_1 - k_0) )
{1\over E_0 - E + E_{k_0{ }' - q} - E_{k_0{ }'} + \omega_q} \} \ \ ,
 & \ & \
\end{eqnarray}
\cap where $E_0 = E_{k_0} + E_{k_0'}$ is the initial energy. If we interpret this matrix element as the matrix
element of a two-body potential $v$ between nucleons, this potential is clearly non-local and energy dependent.
It is convenient to introduce the relative and total momenta of the initial and final two nucleon states
\begin{eqnarray}
 Q\phantom{'} \egu {1\over 2}(k_0{ }' - k_0) \ \ \ ; \ \ \ \,
 P \egu k_0{ }' + k_0 \ & \ & \ \nonumber\\
 Q' \egu {1\over 2}(k_1{ }' - k_1) \ \ \ ; \ \ \
 P' \egu k_1{ }' + k_1 \ & \ & \
\end{eqnarray}
\cap Putting $E = E_0$, the matrix element of $v$ can be written as
\begin{eqnarray}
 \bra k_1 k_1{ }' \vert v \vert k_0 k_0{ }' \ket =
 {1\over 2} {\hbar\over V} G_{s}^2 \delta_K (P - P')
 \times  \ \ \ \ \ \ \ \ \ \ \ \ \ \ \ \ \ \ \ \ \ \ \ \ \ \ \
 \ \ \ \ \ \ \ \ \ \ \ \ \ \ \ \ \ \ & \, & \ \nonumber\\
(-{1 \over \omega_{Q-Q'}}
 ({1\over E_{Q'_{+}}-E_{Q_{+}}+\omega_{Q-Q'} } +
   {1\over E_{Q'_{-}}-E_{Q_{-}}+\omega_{Q-Q'} }) & \, &  \nonumber\\
\phantom{(}+{1 \over \omega_{Q+Q'}}
 ({1\over E_{Q_{-}'}-E_{Q_{+}}+\omega_{Q+Q'} }+
   {1\over E_{Q'_{+}}-E_{Q_{-}}+\omega_{Q+Q'} })) & \, &
\end{eqnarray}
\cap where $Q_{\pm} = \pm Q + P/2$. The fact that a dependence on the total momentum is still present is a
consequence of the already mentioned breaking of Galilei invariance. In the limit of large nucleon mass, the
terms corresponding to the nucleon recoil can be neglected, which is equivalent to put $E_k \approx M$
everywhere in the expression. In this approximation a very simple form is obtained
\begin{eqnarray}
\hspace{-2.6 cm} \bra k_1 k_1{ }' \vert v \vert k_0 k_0{ }' \ket &=&
 {\hbar\over V} G_{s}^2 \delta_K (P - P')  (-{1\over \omega_{Q-Q'}^2}
  +{1\over \omega_{Q+Q'}^2} )\ \ \ \ \ \ \ \ \ \ \ \ \ \ \ \ \ \ \ \  \ \ \ \  \nonumber  \\
  &=& {\hbar\over V} G_{s}^2  \delta_K (P - P') (-{1\over
{(Q-Q)^2c^2 + (mc^2)^2} }
  + {1\over (Q+Q')^2c^2 + (mc^2)^2})
\label{eq:yp} \end{eqnarray} \cap where the explicit form for $\omega(q)$ has been used. The expression is now
Galilei invariant, since it depends only on the relative momenta $Q$ and $Q'$. The expression of equation
(\ref{eq:yp}) can be interpreted as the direct and exchange matrix elements of a local potential. In agreement
with the scalar nature of the exchanged meson, the interaction is independent from the  spins of the nucleons. The
form of such a potential in coordinate representation is the celebrated Yukawa potential
$$
v(r) = -G_{s}^2 {\hbar\over V}\sum_{q}  {1\over q^2c^2 + (mc^2)^2}
 e^{\imath\v{q}\cdot\v{r}/\hbar} = - {G_{s}^2\over 4\pi \hbar^2c^3}
 (mc^2) {e^{-\mu r} \over \mu r}
$$
\cap The range $a = 1/\mu = \hbar/mc$ of this potential is the Compton wavelength of the meson. This means that
heavier mesons produce shorter potential range. The limit of large nucleon mass is equivalent to neglect the
recoil energy of the nucleons involved in the interaction. It is also equivalent to consider the exchange of the
meson as instantaneous, and therefore the approximation is usually referred to as the static approximation. It is
only in this limit that the very concept of potential can be introduced. For the validity of such an approximation
it is essential that the ratio between meson and nucleon masses be small. Unfortunately not all the possible
mesons which can be considered involved in the nucleon-nucleon interaction processes have indeed a mass small
compared to the nucleon one. The potentials derived from heavier meson exchange processes have therefore to be
considered as effective ones, and the corresponding parameters as effective ones. The latter can therefore differ
from the phenomenological ones extracted from meson nucleon scattering. Equation (\ref{eq:mes}) is schematic,
since mesons and nucleons are not point-like particles, and therefore a more refined treatment must introduce
vertex corrections in the interaction processes. Usually these corrections are described by phenomenological
vertex form factors which multiply the expressions of the type of equation (\ref{eq:yp}) for the NN potentials.
The corresponding form in coordinate representation is modified accordingly. From the above results it turns out
that the local potential mediated by a scalar meson is attractive. This is the case of the so-called $\sigma$
meson, which is commonly believed to be responsible of the intermediate range attraction characteristic of the two
nucleon interaction. The lightest known (strongly interacting) meson, the $\pi$ meson, is known to be a
pseudoscalar meson, i.e. a meson with negative internal parity, which is therefore described by a field which
change sign under the parity operation. For the $\pi$ meson the scalar coupling of equation (\ref{eq:mes}) cannot
be used, since the Hamiltonian of strong interaction must be parity invariant. In the non-relativistic limit the
only possibility in this case is a pseudo-vector coupling. Furthermore the $\pi$ meson has three charge states and
it is therefore a vector in isospin space. The simplest non-relativistic coupling is of the form
$$
         H_c \egu G_{pv} {1\over \sqrt{V}}\sum_{k q}
     \sqrt{{\hbar\over 2\omega(q)}}
     ( a^\dagger{}_k (\v{\sigma}\cdot\v{q})\v{\tau} a_{k+q}
     b^\dagger{ }_q
     \plus a^\dagger{}_k (\v{\sigma}\cdot\v{q})\v{\tau}
     a_{k-q} b_q )
$$
\cap where now the $b$ and $b^\dagger$ operators refer to the $\pi$ meson. The quantities
 $\v{\sigma} \, \equiv \, {\sigma_x,\sigma_y,\sigma_z}$ are
the usual Pauli matrices which act on the spin variables of the nucleon creation and annihilation operators. The
particular form ensures rotational invariance. Since the Pauli matrices form a pseudo-vector, the expression for
$H_c$ is indeed a scalar. The matrices $\v{\tau} \, \equiv \, {\tau_x,\tau_y,\tau_z}$ are the Pauli matrices in
isospin spaces and as such they act on the isospin variables of the nucleon operators. The expression includes a
scalar product of these three Pauli matrices, which form a three-vector, with the isospin variables of the meson
operator, namely $\v{\tau} b \, \equiv \, \sum_i \tau_i b_i$ (and analogously for $b^\dagger$), where $i$ labels
the three possible isospin (charge) states of the $\pi$ meson. The scalar product ensures that $H_c$ is scalar
in isospin space, and this is dictated by the charge independence of the nuclear forces, which is
phenomenologically observed to a very high degree of accuracy. Following the same procedure as in the case of a
scalar meson, one gets the following expression for the direct matrix element of the interaction in the static
limit, with $k = Q - Q'$
\beq
 \bra Q' P' \vert v \vert Q P \ket =
 - G_{pv}^2 {\hbar\over V} \delta_K (P - P')
  {(\v{\sigma}_1\cdot\v{k})
  (\v{\sigma}_2\cdot\v{k})\over k^2c^2 + (mc^2)^2}
    (\v{\tau}_1\cdot\v{\tau}_2)
\label{eq:OPE} \eeq
\cap where the matrix elements between spin-isospin states of the corresponding Pauli matrices have to be taken,
i.e. the expression has to be considered still an operator in spin-isospin space. It is customary to introduce
the tensor operator
$$
  S_{12} \egu  3 (\v{\sigma}_1\cdot\v{k})
               (\v{\sigma}_2\cdot\v{k})
               - \v{\sigma}_1\cdot\v{\sigma}_2 k^2
$$
and the expression can be written
\begin{eqnarray}
 \bra Q' P' \vert v \vert Q P \ket =& - {1\over 12} G_{pv}^2 {\hbar\over V} \delta_K (P - P')\cdot \ \ \ \
 \ \ \ \ \ \ \ \ \ \ \ \ \ \ \ \ \ \ \ \ \ \ \ \ \ \ \ \ &\ \nonumber\\
\phantom{\bra Q' P' \vert v \vert Q P \ket =}& [{S_{12}\over k^2c^2 + (mc^2)^2} -
 \v{\sigma}_1\cdot\v{\sigma}_2 {m^2c^2\over k^2c^2 + (mc^2)^2}
 + \v{\sigma}_1\cdot\v{\sigma}_2 ]\v{\tau}_1\cdot\v{\tau}_2
 & \
\end{eqnarray}
\cap A pseudoscalar meson gives rise to a tensor-isospin interaction plus a spin-isospin interaction. The last
term is a contact interaction (a delta function in coordinate space). A more complete treatment should include the
vertex form factors also in this case. It has to be noticed that the coupling constant has a different definition
here than in the relativistic treatment. More complex couplings are possible, and they naturally arise in a
relativistic treatment, which is the framework in which the theory of nuclear forces has ultimately to be
formulated. The interaction is attractive or repulsive according to the quantum numbers of the interacting
nucleons, namely on the two-body channel (including isospin). It turns out that the tensor part is attractive in
the s-wave channels. The $\pi$ meson is responsible of the long range attractive part of the NN interaction.
Another important case is the exchange of a vector meson, namely a meson of spin one. The treatment of this case
is more complex and require a full relativistic treatment. It turns out that a vector meson produces mainly a
repulsive interaction. Therefore, at least part of the repulsive core, characteristic of the NN interaction, can
be described by the exchange of spin-one mesons, like the $\omega$ meson. However, at distances smaller than the
typical core size ($\sim$ 0.4 $fm$) the structure of the nucleons, as described by QCD, starts to play a role and
the meson picture cannot be any more maintained. The meson theory in this range can be regarded as an effective
model for more complex processes and the corresponding coupling constants and cut-off have to be considered as
parameters to be adjusted  to fit the experimental data on NN scattering.\par In all these considerations, one
assumes that only one meson is exchanged at a time, so that the NN interaction is fully determined by the set of
known mesons and by their couplings with nucleons. This is the so-called one boson exchange potential (OBEP). It
turns out, however, that the intermediate range attraction cannot be obtained in this way. As already mentioned,
it is customary then to introduce a fictitious scalar meson, the $\sigma$ meson, with suitable mass and coupling
to reproduce the phenomenological intermediate range attraction. It is usually believed that the hypothetical
$\sigma$ meson simulates the simultaneous exchange of two pions both correlated and uncorrelated. The
phenomenology on pion--pion scattering gives only a broad structure in the s-wave channel, and therefore a fully
satisfactory theoretical basis for the introduction of the $\sigma$ meson is still lacking. For a historical
account of the OBEP theory, see reference \cite{Machjpg}.
\par
\begin{figure} [ht]
 \begin{center}
\includegraphics[bb= 140 0 350 790,angle=90,scale=0.4]{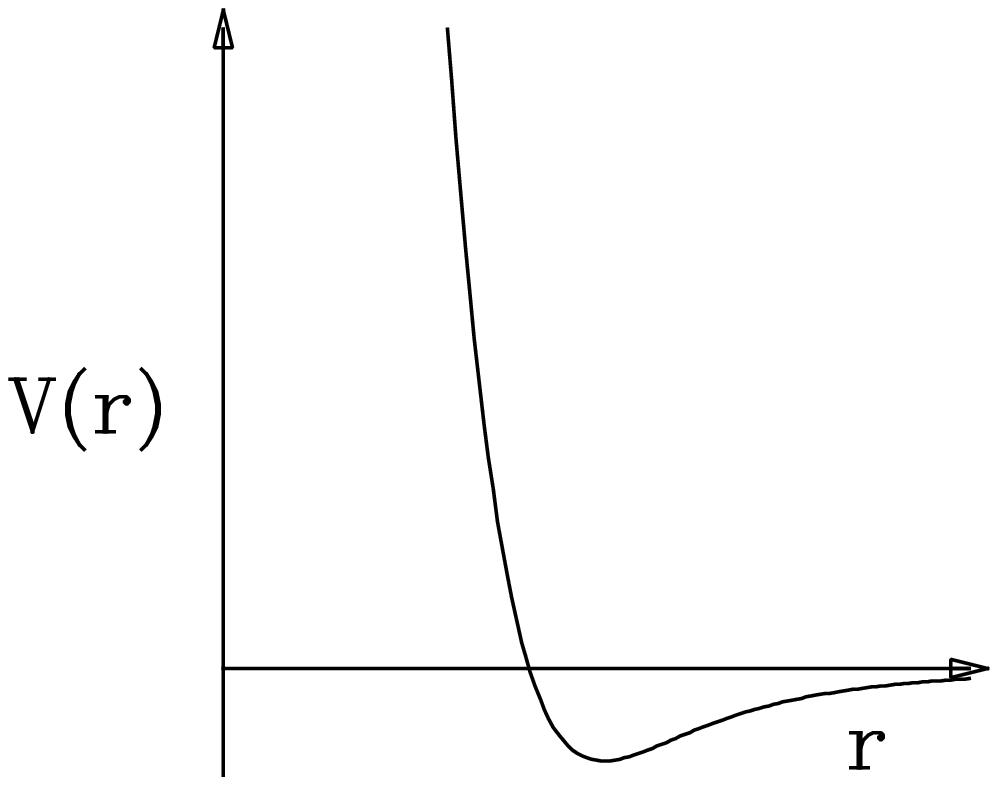}
\end{center}
   \caption{Schematic representation of the nucleon-nucleon interaction
   potential.}
    \label{fig:pot}
\end{figure}

 The main features of the NN interaction, derived from the meson-nucleon model and the phenomenological analysis,
   can be summarized schematically as in figure (\ref{fig:pot}).
 At large distance, $r \geq 1 fm$, the interaction is attractive with an exponential tail. At intermediate
 distance, $ 0.4 \leq r
\leq 1  fm $, a stronger attraction is present, at least once an average is made over the different partial
waves and quantum numbers (i.e., channels). At short distance, $ r \leq 0.4 fm $, a strong repulsive core is in
any case present. The repulsion is so strong that in the early versions of the NN potential an infinite
impenetrable barrier was assumed to exist below about $0.4 fm$. In the more modern versions the repulsive core
is taken finite but very large with respect to the usual nuclear physics energy scale.
The details of the interaction depend on the specific model for NN forces,
but schematic picture of Fig. (\ref{fig:pot}) is in any case valid and the nuclear matter EOS is strongly
influenced by these simple properties.
%
%

\subsection{Theoretical many-body methods}
Once the interaction between two nucleons is established, one can try to solve the many-body problem for the
nuclear matter. However, it is not obvious that the nuclear Hamiltonian includes only two-body forces. Since we
know that the nucleon is not an elementary particle, we can expect that the interaction in a system of nucleons
is not fully additive, namely that it is not simply the sum of the interactions between pairs of nucleons, but
also three or more nucleon forces must be considered. This important issue is discussed later. For the moment we
restrict the treatment to the case of two-body forces, which are expected anyhow to be dominant around
saturation or slightly above.\par

\subsubsection{The Brueckner-Bethe-Goldstone expansion\label{BBG}}
The Brueckner-Bethe-Goldstone (BBG) many-body theory is based on the re-summation of the perturbation expansion of
the ground state energy. The original bare NN interaction is systematically replaced by an effective interaction
that describes the in-medium scattering processes. The in-vacuum $T$-matrix of the general equation
(\ref{eq:Rmat}) is replaced by the so called $G$-matrix, that takes into account the effect of the Pauli principle
on the scattered particles and the in-medium potential $U(k)$ felt by each nucleon.
 The corresponding integral equation for the $G$-matrix can be written
 \beq
 \bar{rl}
 \bra k_1 k_2 \vert G(\omega) \vert k_3 k_4 \ket\!\!\! &\egu
 \bra k_1 k_2 \vert v \vert k_3 k_4 \ket \plus  \\
 &                  \\
 + \sum_{k'_3 k'_4} \bra k_1 k_2 \vert v \vert k'_3 k'_4 \ket\!\!\!
 &{\left(1 - \Theta_F(k'_3)\right) \left(1 - \Theta_F(k'_4)\right)
  \over \omega - e_{k'_3} + e_{k'_4} }
  \, \bra k'_3 k'_4 \vert G(\omega) \vert k_3 k_4 \ket \ \ .
\ear \label{eq:bruin} \eeq
\noindent where the two factors $1 - \Theta_F(k)$ force the intermediate momenta to be above the Fermi momentum
("particle states"), the single particle energy $e_k \, =\, \hbar^2 k^2/ 2m \, +\, U(k)$ and the summation
includes spin-isospin variables.. The $G$-matrix has not any more the hard core of the original bare NN
interaction and is defined even for bare interaction with an infinite hard core. In this way the perturbation
expansion is more manageable. The introduction and choice of the single particle potential are essential to make
the re-summed expansion convergent. In order to incorporate as much as possible higher order correlations the
single particle potential is calculated self-consistently with the $G$-matrix itself
\beq
 U(k) \egu \sum_{k' < k_F}
      \bra k k' \vert G(e_{k_1}+e_{k_2}) \vert k  k' \ket \ \ \ ,
\le{auxu} \eeq
\noindent An account on the diagrammatic method, the degree of convergence of the BBG expansion and a summary of
the results can be found in references \cite{jpg,book}. Here we restrict to indicate the expression of the
correlation energy at the so called Brueckner level ("two hole-line" approximation)
\beq
  \Delta E_2 \egu {1 \over 2} \sum_{k_1,k_2 < k_F}
  \bra k_1 k_2 \vert G(e_{k_1}+e_{k_2}) \vert k_3 k_4 \ket_A
\le{e2h} \ \ \ , \eeq \cap where $\vert k_1 k_2 \ket_A = \vert k_1 k_2 \ket - \vert k_2 k_1 \ket $.
\noindent At this level of approximation, that mainly includes two-body correlations, one can find that the
corresponding ground state wave function $\Psi$ can be written consistently as
\beq
    \vert \Psi \ket \, =\, e^{\hat{S}_2} \vert \Phi \ket \,\,\,\, ,
    \label{eq:ansatz}
    \eeq
    \noindent
where $\Phi$ is the unperturbed free particle ground state and $\hat{S}_2$ is the two-particle correlator
\beq  \hat{S}_2 \,\,\,\,\,  =\, \sum_{k_1 k_2 , k_1' k_2'}
    {1\over 4}
    \bra k_1' k_2' \vert S_n \vert k_1 k_2 \ket
    {a}\creas(k_1'){a} \creas(k_2')  \,
    a(k_2) a(k_1)
\label{eq:S} \eeq \noindent where the $k$ ' s are hole momenta, i.e. inside the Fermi sphere, and the $k'$ ' s
are particle momenta, i.e. outside the Fermi sphere. The function $\hat{S}_2$ is the so called "defect
function". It can be written in term of the $G$-matrix and it is just the difference between the in-medium
interacting and non interacting two-body wave functions \cite{jpg,book}.
A recent systematic study of the dependence of the resulting EoS on the NN interaction can be found in reference
\cite{Compilation}.
\par
\begin{figure} [h]
\vskip -10.5 cm
 \begin{center}
\includegraphics[bb= 140 0 350 790,angle=0,scale=0.6]{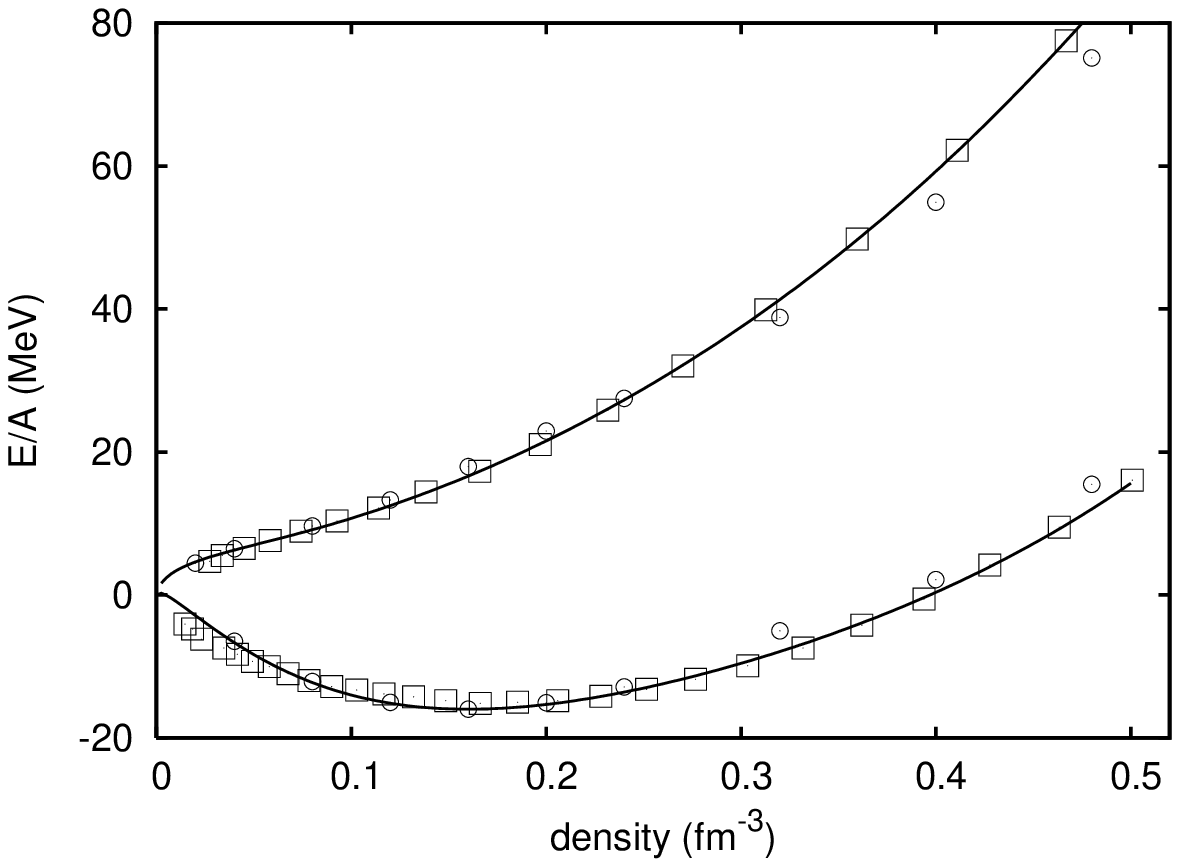}
\end{center}
\vskip -1.5 cm
   \caption{Symmetric and pure neutron matter EOS from BHF scheme including TBF (squares). The full lines
   is a fit to the points. The circles indicate the EoS from reference \cite{AP}. }
    \label{fig:EOS}
\end{figure}

 One of the well known results of all these studies, that lasted for about half a century, is the
need of three-body forces (TBF) in order to get the correct saturation point in symmetric nuclear matter. Once the
TBF are introduced, the resulting EoS, for symmetric matter and pure neutron matter, is reported in figure
(\ref{fig:EOS}) for the two-body interaction $Av_{18}$ (squares). The TBF produce a shift in energy of about $+1$
MeV in energy and of about $-0.01$ fm $^{-3}$ in density. This adjustment is obtained by tuning the two parameters
contained in the TBF, as in references \cite{bbb,NS_TBF,Compilation} and was performed to get an optimal
saturation point (the minimum).
 For comparison is also reported the variational EOS of reference \cite{AP}, that will be discussed
in the next section. The connection between two-body and three-body forces within the meson-nucleon theory of
nuclear interaction is discussed and worked out in references \cite{Andre,Umberto1,Umberto2}. The possible
interplay between two-body interaction and TBF for the resulting EoS is discussed in reference \cite{Aladin}.

\subsubsection{The Variational method}

In the variational method one assumes that the ground state wave function $\Psi$ can be written in a form as in
equation (\ref{eq:ansatz}), i.e. \beq
     \Psi(r_1,r_2,......) \, =\, \Pi_{i<j} f(r_{ij}) \Phi(r_1,r_2,.....)
     \,\,\,\, ,
\le{trial} \eeq \noindent where $\Phi$ is the unperturbed ground state wave function, properly antisymmetrized,
and the product runs over all possible distinct pairs of particles. The correlation factor is here determined by
the variational principle, i.e. by imposing that the mean value of the Hamiltonian gets a minimum (or in general
stationary point) \beq
   {\delta\over \delta f} { {\bra \Psi \vert H \vert \Psi \ket }\over
   {\bra \Psi \vert \Psi \ket} } \,= \, 0 \,\,\, .
\le{euler} \eeq \noindent In principle this is a functional equation for the correlation function $f$, which
however can be written explicitly in a closed form only if additional suitable approximations are introduced. The
function $f(r_{ij})$ is assumed  to converge to $1$ at large distance and to go rapidly to zero as $r_{ij}
\rightarrow  0$, to take into account of the repulsive hard core of the NN interaction. Furthermore, at distance
just above the core radius a possible increase of the correlation function beyond the value $1$ is possible.\par
For nuclear matter it is necessary to introduce a channel dependent correlation factor, which is equivalent to
assume that $f$ is actually a two-body operator $\hat{F}_{ij}$. One then assumes that $\hat{F}$ can be expanded in
the same spin-isospin, spin-orbit and tensor operators appearing in the NN interaction. Momentum dependent
operators, like spin-orbit, are usually treated separately. The product in  equation (\ref{eq:trial}) must be then
symmetrized since the different terms do not commute anymore.
\par
If the two-body NN interaction is local and central, its mean value is directly related to the pair distribution
function $g({\bf r})$ \beq
 < V > \, =\,  {1\over 2}\rho \int d^3r v(r) g({\bf r})  \,\,\,\, ,
\eeq \noindent where \beq
 g({\bf r_1 - r_2}) \, =\, {\int \Pi_{i>2}d^3r_i \vert\Psi(r_1,r_2....)\vert^2
    \over  \int \Pi_{i}d^3r_i \vert\Psi(r_1,r_2....)\vert^2  } \,\,\, .
\le{pairg} \eeq
\par
The main job in the variational method is to relate the pair distribution function to the correlation factors $F$.
Again, in nuclear matter also the pair distribution function must be considered channel dependent and the relation
with the correlation factor becomes more complex. In general this relation cannot be worked out exactly, and one
has to rely on some suitable expansion. Furthermore, three-body or higher correlation function must in general be
introduced, which will depend on three or more particle coordinates and describe higher order correlations in the
medium. Many excellent review papers exist in the literature on the variational method and its extensive use for
the determination of nuclear matter EoS \cite{Navarro,PandaWir}. The best known and most used variational nuclear
matter EoS is the one of reference \cite{AP}, and it is reported in figure (\ref{fig:EOS}). A detailed discussion
on the connection between variational method and BBG expansion can be found in reference \cite{jpg}.
\subsubsection{The Relativistic approach}
One of the deficiencies of the Hamiltonian considered in the previous sections is the use of the
non-relativistic limit. The relativistic framework is of course the framework where the nuclear EoS should be
ultimately  based. The best relativistic treatment developed so far is the Dirac-Brueckner approach. Excellent
review papers on the method can be found in the literature \cite{Machleidt89} and in textbooks
\cite{Brockmann99}. Here we restrict the presentation to the main basic elements of the theory.
\par In the relativistic context the only NN potentials which have been developed are the ones of OBE (one boson
exchange) type. The starting point is the Lagrangian for the nucleon-mesons coupling
\begin{eqnarray}
{\cal L}_{pv}  &=&  -\frac{f_{ps}}{m_{ps}}\barr{\psi}
\gamma^{5}\gamma^{\mu}\psi\partial_{\mu}\varphi^{(ps)}\\
{\cal L}_{s} &=&  +g_{s}\barr{\psi}\psi\varphi^{(s)}\\
{\cal L}_{v} &=&  -g_{v}\barr{\psi}\gamma^{\mu}\psi\varphi^{(v)}_{\mu} -\frac{f_{v}}{4M}
\barr{\psi}\sigma^{\mu\nu}\psi(\partial_{\mu} \varphi_{\nu}^{(v)} -\partial_{\nu}\varphi_{\mu}^{(v)})
\end{eqnarray}
 with $\psi$ the nucleon and $\varphi^{(\alpha)}_{(\mu)}$ the meson fields,
where $\alpha$ indicates the type of meson and $\mu$ the Lorentz component in the case of vector mesons. For
isospin 1 mesons, $\varphi^{(\alpha)}$ is to be replaced by {\boldmath $\tau \cdot \varphi^{(\alpha)}$}, with
$\tau^{l}$ ($l=1,2,3$) the usual Pauli matrices. The labels $ps$, $pv$, $s$, and $v$ denote pseudoscalar,
pseudovector, scalar, and vector coupling/field, respectively.

The one-boson-exchange potential (OBEP) is defined as a sum of one-particle-exchange amplitudes of certain
bosons with given mass and coupling. The main difference with respect to the non-relativistic case is the
introduction of the Dirac-spinor amplitudes. The six non-strange bosons with masses below 1 GeV/c$^2$ are used.
Thus,
\begin{equation}
V_{OBEP}=\sum_{\alpha=\pi,\eta,\rho,\omega,\delta,\sigma} V^{OBE}_{\alpha} \label{31.5}
\end{equation}
with $\pi$ and $\eta$ pseudoscalar, $\sigma$ and $\delta$ scalar, and $\rho$ and $\omega$ vector particles. The
contributions from the isovector bosons $\pi, \delta$ and $\rho$ contain a factor {\boldmath $\tau_{1} \cdot
\tau_{2}$}. In the so called static limit, i.e. treating the nucleons as infinitely heavy (their energy equals the
mass) the usual denominator of the interaction amplitude in momentum space, coming from the meson propagator, is
exactly the same as in the non-relativistic case (since in both cases meson kinematics is relativistic). This
limit is not taken in the relativistic version, noticeably  in the series of Bonn potentials, and the full
expression of the amplitude with the nucleon relativistic (on-shell) energies is included. As an example, let us
consider one pion exchange. In the non-relativistic and static limit the corresponding local potential is reported
in equation (\ref{eq:OPE}). This has to be compared with the complete expression of the matrix element between
nucleonic (positive energy) states \cite{Machleidt2000}. In the center of mass frame it reads
$$
 V_{\pi}^{full}=
 - {g_{\pi}^2  \over 4 M^2} {(E' + M)(E + M) \over k^2c^2 + (mc^2)^2}
 \left( {  \v{\sigma}_1\cdot\v{q}'\over E' + M } - {\v{\sigma}_1\cdot\v{q}\over E +
 M}\right) \times \left( {\v{\sigma}_2\cdot\v{q}'\over E' + M} - {\v{\sigma}_2\cdot\v{q}\over E +
 M}\right)
$$
\par\noindent
where $E , E'$ are the  initial and final nucleon energies. One can see that in this case some non-locality is
present, since the matrix element depends separately on $\v{q}$ and $\v{q}'$. Putting $E = E' = M$, one gets
again the local version. Notice that in any case the two versions coincide on-shell ($E = E'$), and therefore
the non-locality modifies only the off-shell behaviour of the potential. The matrix elements are further
implemented by form factors at the NN-meson vertices to regularize the potential and to take into account the
finite size of the nucleons and the mesons. In applications of the DBHF method usually one version of the
relativistic OBE potential is used, which therefore implies that a certain degree of non-locality is present.
The fully relativistic analogue of the two-body scattering matrix is the covariant Bethe-Salpeter (BS) equation.
In place of the NN non-relativistic potential the sum ${\cal V}$ of all connected two-particle irreducible
diagrams has to be used, together with the relativistic single particle propagators. Explicitly, the BS equation
for the covariant scattering matrix ${\cal T}$ in an arbitrary frame can be written
\begin{equation}
\label{eq:eq20} {\cal T}(q',q|P)={\cal V}(q',q|P)+\int d^{4}k{\cal V}(q',k|P){\cal G}(k|P) {\cal T}(k,q|P) \  ,
\end{equation}
with
\begin{eqnarray}
{\cal G}(k|P)&=&\frac{i}{(2\pi)^{4}} \frac{1}{(\frac{1}{2}\not\!P+\not\!k-M+i\epsilon)^{(1)}}
\frac{1}{(\frac{1}{2}\not\!P-\not\!k-M+i\epsilon)^{(2)}}\\
             &=&\frac{i}{(2\pi)^{4}}
\left[\frac{\frac{1}{2}\not\!P+\not\!k+M} {(\frac{1}{2}P+k)^{2}-M^{2}+i\epsilon}\right]^{(1)}
\left[\frac{\frac{1}{2}\not\!P-\not\!k+M} {(\frac{1}{2}P-k)^{2}-M^{2}+i\epsilon}\right]^{(2)}
\end{eqnarray}
where $q$, $k$, and $q'$ are the initial, intermediate, and final relative four-momenta, respectively (with e.\
g.\ $k=(k_{0},{\bf k})$),
 and $P=(P_0,{\bf P})$ is the total four-momentum;
$\not\!k=\gamma^{\mu}k_{\mu}$.
 The superscripts
refer to particle (1) and (2). Of course all quantities are appropriate matrices in spin (or helicity) and
isospin indices. The use of the OBE potential as the kernel ${\cal V}$ is equivalent to the so-called ladder
approximation, where one meson exchanges occur in disjoint time intervals with respect to each other, i.e. at
any time only one meson is present. Unfortunately, even in the ladder approximation the BS equation is difficult
to solve since ${\cal V}$ is in general non-local in time, or equivalently energy dependent, which means that
the integral equation is four-dimensional. It is even not sure in general if it admits solutions. It is then
customary to reduce the four-dimensional integral equation to a three-dimensional one by approximating properly
the energy dependence of the kernel. In most methods the energy exchange $k_0$ is fixed to zero and the
resulting reduced BS equation is similar to its non-relativistic counterpart. In the Thompson reduction scheme
this equation for matrix elements between positive-energy spinors (c.m.  frame) reads
\begin{equation}
{\cal T}({\bf q'},{\bf q}) = V({\bf q'},{\bf q})+ \int\frac{d^3k}{(2\pi)^3}V({\bf q'},{\bf k})\,
\frac{M^2}{E_{{\bf k}}^2}\, \frac{1} {2 E_{{\bf q}}-2 E_{{\bf k}}+i\epsilon} {\cal T}({\bf k},{\bf q}|{\bf P})
\end{equation}
\noindent where both $V({\bf q'},{\bf q})$ and ${\cal T}$ have to be considered as matrices acting on the
two-particle helicity (or spin) space, and $E_{{\bf k}} = \sqrt{{\bf k}^2 + M^2}$ is the relativistic particle
energy. In the alternative Blankenbecler-Sugar \cite{Machleidt2000} reduction scheme some different relativistic
kinematical factors appear in the kernel. This shows that the reduction is not unique. The partial wave
expansion of the ${\cal T}$--matrix can then be performed starting from the helicity representation. The
corresponding amplitudes include single as well as coupled channels, with the same classification in quantum
numbers $JLS$ as in the non relativistic case and therefore their connection with phase shifts is the same. In
the intermediate states of momentum ${\bf k}$ only the positive energy states are usually considered (by the
proper Dirac projection operator). As in the case of the OBEP potential, again the main difference with respect
to the non-relativistic case is the use of the Dirac spinors.\par The DBHF method can be developed in analogy
with the non-relativistic case. The two-body correlations are described by introducing the in-medium
relativistic $G$-matrix. The DBHF scheme can be formulated as a self-consistent problem between the single
particle self-energy $\Sigma$ and the $G$-matrix. Schematically, the equations can be written
\begin{eqnarray}
G  &=&  V + i\int V Q g g G \nonumber\\
\Sigma &=& -i \int_F(Tr[gG] - gG) \label{eq:sig}
\end{eqnarray}
\noindent where $Q$ is the Pauli operator which projects the intermediate two particle momenta outside the Fermi
sphere, as in the BHF G-matrix equation, and $g$ is the single particle Green' s function. The self consistency
is entailed by the Dyson equation
$$
g = g_0 + g_0 \Sigma g
$$
\noindent where $g_0$ is the (relativistic) single particle Green's function for a free gas of nucleons. The
self-energy is a matrix in spinor indices, and therefore in general it can be expanded in the covariant form
\begin{equation}
\Sigma(k,k_F) = \Sigma_s(k,k_F) - \gamma_0\Sigma_0(k,k_F) + \mbox{\boldmath $\gamma$} \cdot{\bf k}\Sigma_v
\label{eq:sigex}
\end{equation}
\noindent  where $\gamma_\mu$ are the Dirac gamma matrices and the coefficients of the expansion are scalar
functions, which in general depend on the modulus $ |{\bf k}| $ of the three-momentum and on the energy $k_0$.
Of course they also depend on the density, i.e. on the Fermi momentum $k_F$. The free single particle
eigenstates, which determine the spectral representation of the free Green' s function, are solutions of the
Dirac equation
$$
[\,\,\, \gamma_\mu k^\mu \, - \, M\,\,\, ]\, u(k)\,\, =\, 0
$$
\noindent where $u$ is the Dirac spinor at four-momentum $k$. For the full single particle Green's function $g$
the corresponding eigenstates satisfy
$$
[\,\,\, \gamma_\mu k^\mu \, - \, M \, + \, \Sigma \,\,\,]\, u(k)^*\,\, =\, 0
$$
\noindent Inserting the above general expression for $\Sigma$, after a little manipulation, one gets
$$
[\,\,\, \gamma_\mu {k^\mu}^* \, - \, M^*\,\,\, ] u(k)^*\,\, =\, 0
$$
\noindent with \beq
 {k^0}^* \,=\, {k^0 + \Sigma_0\over 1 + \Sigma_v} \,\,\,\,\,\, ;\,\,\,\,\,\, {k^i}^* \,=\,
 k^i
 \,\,\,\,\,\, ; \,\,\,\,\,\, M^* \,=\, {M + \Sigma_s\over 1 + \Sigma_v}
 \label{eq:momen}
\eeq \noindent This is the Dirac equation for a single particle in the medium, and the corresponding solution is
the spinor
\begin{equation} {u}^*({\bf k},s)=\sqrt{\frac{{E}^*_{\bf k}+{M}^*}{2 {M}^*}}
\left( \begin{array}{c} 1\\ \frac{\mbox{\boldmath $\sigma \cdot k$}}{{E}_{\bf k}^*+{M}^*}
\end{array} \right) \chi_{s}  \,\,\,\,\,\ ; \,\,\,\,\,\, {E}^*_{\bf k} = \sqrt{{\bf
k}^2 + {M^*}^2 } \,\,\,  . \label{eq:spino}
\end{equation}
 In line with the Brueckner scheme, within the BBG expansion, in the
self-energy of equation (\ref{eq:sig}) only the contribution of the single particle Green' s function pole is
considered (with strength equal one). Furthermore, negative energy states are neglected and one gets the usual
self--consistent condition between self--energy and scattering $G$--matrix. The functions to be determined are in
this case the three scalar functions appearing in equation (\ref{eq:sigex}). However, to simplify the calculations
these functions are often replaced by their value at the Fermi momentum.
\par In any case, the medium effect on the spinor of equation
(\ref{eq:spino}) is to replace the vacuum value of the nucleon mass and three--momentum with the in--medium values
of equation (\ref{eq:momen}). This means that the in--medium Dirac spinor is ``rotated" with respect to the
corresponding one in vacuum, and a positive (particle) energy state in the medium has some non--zero component on
the negative (anti--particle) energy state in vacuum. In terms of vacuum single nucleon states, the nuclear medium
produces automatically anti--nucleon states which contribute to the self--energy and to the total energy of the
system. It has been shown in reference \cite{Brown87} that this relativistic effect is equivalent to the
introduction of well defined TBF at the non--relativistic level. These TBF turn out to be repulsive and
consequently produce a saturating effect. The DBHF gives indeed in general a better SP than BHF. Of course one can
wonder why these particular TBF should be selected, but anyhow a definite link between DBHF and BHF + TBF is, in
this way, established. Indeed, including in BHF only these particular TBF one gets results close to DBHF
calculations, see e.g. reference \cite{Compilation}.
\par
Despite the DBHF is similar to the non--relativistic BHF, some features of this method are still controversial.
The results depend strongly on the method used to determine the covariant structure of the in--medium $G$--matrix,
which is not unique since only the positive energy states must be included. It has to be stressed that, in
general, the self--energy is better calculated in the matter reference frame, while the  G--matrix is more
naturally calculated in the center of mass of the two interacting nucleons. This implies that the $G$--matrix has
to be Lorentz transformed from one reference frame to the other, and its covariant structure is then crucial.
Formally, the most accurate method appears to be the subtraction scheme of reference \cite{GrossBoelting99}.
Generally speaking, the EoS calculated within the DBHF method turn out to be stiffer above saturation than the
ones calculated from the BHF + TBF method.

\subsubsection{The V$_{low}$ approach \label{Vlow}}
The main effect of the hard core in the NN interaction is to produce scattering to high momenta of the interacting
particles. It is possible to soften the hard core of the NN interaction from the start by integrating out all the
momenta larger than a certain cutoff $\Lambda$ and "renormalize" the interaction to an effective interaction
V$_{low}$ in such a way that it is equivalent to the original interaction for momenta $q \, <\, \Lambda$. By
construction V$_{low}$ must give the same half of the energy shell scattering $T$-matrix or $R$-matrix
$(q'|R(E_q)|q)$ as the original interaction, where $E_q$ is the energy of the initial state at relative momentum
$q$. This can be done in a variety of methods, among which one can mention the Renormalization Group, the low
momenta Effective Theory and the Lee-Suzuky scheme. It is surely outside the scope of the present report to
describe these methods and we refer to recent review and papers where they are extensively discussed and applied
\cite{Bogner_rev,Machleidt_PLB}. All these possible V$_{low}$ interactions are of course much softer, since no
high momentum components are present. The short range repulsion is replaced by the non local structure of the
interaction. It has to be kept in mind that any V$_{low}$ is a legitimate realistic NN interaction. In fact, due
to the mentioned equivalence, they fit exactly the same phase shifts up to an energy corresponding to the cutoff.
The latter is taken above $300$ MeV in the laboratory, corresponding to relative momentum $q \, \approx\,
2.1$fm$^{-1}$, that is the largest energy where the data are established. This means that the data and the NN
interaction are not sensitive to the details of the hard core behavior. Indeed, for the same reason, all the
V$_{low}$, at least their diagonal matrix elements, are almost identical up to the cutoff, provided it is not
taken too large above $2.1$fm$^{-1}$. The fact that V$_{low}$ is soft has the advantage to be much more manageable
than a hard core interaction, in particular it can be used in perturbation expansion and in nuclear structure
calculations in a more efficient way. Again, at purely phenomenological level, a non-local interaction at short
distance is perfectly legitimate as an hard core, since the behavior of the potential at short distance is not
experimentally accessible. In principle, it is only a question of representation. However, as strongly stressed in
reference \cite{Bogner_rev}, the required equivalence implies that, even starting from a local two-body
interaction , the renormalization procedure introduces necessarily three-body forces, which has to be handled in
many-body calculations. This is more apparent if V$_{low}$ is constructed directly in the medium, in which case
the procedure has some similarities with the Brueckner $G$-matrix construction. \par If for the in-vacuum
V$_{low}$ one takes only the two-body component, one finds that nuclear matter does not saturate and actually
seems to collapse towards infinite negative energy. This can be ascribed to the missing three-body forces, which
should provide saturation. According to reference \cite{Bogner_rev} the off-shell effects are not the main
responsible for the saturation mechanism present in the BHF theory. This is probably partly true, but one has to
keep in mind that a BHF procedure without the single particle potential, which is responsible for the presence of
an off-shell energy, would dramatically overestimate the binding energy in nuclear matter.\par In any case, up to
now the consistent three-body force in V$_{low}$ has not be used in nuclear matter calculations or in nuclear
structure. Furthermore the renormalization procedure for a possible three-body forces has not yet been worked out.
Up to now the two-body part of V$_{low}$ has been used in nuclear matter in conjunction with TBF fitted separately
in three and four-nucleon systems. If the TBF are averaged over the quantum numbers of one of the three particles,
this approximate scheme seems to produce a reasonable saturation point. This is line with the fact that in BHF the
repulsive part of the same TBF has to be drastically reduced to get the correct saturation point, since the
two-body interaction gives a saturation point not too far from the empirical one. If the saturation property of
V$_{low}$ and its good performance in few-body nuclear systems will survive to a more consistent treatment of the
renormalization procedure and the corresponding TBF, this will open a new route to the microscopic many-body
theory of the nuclear medium. This point needs still to be clarified.
\subsubsection{Trying a link to QCD : the chiral symmetry approach}
One of the main ambition of nuclear physics is to connect the low energy nuclear physics phenomena with the
underlying more fundamental theory of strong interaction, i.e. QCD and the standard model, based on quark and
gluons degrees of freedom, together with the Weinberg-Salam-Glashow theory of weak interaction. This is quite
difficult because all hadron sector is in the non-perturbative regime, due to confinement. A possible strategy is
the systematic use of the symmetries embodied in the hadronic QCD structure. The main symmetry that still remains
visible in the confined matter is the Chiral Symmetry, the symmetry that QCD possesses if the bare quark masses
are put equal to zero. The symmetry is spontaneously broken in the confined phase, i.e in hadronic matter, but,
according to the general theorem by Goldstone, a zero mass boson should be present. This is indeed the $\pi$
meson, that in the limit of zero quark mass should have also zero mass. This is the main signature of the
underling chiral symmetry.  For non-strange matter, only $u$ and $d$ quarks are relevant, and they indeed are
expected to have a mass of few MeV.  This small explicit breaking of chiral symmetry results in the physical mass
of the pion, that is the lightest meson, even if it is not small at the energy scale of many nuclear phenomena.
All that suggests to treat the pion degrees of freedom explicitly and to describe the short range part by
structureless contact terms. Along this line Weinberg \cite{Wei1,Wei2,Wei3} proposed a scheme for the expansion of
the NN interaction in the ratio $Q/M$ between the relative momenta and the nucleon mass. The pion exchange term is
still treated explicitly, and is considered the lowest order (LO) term of the expansion. Then contact terms are
added and a power counting scheme is introduced. These terms can be considered as expansion of the nucleonic
loops, that are the ones that give the largest contributions. At the same time the contact terms can be considered
as counter-terms to regularize and renormalize the divergences coming from the loop integrals appearing as the
order of the expansion increases. This procedure of renormalization is common in Quantum Field Theory, like
Quantum Electrodynamics (QED). In renormalizable field theory the number of counter-terms are finite and their
strengths are fixed by demanding that some quantities have their physical values. In QED they are fixed by
imposing the physical values of the electron charge and mass. Then the perturbation expansion terms are all
finite. In the case of nuclear physics one demands that the phase shifts in some channels and specific energies
are reproduced correctly. However in this case at any order new counter-terms must be introduced. This
renormalized expansion, the Chiral Perturbation  Expansion (ChPE), can be used to construct  NN interactions that
are of reasonably good quality \cite{Machleidt_PLB,Kaiser} in reproducing the two-body data. They contain a set of
parameters, and therefore they are still models, whose connection with QCD is still a little loose. To tighten the
QCD link a non-perturbative regularization and renormalization has been tried, where the results are shown
numerically to be independent on the single cut-off used in the renormalization procedure. This property assures a
clear-cut separation between the long range pion exchange processes and the unresolved short range part of the
interaction. Along this line progress has been made recently \cite{Mach_cut}, but still a reasonable realistic
interaction has not been constructed.\par To summarize, the ambitious program of connecting the NN interaction
with the underlying QCD theory is still a work in progress, but the connection is becoming stronger and solid. It
requires the development of a quite complex formalism, that is hoped will be able to shed light on the origin of
the NN interaction, including few-body interactions.

\section{Nuclear Matter as a Fermi liquid \label{landau}} The many fermion systems are characterized by a sharp Fermi surface.
If the interaction is not too peculiar, this is a general result known as  Migdal' s theorem \cite{Mig}. The
nuclear medium is not an exception, and many properties of both finite nuclei and nuclear matter are strongly
affected by this feature. In particular, low energy excitations and low temperature thermal properties can involve
only particles close to the Fermi surface. For the same reason transport phenomena are determined by the
scattering processes that occur close to the Fermi surface.\par Landau theory of (normal) Fermi liquids exploits
systematically this feature to develop a semi-phenomenological treatment of most of the low energy phenomena in
homogeneous Fermi systems. Nuclear matter can be treated along the same lines, provided the isospin degrees of
freedom is properly included. Migdal and collaborators \cite{Mig,Eduard} have extended the approach to finite
systems, in particular nuclei, where the so called Finite Fermi System Theory (FFST) has been extensively applied.
Excellent and pedagogical expositions of Landau theory can be found in textbooks \cite{Mig,Peth_Gord}. Here we
limit to remind the main concepts and some basic applications. At the basis of the theory is the introduction of
quasi-particle states. The suggestion comes form the so-called adiabatic switching on of the interaction in a
many-body system. The Gell-Mann and Low \cite{GL} theorem states that if one evolves a many-particle system,
starting from an independent particle eigen-state, by switching on the interaction adiabatically, i.e. with an
infinitely slow variation, then one obtains a state of the interacting system. Since for independent particles
eigen-states are identified by the values of the occupation number $n(p)$ for each single particle state $p \equiv
({\vec p},\sigma,\tau)$, so will be the corresponding state of the interacting system. The statement is correct if
no phase transition or cluster formation occur during the switching on of the interaction, and this is what we
assume for the moment. It follows that the ground state of the interacting system will be characterized by a
distribution of occupation numbers as the non-interacting one, i.e. a sharp Fermi distribution (zero temperature).
If we consider the excitation of the non interacting system that is obtained by adding a particle or forming a
hole (i.e. subtracting a particle) at the single particle state $p$, these excitations will be called
quasi-particle and quasi-hole in the interacting system. The corresponding variation in energy of the interacting
many-particle system is called quasi-particle or quasi-hole energy, denoted by $\epsilon(p)$. In general, if we
vary the distribution of occupation numbers in a smooth way, or we average the distribution within neighboring
states, the general variation in energy $\delta E$ of the system to first order in the variation $\delta n(p)$ of
the occupations can be written
\beq
\delta E \, =\, {1 \over \Omega} \sum_{p} \epsilon(p) \delta n(p)
 \label{eq:qp}\eeq
\noindent where $\Omega$ is the volume of the system. It is clear that the quasi-particles are Fermions. However,
unlike in the non-interacting system, they do interact, since the variation of the total energy will be a complex
function of the variation $\delta n(p)$ and not just an additive linear function. To second order the energy
variation will be
\beq
\delta E \, =\, {1 \over \Omega} \sum_{p} \epsilon(p) \delta n(p) \, +\,
 {1 \over 2} {1 \over \Omega^2} \sum_{p,p'} f(p,p') \delta n(p) \delta n(p')
\label{eq:exp2}\eeq
\noindent where
\beq
f(p,p') \, =\, \Omega^2 {\delta^2 E\over \delta n(p) \delta(p')} \, =\, \Omega {\delta \epsilon(p) \over \delta
n(p') }
\label{eq:lanint}\eeq
\noindent is the quasi-particle interaction, since it describes the variation of a quasi-particle energy due to
the presence of the other quasi-particles. More precisely, if the quasi-particle distribution is changed by the
amount $\delta n(p')$ for each $p'$, the energy of the quasi-particle of momentum $p$ changes by an amount given
by the expression
\beq
\delta \epsilon(p) \,=\, \sum_{p'} f(p,p') \delta n(p')
\label{eq:qpe} \eeq
\noindent However, the Gell-Mann and Low theorem is valid only if the perturbative expansion is convergent, since
its demonstration is developed by considering each term of the perturbation expansion. This is not necessarily a
valid assumption, and the quasi-particle state so constructed can be at best only approximate eigen-states of the
system. This is indeed what happens, and the quasi-particle states have a finite lifetime and they actually decay.
This can be best seen in the Green' s function formalism, where the Landau theory can be more rigorously
formulated \cite{Nozieres,Mig}. It can be seen that the quasi-particles has an infinite lifetime only exactly at
the Fermi surface, while they have a decreasing lifetime as one moves away from the Fermi surface. This is a
general property based only on phase space argument. In a perturbative picture, a single quasi-particle can decay
into two particles - one hole state (for momenta above the Fermi momentum), or in a two hole - one particle state
(below the Fermi momentum), and one can easily see that the possible phase space vanishes exactly at the Fermi
momentum. This remains true if one considers more complicated decay states (multi particles - multi holes). Once
the quasi-particles are introduced, these considerations hold for the scattering processes between
quasi-particles, which are then responsible for the decay. If the quasi-particle have no decay width at the Fermi
surface, it is reasonable to expect that close enough to the Fermi surface the width will remain small, and there
will be a region around the Fermi surface where the quasi-particles can be considered as stable, provided the
phenomena that are considered have characteristic time scales shorter than the lifetime of the quasi-particles
involved. \par Again in a more formal language, a quasi-particle corresponds to the singular part of the single
particle Green's functions, i.e. a pole in the complex energy plane at a given momentum. The non-singular part
should give a negligible contribution in many dynamical processes, since it is expected to be highly incoherent.
However the non-singular part has the effect of renormalizing the properties of the quasi-particles. In summary,
it is the pole contribution that behaves like a particle, with properly renormalized physical parameters. In this
sense a quasi-particle can be viewed as a particle dressed by the interaction with the other particles. The pole
moves from one sheet of the complex energy plane to the other, when the momentum moves from below to above the
Fermi surface. This means that the occupation number has a jump at the Fermi surface, a property anticipated at
the beginning of this section. If we neglect the non-singular part, the quasi-particles distribution in the ground
state has to be considered as an unperturbed Fermi distribution, i.e. with occupation number 1 and 0 (at zero
temperature), so they are fermion particles. \par In summary, the dynamical processes that involve excitations
close to the Fermi surface can be described in terms of quasi-particles kinetics, whose dynamics can be treated as
particles (fermions), but with renormalized properties. In the semi-classical regime, valid in the long
wave-length limit, the kinetic equations for the quasi-particle distribution $n(\mathbf{r},p,t)$ must follows
equations (\ref{eq:BUU},\ref{eq:Upot}), where the effective NN interaction, in the momentum representation, is
just the Landau effective interaction $f(p,p')$ of equation (\ref{eq:lanint}).
\par
One has to be aware that the distribution function $n(\mathbf{r},p,t)$ that appears in the equations
(\ref{eq:BUU},\ref{eq:Upot}) is essentially the semi-classical limit of the quantal density matrix $<\psi^\dagger
(r',t)\psi(r,t)>$ (to be precise, its Wigner transform) \cite{baym_K}. If we consider a perturbation with total
momentum $\mathbf{q}$, one has to consider the Fourier transform of the distribution function, which is equivalent
to the density matrix in momentum representation $<\psi^\dagger (p+q,t)\psi(p,t)>$. Therefore, the momenta $p+q$
and $p$ must form a particle-hole pair, i.e. they must lie on opposite sides of the Fermi surface.
In fact, any perturbation of a Fermi liquid must imply the promotion of a particle from below to above the Fermi
surface, and if the particles are removed from a position slightly different from the position where they are
promoted the process involves a {\it variation} $\delta n(\mathbf{r},p,t)$ of the distribution function at each
point $\mathbf{r}$ which satisfies the kinetic equations (\ref{eq:BUU},\ref{eq:Upot}).
In the long wave-length limit, i.e. when $|\mathbf{q}|$ is much smaller then the Fermi momentum, the
quasi-particle momentum $p$ must lie close to the Fermi surface. Since the effective interaction is expected to be
a smooth function, it can be then calculated for values of the momenta just on the Fermi surface, i.e. for $|{\bf
p}| \,=\, |{\bf p}'| \,=\, p_F$, where $p_F$ is the Fermi momentum. For rotational invariance $f$ must depend only
on $|{\bf p} \,-\, {\bf p}'|$. The dependence on the angle $\theta$ between ${\bf p}$ and ${\bf p}'$ can be
expanded in Legendre polynomials $P_l(\cos \theta)$. Of course one should include also the spin and isospin
dependence, and the explicit expression for $f$ is usually written, following Landau,
\beq
f(p,p') \,=\, f \,+\, g \sigma\sigma' \,+\, f' \tau\tau' \,+\, g' \sigma\sigma' \tau\tau'
\label{eq:st} \eeq
\noindent and each coefficient can be now expanded, e.g.
\beq
f \,=\, \sum_{l} f_l P_l(\cos \theta)
\label{eq:expan}\eeq
\noindent and similar. This defines four sets of parameters, $\{f_l\}, \{g_l\}, \{f_l'\}, \{g_l'\}$. They are
basic quantities in the Landau theory. One has not to confuse this expansion with the expansion in partial waves
of the NN interaction (effective or not), but rather the different terms in $l$ are connected with the range in
non-locality of the particle-hole interaction, or, equivalently, to its momentum dependence. In most applications
these parameters are multiplied by the single particle density of state $N$ at the Fermi surface, and one then
introduces the dimensionless parameters $F_l = N f_l$ and similar. The other basic quantity of the Landau theory
is the collision integral. It can be worked out \cite{Peth_Gord} for two-body collisions, that are assumed to
dominate for not too high density. The two-body collision probability must contain two factors $n(k)$ (Fermi
functions) to weight the two occupied initial single particle states and two factors $1 - n(k)$ for the Pauli
blocking of the two unoccupied final single particle states. Explicitly, the collision integral that has to be
included at the second member of the kinetic equations (\ref{eq:BUU},\ref{eq:Upot}), reads \cite{Peth_Gord}
\beq
\begin{array} {rl}
I(p_1) \,= &{2\pi \over \hbar^2} \sum_{p_2,p_3,p_4} | <p_3 p_4 | T |p_1 p_2> |^2
 \nonumber  \\
\  &\delta(p_1 + p_2 -p_3 -p_4) \delta(\omega_1 + \omega_2 -\omega_3 -\omega_4) \nonumber \\
\  &\big[ n_3 n_4(1-n_1)(1-n_2) - n_1 n_2 (1-n_3)(1-n_4) \big]
\end{array} \label{eq:coll}\eeq
\noindent where $T$ is the scattering $T$-matrix (in the medium !), $n_i \,=\, n(p_i)$ and $\omega_i$ is the
single particle energy of momentum $\mathbf{p}_i$. Remind that in the adopted notation $p_i \,=\,
(\mathbf{p}_i,\sigma_i,\tau_i)$. This collision integral gives the probability that a particle of a given momentum
$p_1$ scatters with a particle of any momentum $p_2$ to all possible final momenta $p_3, p_4$ (first term in the
square bracket) and the probability that two particles of momenta $p_3, p_4$ scatters to the final momentum $p_1$
and any other momentum $p_2$ (second term). The two terms correspond to the loss and gain processes, for the state
of momentum $p_1$, that can occur in the medium. We have not indicated explicitly that actually all the
distribution functions appearing in the collision integral are calculated at a given space-time point. This is
justified if the range of the interaction producing the collisions is much smaller than the average distance
between particles. In this sense the theory is valid for low density, which in this context means that the average
number of quasi-particles must be not too large.
The collision integrals is the key quantity that determines the different transport coefficients, because if the
particle distribution is macroscopically perturbed by an external action, it is through collisions that the system
reacts to bring back the distribution to the equilibrium one. The particles through collisions transport different
physical quantities, like momentum, energy, and so on, and the frequency of collisions is the main features that
fix the speed of this restoration and therefore the corresponding transport coefficients or related quantities.
\par On the other hand, the set of Landau parameters that characterize the interaction are more related to the
mechanical properties of the medium or to the dynamical microscopic processes that can take place.
\subsection{Effective mass}
We have introduced the concept of quasi-particles, and we have anticipated that their physical parameters have to
be renormalized. If the interaction $f(p,p')$ depends explicitly on momentum, it changes the relationship between
energy and momentum of the quasi-particle due to the dragging effect, see equation (\ref{eq:qpe}). The standard
result is that the quasi-particle velocity $v_k$ for the momentum $k$ can be written
\beq
v_k \,=\, {d \epsilon(k) \over d k} \,=\, {k \over m^*}
\label{eq:mstar1}\eeq
\noindent where the effective mass $m^*$ is given by
\beq
{m^* \over m} \,=\, 1 \,+\, {F_1 \over 3}
\label{eq:mstar2}\eeq
\noindent Here $m$ is the bare mass, and $F_1$ is the dimensionless constant previously introduced. Only the term
$l = 1$ of the expansion (\ref{eq:expan}) contributes. The relation (\ref{eq:mstar2}) is a consequence of Galilei
invariance \cite{Peth_Gord}. The concept of effective mass can be extended also to finite nuclei. It is
extensively used in Energy Density Functional schemes or Skyrme forces (see section \ref{EDF}) as a parameter,
possibly density dependent. Physical quantities that are particularly sensitive to its value are the energy of
different Giant Resonances, notably the monopole one, see section \ref{giant}, and the single particle density of
states. The latter is mainly proportional to the effective mass.
\par The canonical value of the effective mass at the saturation density that appears in most of the Skyrme forces
is close to $0.7 m$. However, the density of state close to the Fermi energy extracted phenomenologically in
finite nuclei seems to require a value close to $m$. This discrepancy can be explained and understood if one
introduces dispersive effects in the single particle spectrum \cite{Francesco}. In fact in the nuclear medium it
is essential to distinguish between the so called $k$-mass $m_k$ and $\omega$-mass $m_\omega$. If one considers
the single particle self-energy $M(k,\omega)$, the total effective mass can be written \cite{NegOrl}
\beq
\begin{array} {rl}
{m_\omega \over m} &\,=\, ( 1 \,-\, {\partial M\over \partial \omega} ) \nonumber \\
 \  &\                     \nonumber \\
{m_k \over m} &\,=\,  ( 1 \,+\, {m\over k}{\partial M\over \partial k} )^{-1} \nonumber \\
 \  &\                     \nonumber \\
{m^*\over m} &\,=\, ({m_\omega\over m})({m_k\over m})
\end{array} \label{eq:omm}\eeq
\noindent The $\omega$-mass is due to the energy dependence of the self-energy. In the calculation of the ground
state energy and wave function with Skyrme forces the mean field is directly related to the $k$-mass. In the
density of states what is involved is the single particle dynamics, that must include also the $\omega$-mass, and
therefore the total mass must be used. Extensive calculations \cite{Francesco} of the single particle levels
confirm that indeed the total effective mass around the Fermi energy is close to the bare mass $m$. Similar
results are obtained in microscopic calculations of symmetric nuclear matter \cite{Mahaux}.
\par In a general treatment of nuclear structure based on Energy Density Functional method, to be discussed in
section \ref{EDF}, the effective mass is a parameter to be fixed or fitted to the experimental mass table, and not
necessarily the distinction between the $\omega$-mass and $k$-mass is apparent or displayed. Therefore the
effective mass in nuclear structure is not a well defined concept, nor it can be given a well defined value. In
other words, its value depends on the theoretical scheme and on the physical quantity that is considered.
\subsection{Static and equilibrium properties \label{stat}}

Since Landau theory introduces the interaction between quasi-particles, it can be used to calculate some static
and equilibrium properties of an interacting Fermi liquid with respect to a free gas.
\par In particular the incompressibility of equation (\ref{eq:com}) is modified, as follows by simple arguments.
If the system is compressed, the Fermi energy $\epsilon_F$ increases, but the quasi-particles filling the new
available states interact among each other, so that, according to equation (\ref{eq:qpe}) the variation in the
quasi-particle energy is
\beq
 \delta \epsilon(p) \,=\, f_0 {1\over V} \sum_p \delta n(p) \,=\, f_0 \delta n
\label{eq:deps}\eeq
\noindent where the spherical symmetry of the distribution has been used. Then the Fermi energy will get an
additional variation, with respect to a free gas model, given by equation (\ref{eq:deps}). Since $\delta P =
n\delta \epsilon_F$, this means that the incompressibility is multiplied by the factor $(1 \,+\, F_0)$, where
$F_0$ is the Landau dimensionless constant for $l = 0$. At the same time the mass should be substituted by the
effective mass. Then, referring to equation (\ref{eq:com}), the incompressibility $K$ is related to the free gas
one $K_0$ by
\beq
 K \,=\, K_0 \left({m \over m^*}\right) (1 \,+\, F_0)
\eeq
\noindent Other bulk properties are the specific heath $c_V$ and entropy $s$. They require the formulation of the
Landau theory at finite temperature \cite{Peth_Gord}. In the low temperature limit the standard result is
\beq
c_V \,=\,{m^* \over m} c_V^0 \ \ \ \ \ \ \ \ \ s \,=\, {m^* \over m} s^0
\eeq
\noindent where $c_V^0$ and $s^0$ are the corresponding values for a free gas. The interaction changes just the
density of states at the Fermi surface. These quantities are relevant for the thermodynamic evolution of neutron
stars. For supernovae the temperature is much higher, typically few tens of MeV. Then the Landau theory cannot be
applied and other methods must be used, either phenomenological (e.g. Skyrme functionals) or microscopic, see
section \ref{finiteT}. The thermodynamical properties of nuclear matter is only partially known at such
temperatures. A clarification of this subject will be of great value for the detailed description of supernovae
evolution.

\subsection{Transport coefficients and macroscopic dynamics \label{coeff}}
Transport coefficients are fundamental properties of a Fermi liquid in general and of the nuclear medium in
particular. They are physical parameters that describe the dynamical behavior of the system at macroscopic level
and as such their main interest for the nuclear medium is mainly related to astrophysical phenomena in compact
objects. Shear viscosity, that can be viewed as momentum transport coefficient, is essential for understanding the
damping of neutron star oscillations and supernovae evolution. Heat diffusion coefficient determines the early
cooling evolution of neutron stars. The evaluation of these physical parameters can be done within the Landau
theory by assuming a macroscopic deviation from equilibrium and solving the kinetic equation in stationary
condition. The (linear) relation between the quantity that describes the deviation from equilibrium and the
corresponding flux of the quantity that is driven by the deviation. A temperature gradient drives a heath flux, a
fluid velocity gradient produces a transmission of momentum perpendicularly to the gradient.
\par Generally speaking, if local thermodynamical equilibrium is established in a Fermi system of particles,
the collision integral vanishes, see equation \ref{eq:coll}, no flux can flow and the driving quantity is the
deviation from local equilibrium $\delta^{leq} n(\epsilon,r)$. However, in a system of interacting quasi-particles
one has to consider the true quasi-particle energies, that depends in turn on the quasi-particle distribution
itself, see equation (\ref{eq:qpe}). Therefore the collision integral is expanded with respect to the deviation of
the quasi-particle distribution from the particle local equilibrium distribution.
The profile of this deviation is then determined to first order by solving the kinetic equation that includes the
collision term. In stationary conditions the kinetic equation becomes a linear integral equation for the
quasi-particle local equilibrium deviation, where the kernel of the integral is determined by the collision
integral under the specific physical situation. Along these lines various approximations to solve the integral
equations have been developed, until the works by Brooker and Sykes \cite{BS} and Jensen, Smith and Wilkins
\cite{JSW1,JSW2}, who supplied the exact analytical solutions for both the viscosity $\eta$ and the thermal
conductivity $\kk$, besides the spin diffusion coefficient. The details of the derivation can be found in the
original work or in reference \cite{Peth_Gord}. The results involve in all cases the same angular integral over
the probability $W$ for elastic scattering of two quasi-particles at the Fermi surface
\beq
I_W \,=\, \int {d\Omega \over 4\pi} {W(\theta,\phi)\over \cos(\theta/2)}
\label{eq:IW}\eeq
\noindent where $W$ is proportional to the square of the matrix element of the scattering $T$-matrix, see equation
(\ref{eq:coll}). Because of conservation of momenta and energy, $W$ depends only on two angles, here taken
according to the Abrikosov-Khalatnikov convention, see references \cite{AK,Peth_Gord}. The results for the thermal
conductivity and the shear viscosity read
\beq
\begin{array} {rl}
\kk &\,=\, {1\over 2\pi^2} C_V v_F^2 \tau A_K \nonumber \\
 \  &\                     \nonumber \\
\eta &\,=\, {1\over 5} p_F v_F^2 \tau A_\eta
\end{array} \label{eq:Keta}\eeq
\noindent where $v_F$ is the Fermi velocity, $C_V$ the specific heath, $A_K$, $A_\eta$ are numerical factor that
can be expressed in terms of numerical series, and $\tau$ is related to $I_W$ by
\beq
\tau \,=\, 8\pi^4\hbar^6/(m^{*3} I_W T^2)
\eeq
\noindent that has the meaning of a relaxation time. Actually the factors $A_K$ and $A_\eta$ depend also on the
probability $W$ through an additional integral, characteristic of each one of them. This must be carefully
considered when comparing different models for $W$.
\par
Several estimates of the shear viscosity have been presented in the literature, based on different models for $W$
and different approximation schemes. The most recent microscopic calculations \cite{omar1,omar2,umbe}, where
references to previous works can be found, stress the necessity of a treatment consistent with the nuclear matter
EoS. These microscopic calculations show a fair agreement. However, further studies are needed to establish on a
firm basis the value of $\eta$ and its density dependence. In addition, a complete treatment in the asymmetric
matter present in neutron stars is still missing.
\par As already mentioned in section \ref{giant}, the shear viscosity is not relevant in the dynamics of finite
nuclei at low excitation energy. The use of dissipative hydrodynamics, that involves the value of $\eta$, in heavy
ion collisions has been rather limited. Collision dynamics appears too complex to allow any study on the relevance
of the shear viscosity, not to speak of its possible value.
\par
The thermal conductivity in neutron stars is dominated by the electron component. The specific heath of baryon
matter can play a role in the cooling process. Only recently microscopic calculations have been developed,
 but the superfluid properties of neutron matter must be included \cite{Nicu1,Nicu2}.  Still much work has to be
done in this field.
\par
Finally, one should consider also the bulk viscosity of the nuclear medium. However, the main contribution to the
bulk viscosity in the nuclear matter present in compact stars is coming from the weak processes, and therefore it
falls outside the scope of the present review. In any case it has been extensively studied, with somehow
controversial results to be clarified \cite{Alf_bulk,Haen_bulk}. Furthermore, the possible hyperon component can
be of decisive relevance \cite{Jha}

\subsection{Collective excitations \label{collective}}
One of the fundamental results of the Landau theory of Fermi liquid, and nuclear matter in particular, is the
possibility of collective microscopic excitations. Indeed, since the foundation of the theory, it has been shown
by Landau that at low excitation energy the quasi-particle interactions can produce a concentration of the
excitation strength at a particular value of the energy, where the response function displays a pole, i.e. a
resonant behavior, that physically corresponds to a coherent motion of quasi-particles. Microscopically, the
collective excitation is produced by the particle-hole interaction, as displayed in equation (\ref{eq:st}). If we
limit the interaction to such a form, the possible excitations can be classified according to the values of the
total spin $S$ and isospin $T$. The $S \,=\, T \,=\, 0$ excitations correspond to density oscillation of the
system. In finite nuclei they correspond to the iso-scalar monopole giant resonance. The function $f$ in equation
(\ref{eq:st}) determines the position and strength of the excitation. If the interaction is zero range (contact
interaction), then only the value of $f_0$, i.e. $f_l$ for $l \,=\, 0$, is different from zero. The value of $f_0$
is therefore a fundamental constant that characterizes the nuclear medium. The onset of this collective motion can
be seen by linearizing the kinetic equation (\ref{eq:BUU}) with respect to the variation of the quasi-particle
distribution function, since we are considering small oscillations of the system. We put then $n(p,r,t) \,=\,
n_0(p) \,+\, \delta n(r,p,t)$, where $n_0$ is the ground state quasi-particle distribution, i.e. a sharp Fermi
distribution, and neglect any quadratic term in $\delta n$. We also neglect the collision integral $I$, that is we
assume that the collision frequency is small with respect of the oscillation frequency. This means also that the
quasi-particles decay can be neglected. After Fourier transform of the equation at the momentum $\mathbf{q}$ and
frequency $\omega$ and simple manipulations, one gets, in the long wave-length limit ($q \rightarrow 0$)
\beq
 \left( \omega \,-\, \mathbf{q}\cdot \mathbf{v_p} \right) \delta n(\mathbf{q},\mathbf{p},\omega)
 \,+\, {\partial n_0 \over \partial \epsilon_p} (\mathbf{q}\cdot \mathbf{v_p})
  \sum_\mathbf{p'} f(\mathbf{p'} - \mathbf{p}) \delta n(\mathbf{q},\mathbf{p},\omega)
\label{eq:zsound1}\eeq
\noindent This is an eigenvalue equation for the frequency $\omega$ at the momentum $\mathbf{q}$. Actually the
equation depends only on the ratio $\omega/q$, which is the propagation velocity of the oscillatory wave. The
corresponding distribution distortion $\delta n$ can be obtained by noticing that the derivative ${\partial n_0
\over \partial \epsilon_p}$ equals $- \delta(\epsilon - \epsilon_F)$, since $n_0$ is just a sharp Fermi
distribution. Then also $\delta n$ is proportional to the delta function
\beq
\delta n(\mathbf{q},\mathbf{p},\omega) \,=\, \delta(\epsilon - \epsilon_F) \xi(\mathbf{q},\mathbf{p},\omega)
\label{eq:csi}\eeq
\noindent If the interaction has only the $l = 0$ term different from zero, then $\xi$ depends only on the modulus
of $\mathbf{p}$, and substituting equation (\ref{eq:csi}) in equation (\ref{eq:zsound1}) one gets an explicit
eigenvalue equation for $\omega$. Putting $s \,=\, \omega/qv_F$, it reads
\beq
 1 \,+\, {1\over 2} F_0 \int_{-1}^1 d(cos\theta) {cos\theta \over cos\theta - s} \,=\, 0
\label{eq:zsound2}\eeq
\noindent where $F_0 = N f_0$. If $s > 1$, corresponding to a repulsive interaction $f_0 > 0$, the integral is non
singular and one gets a dispersion relation for $s$
\beq
1 \,+\, {1\over 2} F_0 \left( 1 \,+\, \ln {s - 1\over s +1} \right) \,=\, 0
\label{eq:disp}\eeq
\noindent This solution is called "zero sound". It is an excitation mode of purely quantal origin, typical of any
Fermion liquid with a repulsive interaction. As $F_0$ increases from $0$ to large values, much larger than $1$,
the solution $s_0$ of the dispersion relation varies from $1$ to $\sqrt{F_0/3}$. As already anticipated, this mode
has an analogous mode in finite nuclei in the isoscalar monopole giant resonance, that can be then considered as
the zero sound in finite nuclei. In nuclear matter other types of zero sounds can exist, corresponding to
different spin-isospin total quantum number. The $T = 1$ and $S = 0$ corresponds to the dipole giant resonance in
nuclei, the $T = 0, S = 1$ to the spin magnetic mode, the $T = 1, S =1$ to the Gamow-Teller resonance. They have
been all extensively studied in nuclei. If one keeps the correspondence with finite nuclei, it is possible to have
phenomenological indications on the Landau parameters from the positions of the giant resonances in nuclei. It
turns out, following this line, that the Landau parameter $F_0$ must be slightly negative, approximately  $-0.4 <
F_0 < 0$. The analogous Landau parameter $G_0 = N g_0$ should be very small, due to the observed lack of
collectivity of the spin mode in nuclei. The value of $F_0'$ for the dipole mode should be also slightly negative,
but the extrapolation from nuclear matter to nuclei is questionable, because in nuclei there is a substantial
contribution of the surface to this mode. The Landau parameter for the Gamow-Teller mode is much less known due to
the substantial contribution of the $\Delta$ excitation, not included in the usual Landau treatment. The Landau
parameters characterize the nuclear medium at fundamental level, but they are only partially known.
\par
Going back to the dispersion relation (\ref{eq:disp}), let us consider the case $s < 1$, that should appear when
$F_0 < 0$. The integral is then singular and one has to specify how to handle the singularity. Physically
speaking, if $s < 1$ the possible eigenfrequency falls inside the unperturbed particle-hole continuum. In fact the
unperturbed continuum, in the long wavelength limit, spans the excitation energy up to $+qv_F$. This means that
the mode couples directly to the particle-hole continuum and can decay. This implies that one should look for a
complex solution of the dispersion relation, whose imaginary part will provide an estimate of the damping decay
time of the mode. This decay mechanism is called "Landau damping" and is a phenomenon characteristic of Fermi
liquids in general. It is not related to the collisions between quasi-particles, and therefore it has no
connection to any sort of viscosity, being a purely quantal effect. It has to be stressed that when Landau damping
is active, the excitation mode is strongly damped and in practice it disappears.
\par
One could ask if Landau damping occurs in finite nuclei. In general the main strength of the giant resonances
falls in the single particle continuum, that is at excitation energy where particle can escape, and therefore
where the unperturbed particle-hole spectrum is continuous. In principle the Landau damping can therefore be
present, however at the same time also the so called spreading width is present, that is the coupling with more
complex configurations, like two particles-two holes states. In nuclear matter the latter turns out to be very
small with respect to the Landau damping, just due to phase-space restrictions. In finite nuclei Landau damping is
not so strong as in nuclear matter, due to the relatively small density of states at the excitation energy where
they are located. The width of giant resonances, as already mentioned, is therefore mainly a nuclear structure
problem, not related to the gross properties of the nuclear medium.
\par
\begin{figure}
\vskip -5. cm
 \begin{center}
\includegraphics[bb= 200 0 400 790,angle=0,scale=0.55]{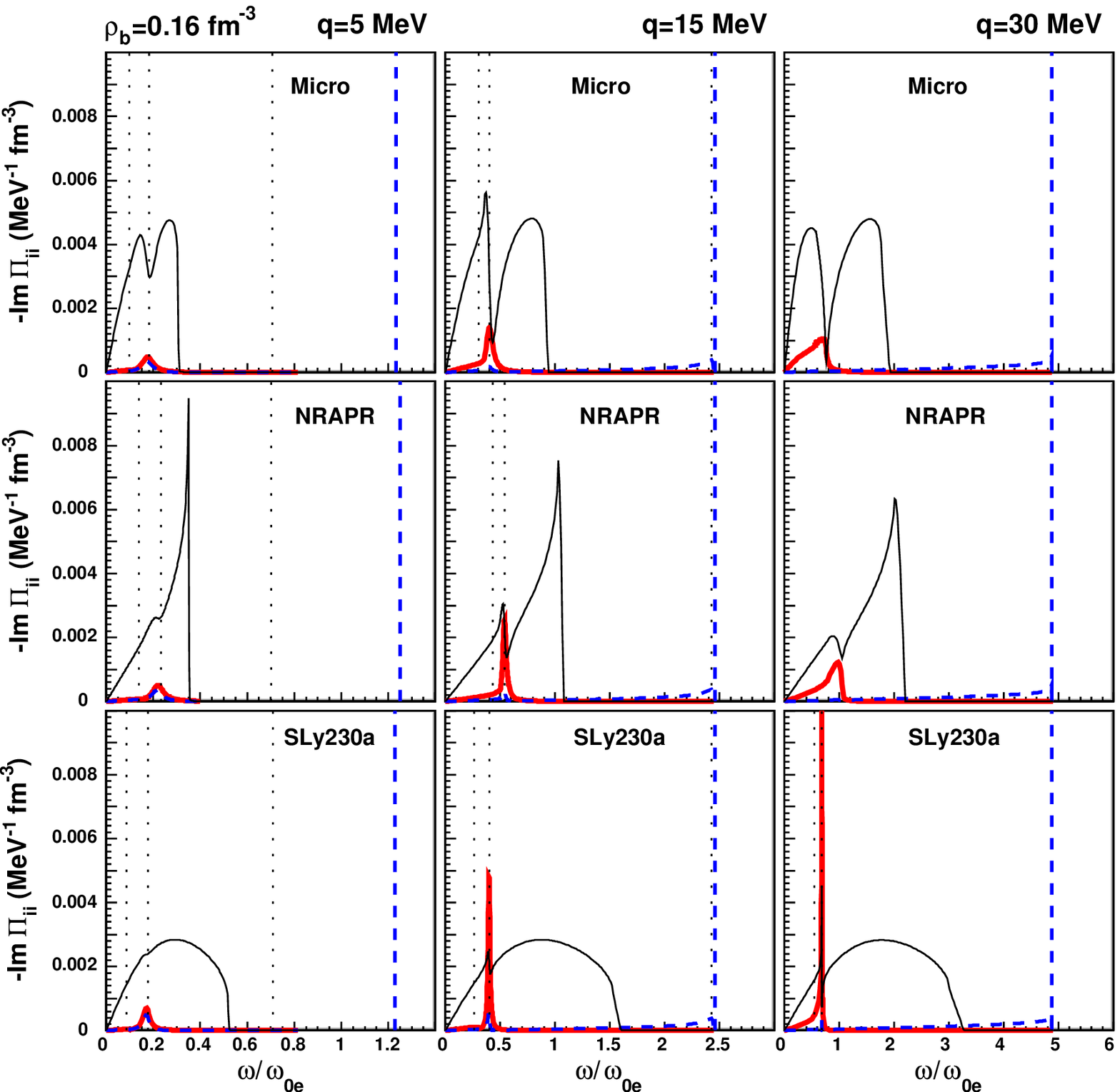}
\end{center}
\vskip -0.5 cm
   \caption{Strength functions in Neutron Star matter from reference \cite{paper1}. The thin line corresponds
   to the neutron component, the thick one to the proton component. The vertical dashed lines indicate
   the position of the excitation branches, in particular the highest in energy is the electron plasma excitation.
   For detail see the reference \cite{paper1}. }
    \label{fig:excit}
\end{figure}

The elementary excitations in the nuclear medium have relevance for the physics of neutron stars. In the
homogeneous region of the star these excitations affect the emission and propagation of neutrinos and the specific
heath of matter. They are expected to be present also in the crust region \cite{Khan}. For illustration we report
in figure (\ref{fig:excit}), taken form reference \cite{paper1}, the spectral functions of neutrons, protons and
electrons at a nucleonic density equal to the saturation one. The neutron density is much higher and therefore
also the strength function is much larger. One can see that the neutron strength is quite spread, showing that the
excitations are in the region of Landau damping. One distinguish the sharp electron peak corresponding to the
plasma mode. Neutron and proton excitations are mixed, but their interaction can be considered weak, because the
proton strength looks rather localized and not so much affected by the Landau damping. In this calculations three
possible effective interactions have been taken, corresponding to the Landau parameter $F_0$. Since matter is
asymmetric, in this case we have three different $F_0$ parameters, corresponding to neutron-neutron, proton-proton
and neutron-proton interactions. The first choice was taken from BHF calculations, the other two from particular
Skyrme forces. The matter was assumed to be normal. Superfluidity can change partly this picture
\cite{paper_super}. In any case the overall picture that comes from these analysis characterizes some of the
fundamental properties of the nuclear medium.
\par Finally, we mention that in nuclear matter, when the characteristic frequencies of the motion become very small,
the number of
quasi-particles collisions per period of the oscillations can become so large that local equilibrium can be
reached and the hydrodynamical regime is then established. Density oscillations can be still present and we have
in this case the "first sound" mode. This is virtually identical to the sound in classical fluid, like air, in the
same dynamical limit. In the hydrodynamical regime macroscopic motion are determined just by the conservation laws
and the EoS of the nuclear medium. Macroscopic physical parameters, as the ones discussed in sections
(\ref{stat},\ref{coeff}), then play the central role.
\subsection{Theoretical challenges}
We have seen that the fundamental parameters of Landau theory of Fermi liquid, as applied to the nuclear medium,
are only poorly known on the basis of laboratory experiments on nuclei. Astrophysical observations can also give
information on their values, but the data are still scarce and difficult to interpret. They are associated mainly
with observations on neutron star oscillation or supernova evolution. It would be therefore quite desirable to
have theoretical evaluations of their values and behavior with density on the basis of sound microscopical
many-body theories.
\par We have already discussed the scattering matrix terms which appear in the collision integral and the present
state of the art.
There is a vast literature on the theoretical evaluation of the Landau parameters $\{f_l, g_l, f_l', g_l'\}$ of
the quasi-particle interaction. Their {\it ab initio} theoretical determination is extremely difficult because of
the hard core character of the bare NN interaction. Among the different schemes that have been employed we can
mention the self-consistent Babu-Brown equations \cite{babub,backman}, the self-consistent Green' s functions
expansion \cite{Pines,Hans_NS} and more recently the Renormalization Group method \cite{Schwenk}. The main
difficulty in all methods lies in the estimate of the role of the screening processes. The quasi-particles can
exchange collective excitations, as the ones discussed in section \ref{collective}, where again the quasi-particle
interaction appears. This entails a self-consistent procedure to calculate the interaction itself. However the
results are sensitive to the scheme and approximations employed in the procedure and conclusive results are not
yet available. Furthermore, for asymmetric matter, as the one present in neutron stars, calculations are quite
scarce.

\section{Neutron Matter at very low density. An exercise in many-body theory. \label{low}}
 The low density region of pure neutron matter, as present in the inner crust of Neutron Stars, is less trivial than one
could expect at a first sight since the neutron-neutron scattering length is extremely large, about $-18$ fm, due
to the well known virtual state in the $^1S_0$ channel, and therefore even at very low density one cannot assume
the neutrons to be uncorrelated. These considerations have also stimulated a great interest in the so called
unitary limit, i.e. the limit of infinite (negative) scattering length of a gas of fermions at vanishingly small
density. A series of works \cite{carl,boro,bulg1} have been presented in the literature based on various
approximations, and a recent Monte Carlo calculation \cite{bulg2} on a related physical system has shown that the
unitary limit can present a quite complex structure, involving both fermionic and bosonic effective degrees of
freedom, which has still to be elucidated. Variational \cite{panda1} and finite volume Green's function Monte
Carlo calculations  \cite{panda2} for neutron matter at relatively low density have shown that the EOS, in a
definite density range, can be written as the free gas EOS multiplied by a factor $\xi$, which turns out to be
close to $0.5$. This is actually what one could expect in the unitary limit regime, since no scale exists in this
case, except the Fermi momentum. Monte-Carlo calculations \cite{carl,boro,bulg1} with schematic forces in a regime
close to the unitary limit have found a factor $\xi \approx 0.44$. The connection between the variational results
and the unitary limit has been studied in reference \cite{Pethick} by means of effective theory methods.
\subsection{A single G-matrix problem}
Since the scattering length $a$ and effective range $r_0$ in the $^1S_0$ channel of the neutron-neutron
interaction differ by about a factor 6, there is no density interval where the unitary limit can be considered
strictly valid. However, in the range  $r_0 < d < |a|$, where $d$ is the average inter-particle distance, the
physical situation should be the ``closest" possible to the unitary limit. This range corresponds to Fermi
momentum range $0.4$ fm$^{-1}$ $<  k_F  <$  $0.8$  fm$^{-1}$, which corresponds to densities between about 1/50
and 1/5 of the saturation density. Let us choose as realistic nucleon-nucleon potential the Argonne v$_{18}$
interaction \cite{wir}. The first finding is that the three-body forces of the Urbana model, adjusted to
reproduce the correct saturation point \cite{bbb}, give a contribution which is less than $0.01$ MeV, and
therefore we can neglect three-body forces to a good approximation.
 The second finding is that the single particle potential is very small in this density range, and its effect
 can be neglected. It affects the energy per particle less than $0.1$ MeV.
 \par
\begin{figure}
\vskip -4.5 cm
 \begin{center}
\includegraphics[bb= 200 0 400 790,angle=0,scale=0.55]{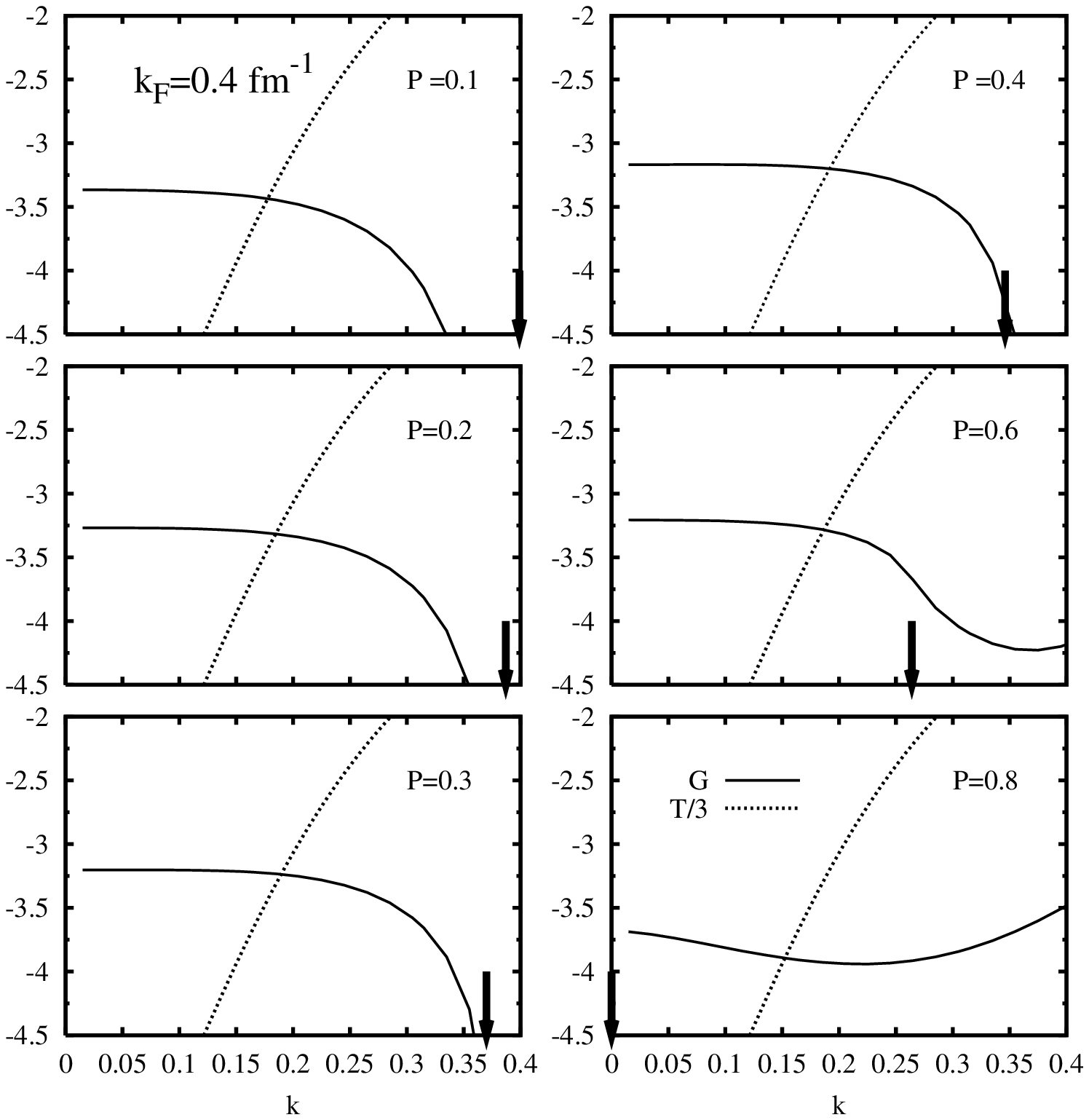}
\end{center}
\vskip -1.7 cm
   \caption{Plots of the free K-matrix (T), divided by 3 for convenience, in comparison with the
   in-medium K-matrix (G), at the indicated Fermi momentum, for different total momentum $p$, as a function
   of the relative momentum k. All the momenta are in fm$^{-1}$. The results are for the $^1S_0$ channel
   in pure neutron matter. The arrow indicates the maximum momentum needed in the calculation of the
   EoS. See reference \cite{low} for detail. }
    \label{fig:gt}
\end{figure}

 It is enlightening to compare the in-medium G-matrix with the free $K$-matrix  in the $^1S_0$
channel, reported in figure (\ref{fig:gt}), taken from reference \cite{low}, at selected values of the relative
momentum $k$ and total momentum $P$ (in fm$^{-1}$) at the Fermi momentum $k_F = 0.4$ fm$^{-1}$. For sake of
comparison the free $K$-matrix has been divided by 3.
Due to Galilei invariance, the free $K$-matrix is independent of $P$. Despite the Fermi momentum is quite small, a
drastic difference between the two scattering matrices is apparent, not only in shape but also in absolute value.
The Pauli operator effect is enhanced in this particular channel since the virtual state is suppressed in the
medium. This illustrates the dramatic difference that can exist between the in-medium effective interaction and
the free bare interaction.
The large enhancement at the Fermi momentum and for small total momentum $P$ is due to the pairing singularity, to
be discussed in section \ref{super}
The BBG expansion relies on the basic idea that the contributions of the diagrams of the expansion decrease with
increasing number of hole-lines which are included. Although the BBG scheme is essentially a low density
expansion, it has been found \cite{bal1,bal2} that the convergence is valid up to densities as high as few times
saturation density in symmetric nuclear matter and even better in neutron matter. It is then likely that at the
low densities we are considering this convergence should be even faster. This is indeed confirmed by explicit
numerical calculations \cite{low}, and indeed the third finding is that the three hole-line contribution is at
most $0.15$ MeV at the highest density considered and rapidly decreasing to vanishingly small values as the
density decreases. Finally, the fourth finding is that higher partial waves give a negligible contribution both at
the two hole-lines (Brueckner) and three hole lines level. These four findings, all together, point out that the
many-body problem of neutron matter at low density is reduced to a single G-matrix problem, i.e. to the
calculation of the $^1S_0$ G-matrix.
\subsection{The "exact" EoS}
\begin{figure}
\vskip -11. cm
 \begin{center}
\includegraphics[bb= 100 0 400 790,angle=0,scale=0.65]{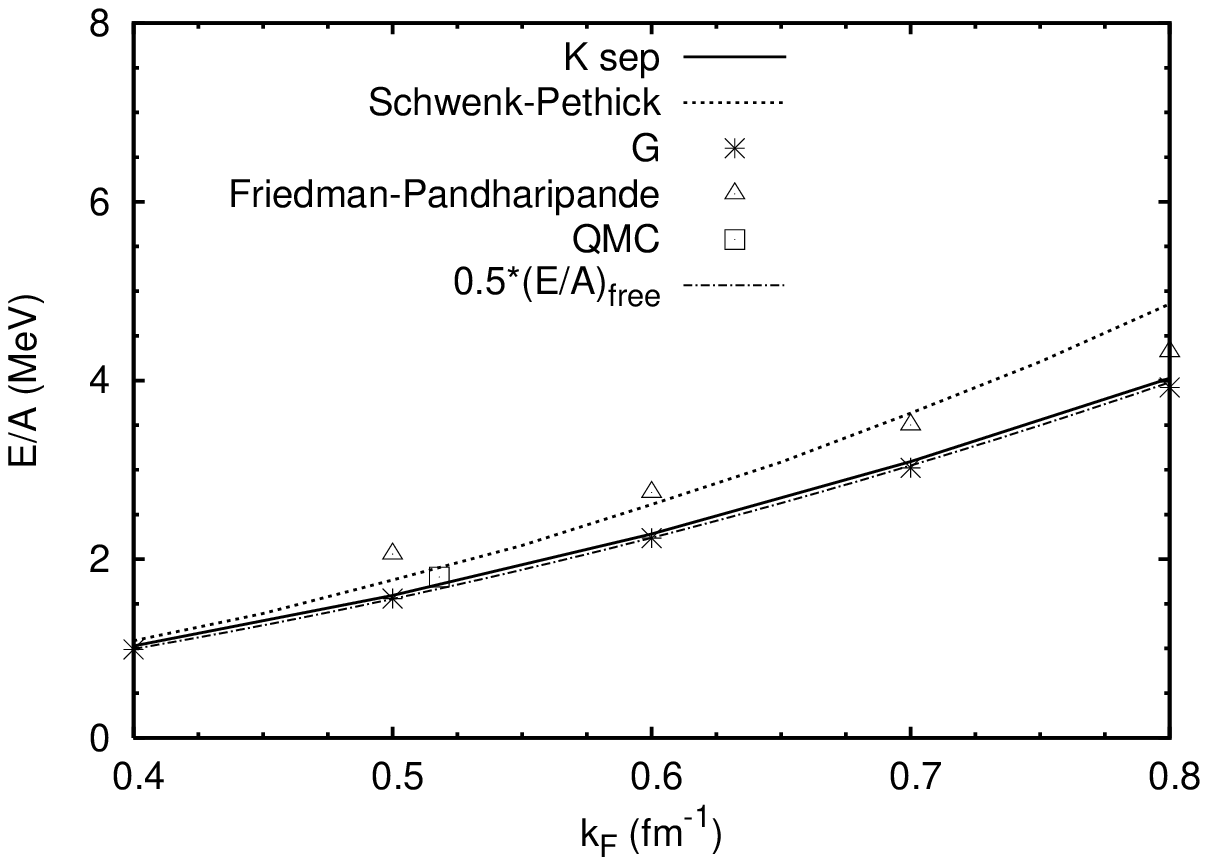}
\end{center}
\vskip -1.7 cm
   \caption{Neutron matter EOS calculated within the BBG method (label G),
 within the variational method of ref. \cite{panda1} (triangles), according to the
 estimate of ref. \cite{Peth_eff} (dotted line) and with the separable representation of the
 G-matrix (label G sep). The dash-dotted line is one half of the
 free gas EOS. The square represents the result of the Quantum Monte-Carlo calculation of reference \cite{Carlson1}.}
    \label{fig:eoslow}
\end{figure}

\begin{figure}
\vskip -11.5 cm
    [hb]
 \begin{center}
\includegraphics[bb= 100 0 400 790,angle=0,scale=0.65]{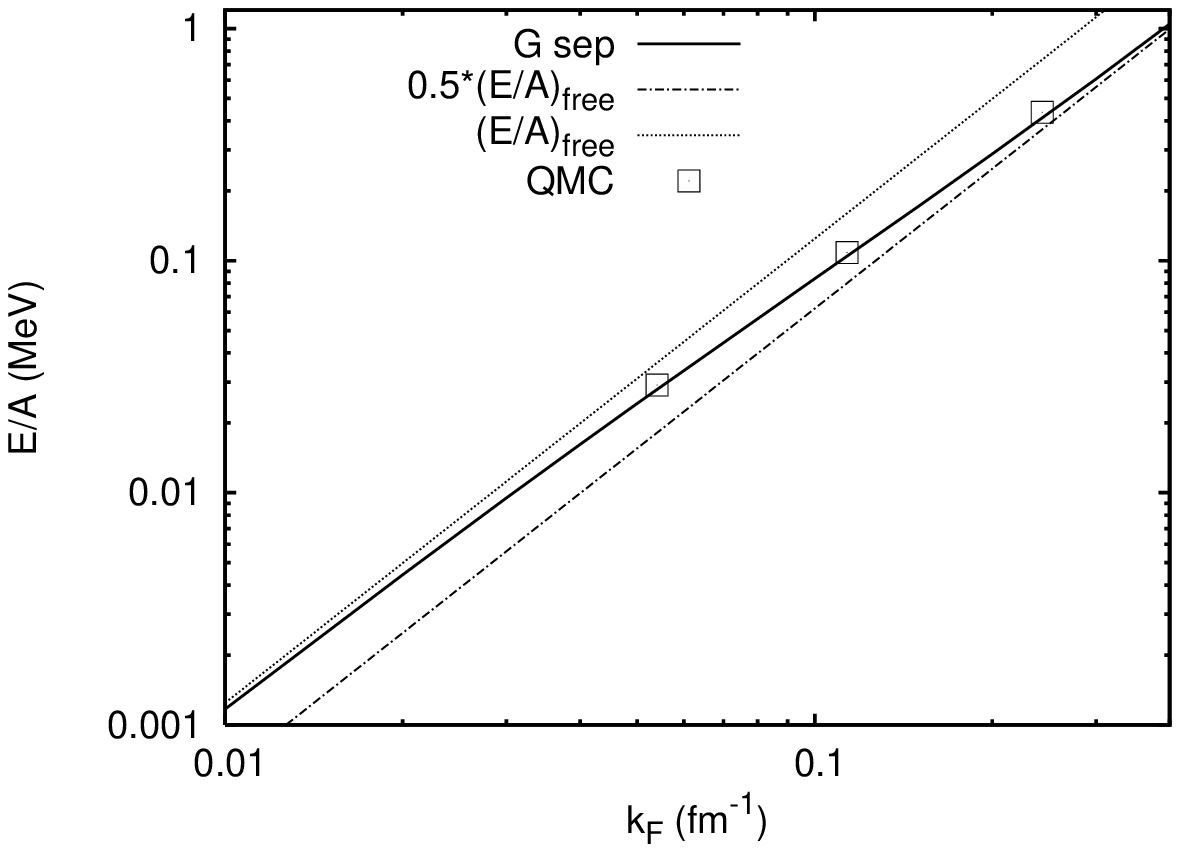}
\end{center}
\vskip -1.7 cm
   \caption{Neutron matter EOS calculated with the separable representation of the G-matrix,
   in comparison with the free Fermi gas EoS and one half of it.
   The squares represent the results of the Quantum Monte-Carlo
   calculation of reference \cite{Carlson1}. }
    \label{fig:eosvlow}
\end{figure}

The two EoS, one calculated within the full BBG expansion up to the three hole-lines contributions and the other
calculated with the single $^1S_0$ G-matrix, are compared in figure (\ref{fig:eoslow}), taken from reference
\cite{low}.
They are mainly indistinguishable. The energy per particle is very close to $1/2$ of the kinetic energy. It turns
out that the G-matrix is fully determined by the scattering length and effective range. One can construct a
rank-one separable interaction \cite{low}
\beq
 (k'|v|k)\, =\, \lambda\, \phi(k') \phi(k)
\label{eq:vsep}
\eeq
\noindent with a simple form factor
\beq \phi (k) \,=\,1 / (k^2 \,+\, b^2) \label{eq:form} \eeq
\noindent where the parameters $\lambda$ and $\beta$ are determined by imposing that the scattering length and
effective range are reproduced. Then the G-matrix can be obtained analytically and the corresponding EoS by simple
numerical integration. The procedure is equivalent to an Effective Theory with smooth cut-off and its accuracy is
shown in figure (\ref{fig:eoslow}). The calculation can be extended to very small density, as reported in figure
(\ref{fig:eosvlow}), where one can see that the EoS approaches the one for a free gas, as it must be for $k_F <
1/|a|$. In both figures (\ref{fig:eoslow},\ref{fig:eosvlow}) the squares indicate the results of the Monte Carlo
calculations of reference \cite{Carlson1}. They agree fairly well with the BBG results up to the density where the
Monte Carlo calculation can be performed.
\par What is missing in the BBG calculations is the pairing correlations.
The contribution to the EoS of pairing is not expected to be relevant, but it is important to know the value of
the gap for many phenomena in Neutron Stars and as an indication for finite nuclei. The subject will be taken in
sections \ref{gapnu} and \ref{super}.

\section{Bulk properties of Nuclear Matter}
In this section we try to illustrate our knowledge on some of the physical parameters that characterize the
properties of the nuclear medium. Some of them have been discussed in previous sections for matter close to
saturation density, so we concentrate mainly on their density dependence. A subsection is devoted to the possible
superfluid phases, that have not been discussed up to now.

\subsection{Density dependence of the symmetry energy}
Hints about the symmetry energy below saturation density have been claimed to come from experimental data on heavy
ion reactions at intermediate energy. The isoscaling regularity \cite{Betty1,Betty2} appears to be present in many
experiments on multi-fragmentation, from which an estimate of the density dependence of the symmetry energy can be
deduced \cite{Betty3}. Another possibility has been advocated \cite{Shi,Bao} to be the data on the so called
"isospin diffusion" in heavy ion reaction at moderate energy.\par  Microscopic calculations show a fair agreement
with these "data" and among each other up to saturation and slightly above. Comparison with results from different
Skyrme forces show that only few of them are compatible with this general behavior. In figure (\ref{fig:densy}),
taken from reference \cite{BCP-NPA}, is shown the symmetry energy in the low density region calculated in BHF
scheme in comparison with the few Skyrme forces that show a fair agreement with the microscopic calculations. The
most modern Skyrme forces are constructed in such a way to fit the microscopic results, and the agreement is
therefore enforced.
    \begin{figure}
\vskip 1.8 cm
 \begin{center}
\includegraphics[bb= 200 0 400 790,angle=-90,scale=0.35]{figures/densy.ps}
\end{center}
\vskip 1.8 cm
   \caption{The symmetry energy at low density. The symbols correspond to the Brueckner
   calculations with realistic forces, Argonne v$_{14}$ (diamonds), Argonne
v$_{18}$ (small open circles which correspond to the polynomial fit ) and Paris potentials (open squares). The
lines correspond to phenomenological nucleon-nucleon forces, the SkM$^*$ (solid line) and the Sly4 Skyrme forces
(short dashed line), and the Gogny force (long dashed line}
    \label{fig:densy}
\end{figure}
\noindent Above saturation the situation is less under control. One one hand the microscopic calculations need to
extend the use of three-body forces to density where they are not well known. On the other hand experimental data
on heavy ion collisions can provide hints on the density dependence of the symmetry energy only very indirectly
through extensive simulations. This approach is extensively reviewed in reference \cite{Bao_PR}, where the
prospectives in this line of research are presented in detail, in particular in connection with the development of
the facilities for exotic nuclei.
\par
Astrophysical observations on the mass of neutron stars are also indirectly testing the symmetry energy at high
density, because the value of the symmetry energy can change the value of the incompressibility of the very
asymmetric matter at the center of compact stars. As already mentioned, the interplay of the data extracted from
heavy ion reaction at intermediate energy, that test the EoS of almost symmetric matter, and the analysis of the
observational data on neutron stars, where the matter is highly asymmetric, can be of great help to clarify this
difficult but fundamental issue.

\subsection{Incompressibility}
The incompressibility near saturation has been already discussed. Below saturation density usually the EoS of
symmetric matter and neutron matter are assumed to be simple low order polynomial functions. This can be a
delicate point, since the small detail of the EoS behavior at low density, especially for symmetric matter, can be
relevant for the construction of accurate Energy Density functionals to be used in the calculations of the mass
table. Microscopic calculations look in close agreement in this density region, see figure (\ref{fig:EOS}). At
higher density, well above saturation, the microscopic theories face the problem, already mentioned, of the
three-body forces, that give a very large contribution but are not well known. As a reference density above which
this problem can be serious, one can take a value around 3-4 times saturation density. Also in this case the
interplay between theory and experiments is essential to make progress in this issue.

\subsection{Viscosity}
 The notion of viscosity has been already discussed within the Landau theory of Fermi liquid. Here the effective
 interaction that enter, in a way or in another, in the calculation of the EoS must be the basic quantity to be
 used to calculate also the scattering probability $W$ of section \ref{coeff}. For instance the G-matrix of the
 BHF scheme should be used as the basic quantity. Of course, at high density we meet the same problems as for the
 EoS.

\subsection{Superfluidity \label{super}}
\begin{figure}
\vskip -2.3 cm
 \begin{center}
\includegraphics[bb= 200 0 400 790,angle=0,scale=0.65]{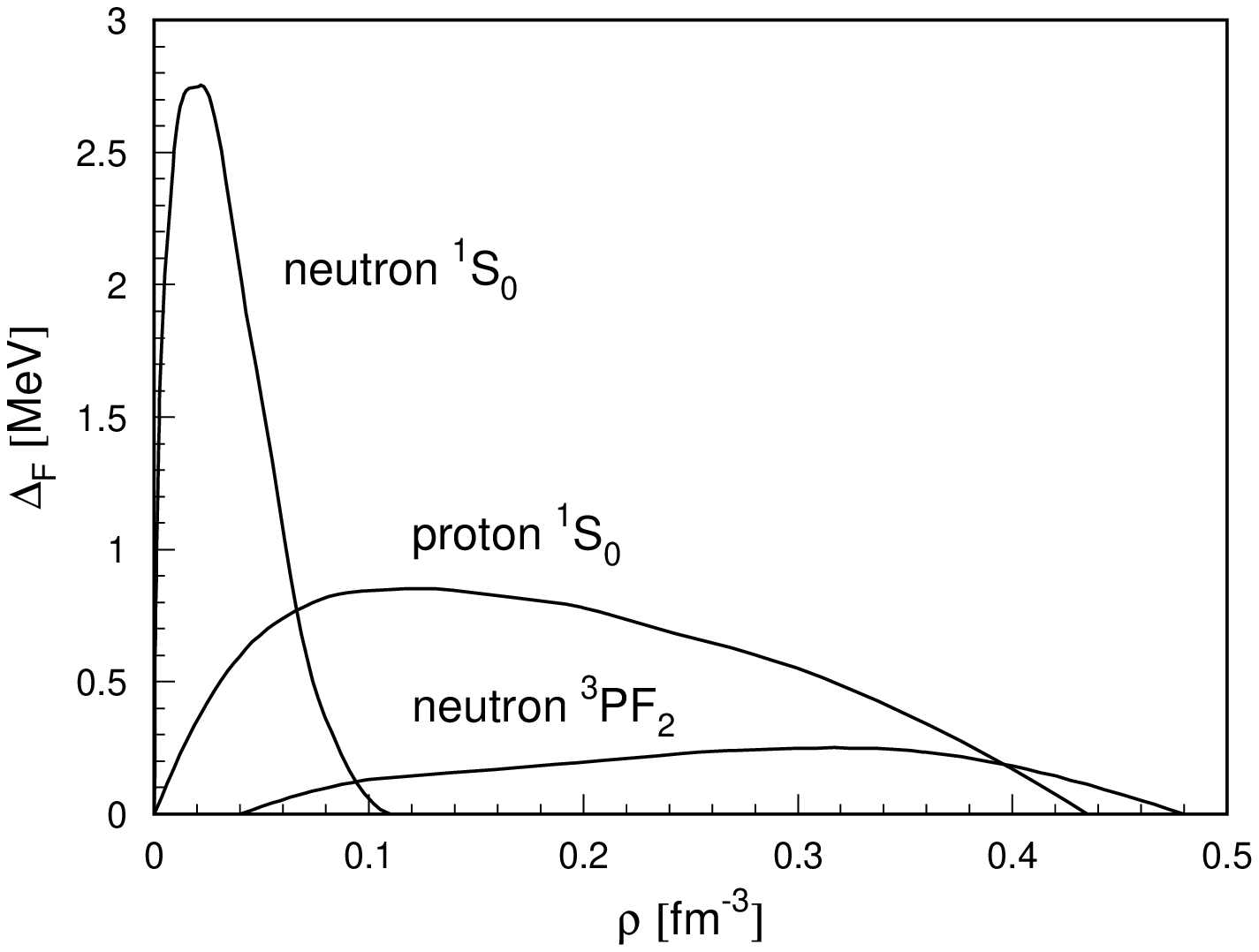}
\end{center}
\vskip -9.2 cm
   \caption{Different pairing gaps as a function of density in Neutron Star matter.}
    \label{fig:gaps}
\end{figure}

It has been argued a long ago \cite{Alp1,Alp2} that to explain the observed glitch phenomenon in several pulsars
it would be natural to assume that nuclear matter is superfluid in the interior of neutron stars. In fact, the
sudden increase of the rotational frequency and the long time needed to recover the initial rotational frequency
suggest the presence in the crust of an almost decoupled component with low viscosity. Since then a vast
literature has developed on the subject, and different superfluid phases have been found theoretically at the
physical conditions expected inside neutron stars. It is well known that theory of superfluidity, or better
superconductivity, was first formulated by Bardeen, Cooper and Schiffrer \cite{BCS}. In a more general
formulation, the constitutive equation for the onset of a superfluid phase is the gap equation \cite{Nozieres,Mig}
\beq
\Delta(\mathbf{k}) \,=\, - {1\over 2}\sum_\mathbf{k'} {\cal V}(\mathbf{k},\mathbf{k'}) {\Delta(\mathbf{k})\over
E_k}
\label{eq:gap} \eeq
\noindent Here the gap function $\Delta$ is related to the pairing correlation function, or "pair wave function"
by
\beq
{\Delta(\mathbf{k})\over 2 E_k} \,=\, < \psi^\dagger(\mathbf{k}) \psi(-\mathbf{k}) >
\eeq
\noindent For simplicity we assume pairing in the s-wave and the particles that form the Cooper pair have opposite
momentum and spin. Then the so called quasi-particle energy is given by
\beq
E_k \,=\, \sqrt{e_k \,+\, \Delta_k^2}
\eeq
\noindent where $e_k$ is the single particle spectrum in absence of pairing interaction. In this case all
quantities depend only on the modulus of $\mathbf{k},\mathbf{k'}$. The kernel ${\cal V}$ is the irreducible
particle-particle interaction. In the many-body language, it is the sum of all diagrams that describe the
interaction of two particles and that cannot be divided into two disconnected parts by cutting two particle lines.
For strong pairing correlation it also depends on the pairing gap $\Delta$. The exact equation for the pairing gap
is more complex than equation (\ref{eq:gap}). It can be formulated in terms of single particle Green' s functions
and it can be found in textbook \cite{Nozieres,Mig}. We will discuss later the possible modifications, but
equation (\ref{eq:gap}) already contains some of the main features of the pairing problem for the nuclear medium.
For neutron-neutron or proton-proton pairing the only s-wave channel is the $^1S_0$ two-body channel, and we will
first discuss this case. Equation (\ref{eq:gap}) is a non-linear equation for the gap function $\Delta(k)$, due to
its presence in $E_k$ in the denominator. The onset of the superfluid phase of the nuclear medium is indicated by
a non-zero solution for the gap. The zero solution $\Delta(k) = 0$ is always present, but it can be shown that the
energy of the superfluid phase, if it exists, is always lower than the normal phase. This is of course a general
feature. What characterizes the pairing correlation in the nuclear medium is the value of the gap, that is
estimated to be a substantial fraction of the Fermi energy. Another feature is that the effective interaction is
not concentrated around the Fermi surface, like in ordinary superconductor or in superfluid $^3He$. Although a
large contribution in the gap equation comes from the interaction at the Fermi momentum, also momenta that are far
away from the Fermi surface are essential and the pairing gap function $\Delta(k)$ is momentum dependent. All that
can be seen by considering the simplest approximation for the pairing interaction, the bare NN interaction. It
turns out that the bare NN interaction is able to produce a substantial pairing gap at low density, but, due to
the hard core of the interaction, high momentum components must be taken into account. The approach can be
extended to other channels, and the possible pairing gaps at the Fermi surface in this approximation as a function
of density in neutron star matter is depicted in figure (\ref{fig:gaps}), taken from reference \cite{hanumbe}. In
this evaluation of the different pairing gaps the single particle spectrum has been taken form BHF calculations
for the neutron star matter.
\noindent Accordingly, the concentration of protons has been calculated at the beta equilibrium and the nuclear
matter EoS from the BHF theory. Other microscopic EoS would give anyhow similar results. This is mainly equivalent
to the use of the effective mass. In this case the effective mass is smaller than the bare one. This reduces the
pairing gap because it increases the kinetic energy and reduces the density of states. The effect is larger at
higher density, such as for the proton and the $^3PF_2$ neutron pairing, and it is almost negligible for the
$^1S_0$ neutron pairing. The proton-proton pairing is shifted at higher density with respect to neutron-neutron
one just because the fraction of proton is varying smoothly from few percents to about 10-15 \%, according to
adopted EoS. The p-wave neutron pairing is present only at high density because the phase shifts of the $^3PF_2$
channel becomes appreciable only at higher momentum, and so at higher Fermi energy. \par This overall picture is
very important for many phenomena that occur in neutron stars, in particular for the cooling process and the
glitches events. However the quantitative description of the observational data requires an accurate prediction of
all these pairing gaps. This turns out to be an extremely difficult task. First of all the pairing gap is a quite
sensitive quantity and small variations of the interaction can change substantially its value. This can be seen in
the so called weak coupling limit, that corresponds to assume the interaction independent of the momentum with a
cutoff and approximate as constant the density of states, taken at the Fermi momentum. In this limit the pairing
gap can be obtained from a simple integration and it is a constant. It depends exponentially and non-analytically
on the interaction strength
\beq
\Delta \,=\, E_0 \exp ( 1/{\cal V}k_F m)
\label{eq:weak}\eeq
\noindent where the factor $E_0$ is related to the cutoff, but it varies smoothly and has a logarithmic dependence
on it. The non-analyticity is a manifestation that a phase transition takes place, for any small value of an
attractive interaction at the Fermi surface. \par
\begin{figure}
\vskip -1.3 cm
 \begin{center}
\includegraphics[bb= 200 0 400 790,angle=0,scale=0.62]{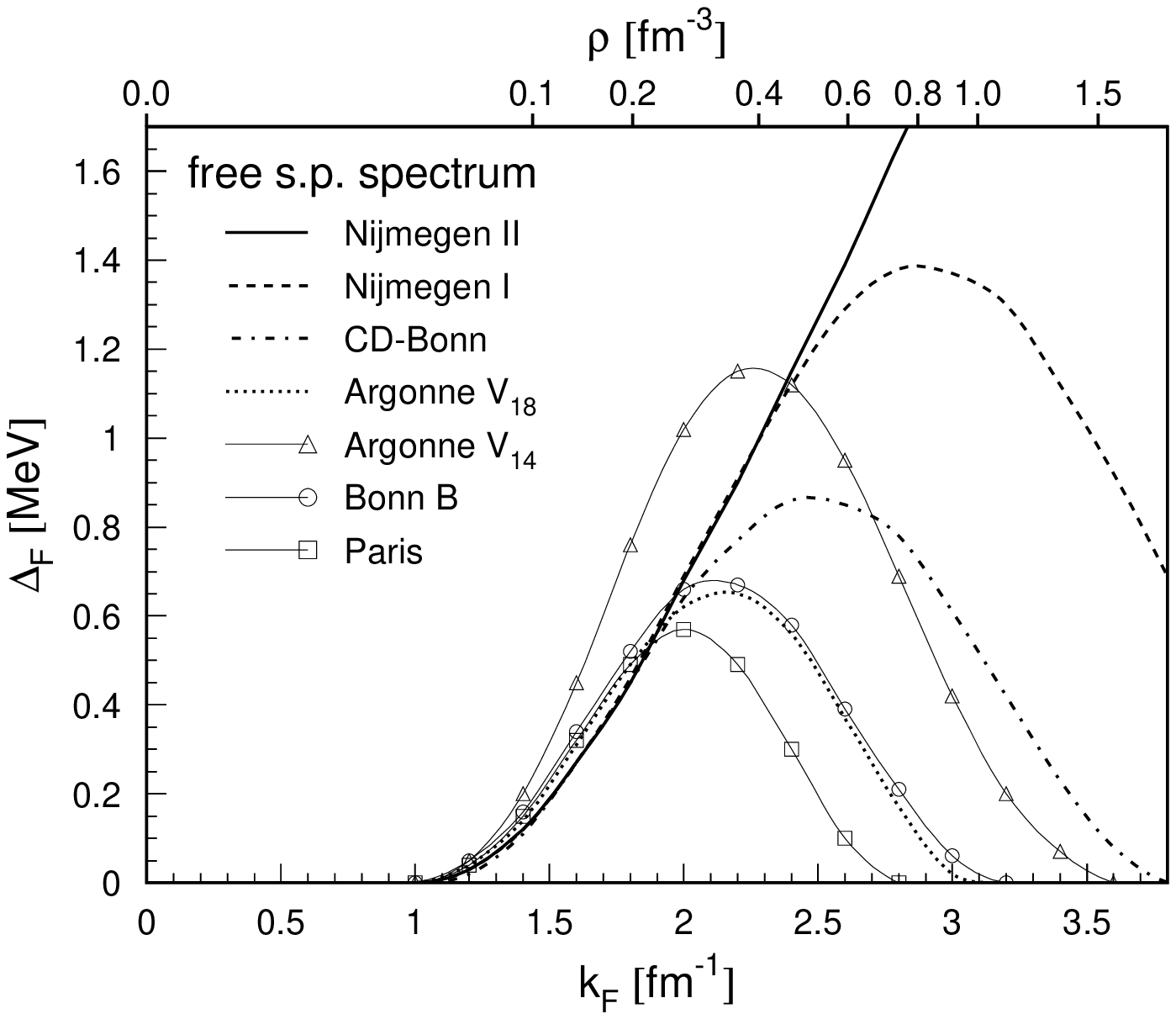}
\end{center}
\vskip -8.7 cm
   \caption{Pairing gap in the $^3PF_2$ channel as a function of density in neutron matter.}
    \label{fig:3p2}
\end{figure}

 Then the question arises if we know accurately enough the bare
NN interaction that the pairing gap so calculated is essentially independent on the particular realistic NN
interaction. This turns out to be true for the $^1S_0$ neutron and proton pairing gap, but it is invalid for the
$^3PF_2$ channel at the higher density. This is because the phase shifts are known up to about relative momentum
$k = 2$fm$^{-1}$, above which any potential give essentially extrapolated values. This can be dramatically seen in
figure (\ref{fig:3p2}), where the pairing gap in the $^3PF_2$ channel, calculated in the BCS approximation, is
reported for different interactions. Above $k_f = 2$fm$^{-1}$ the diverging results is a manifestation of this
uncertainty. Further uncertainty is introduced at the higher density, where three-body forces start to play a
role.
Besides this basic uncertainty, not present for the lower density, the BCS approximation must be implemented by
including the many-body effects in the gap equation. Correlations affect the single particle motion. As already
mentioned, the momentum dependence of the normal part of the self-energy introduces the single particle potential,
that can be approximated by substituting the bare mass with the effective mass. Besides that, the energy
dependence introduces the so-called quasi-particle strength or Z-factor, that gives the weight with which the
quasi-particles can take part to a Cooper pair. More generally, if one assumes that the effective pairing
interaction is energy independent, the gap equation can be written \cite{Mig}
\beq
\bar {rl}
\Delta(k)\, &=\, - \sum_{k'} {\cal V}(k,k') {\Delta(k)\over {2{\cal E}(k)}} \nonumber \\
 \    \  \nonumber \\
{1 \over {2{\cal E}(k)}}\, &=\, {1\over \pi}\int d \omega'{\rm Im}({1\over D(k,\omega')}) \nonumber \\
\     \   \nonumber \\
 D(k,\omega)\, &=\, (\tilde{\epsilon}_{k} - \omega + M(k,\omega))\cdot
         (\tilde{\epsilon}_{k} + \omega + M(k,-\omega)) +
         \Delta (k)^2
\ear
\label{eq:dispsig}\eeq
\noindent where $\tilde{\epsilon}_{k} = e_k - \mu$, being $e_k$ the free spectrum and $\mu$ the chemical
potential, and $M(k,\omega))$ is the normal component of the self-energy. It can be calculated in normal
(non-superfluid) nuclear matter, e.g. within the BHF scheme, since the pairing gap is small with respect to the
Fermi energy. It contains both a real and imaginary part. The factor ${\rm Im}(1 / D(k,\omega'))$ appearing in
this gap equation is the single particle strength function, that describes the distribution in energy of the
single particle at momentum $k$ inside the medium. The strength function is an even function of the energy and has
two poles at positions symmetrical with respect to the imaginary axis. If one takes only their contributions to
the energy integral, which corresponds to the quasi-particle approximation, the gap equation reduces to
\beq
\!\!\!\!\!\!\!\!\!\!\!\!\!\!\!\!\!\!\!\!\!
\Delta (k) \, = \, -\sum_{k'} {\cal V}(k,k') Z_{k'}
{\Delta_{k'}\over
 2 \sqrt{ \left[ \tilde{\epsilon}_k +
 {1\over 2} {\rm Re}(M(k,-E_k) + M(k,E_k))\right]^2
 + \Delta(k)^2 } }
\eeq
\noindent where $E_k$ and $-E_k$ are the real parts of the two poles energy, being $E_k$ the quasi-particle energy
in the superfluid nuclear medium, the factor $Z(k')$ is the anticipated residue at the poles, and ${\rm Re}$ means
real part. In the weak coupling limit, the presence of the self energy can be approximated by introducing the
effective mass, while the $Z$ factor (always smaller than 1) reduces the strength of the interaction. Equation
(\ref{eq:weak}) becomes
\beq
\Delta \,=\, E_0 \exp ( 1/{\cal V}k_F Z(k_F)^2 m^*)
\label{eq:mz}\eeq
\noindent Due to the exponential dependence, the $Z$ factor can substantially reduce the pairing gap. The total
effective mass at the Fermi momentum, when dispersive effects are included, turns out to be close to 1 in the
density range where neutron matter is expected to be superfluid. These effects have been studied by several
authors, both using the more general equation (\ref{eq:dispsig}) and the pole approximation
\cite{Alb1,Alb2,Cao,Bozek1,Bozek2} of equation (\ref{eq:mz}). The reduction of the gap looks to be moderate,
approximately 20-30\%.
\par The estimate of the many-body effects on the pairing interaction ${\cal V}$ turns out to be the most difficult
task. The main problem seems to be the inclusion of the medium polarization effects on the effective interaction.
This problem has been approached by a variety of techniques and approximations. The results are summarized in
reference \cite{hanumbe}. They all display a strong reduction of the pairing gap in the neutron $^1S_0$ channel,
but they do not agree on the size of the reduction, that has a value between 0.5 and 0.2 or even smaller. The
discrepancy is again related to the extreme sensitivity of the pairing gap to the value of the effective pairing
interaction, that requires an accuracy difficult to reach in this many-body problem. More recently the
Renormalization Group method has been used \cite{Schwenk}. The results look similar to the previous calculation in
reference \cite{Pines}. In the region of lowest density also Monte-Carlo calculations are available
\cite{Carlson1,Carlson2,Fantoni}. Despite some discrepancies among the results, they seem to indicate a much
smaller reduction of the pairing gap for this channel. If this is due to some limitation of the Monte-Carlo
calculations or to a drawback of the other microscopic many-body methods is a question that is waiting for an
answer.
\par
\begin{figure}
 \begin{center}
\includegraphics[bb= 200 0 400 790,angle=0,scale=0.55]{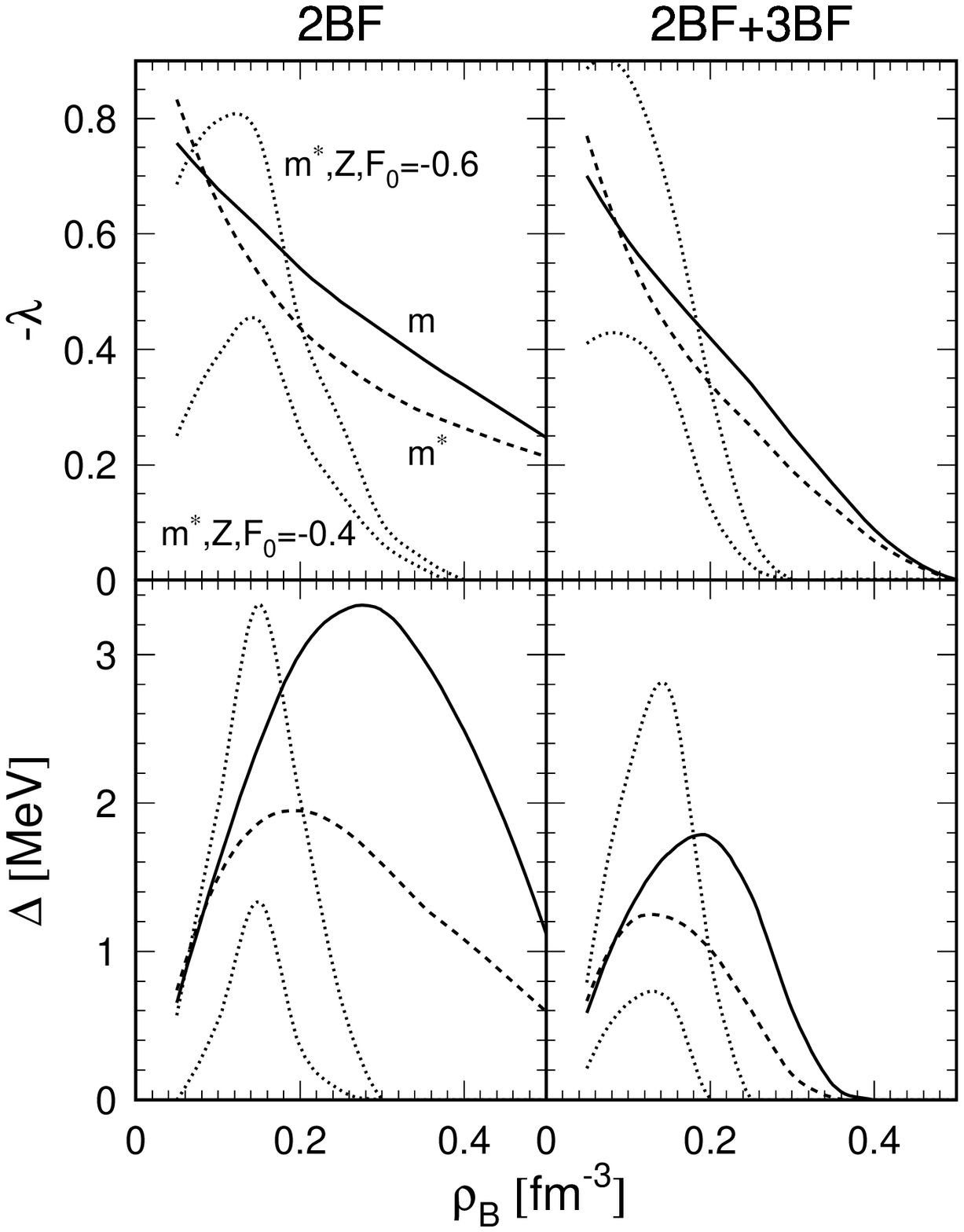}
\end{center}
\vskip -5.5 cm
   \caption{Weak coupling parameter $\lambda$ (top panel) and gap $\Delta$ (bottom panel) in several approximation.
   Here $\Delta \sim \exp(1/\lambda)$, see equation (\ref{eq:weak}). The solid curves show BCS results without
   any in-medium effects, the dashed curves include the modification of the effective mass $m^*$, and the
   dotted curves take account of the Z-factor and the polarization corrections in addition. The left panels show
   results with only two-body forces in the interaction kernel of the gap equation, and the right panels
   include also three-body forces.}
    \label{fig:pppair}
\end{figure}

The $^1S_0$ proton pairing gap is more difficult to treat, because the much higher density of neutrons can have a
dominant effect. The most recent calculation \cite{hanma} indicates that the effect of the neutron component
strongly enhance the effective pairing interaction through the tensor component of the NN interaction, but this
effect is in competition with the strong reduction due to the effective mass and $Z$-factor. To illustrate how
difficult is to control all these competing effects in figure (\ref{fig:pppair}) is reported the pairing gap when
each one of these many-body contributions is introduced. Notice that without the neutron polarization effect the
gap would be zero. The final gap turns out to be more concentrated at lower density with a smaller strength.
Further studies are surely needed to confirm these results.
\par
The $^3PF_2$ neutron pairing gap turns out to be much smaller and relevant only in a deeper region of a neutron
star. Application \cite{AkimPRL} of the Renormalization Group method to this channel brings the gap to a very
small value, in the range of few tens of KeV, even compatible with a vanishing one.
\par
Observational data on neutron star cooling can indirectly give indications about the values of all these pairing
gap. Analysis based on the assumption of pure nucleonic matter \cite{dimachri} constraints the gaps to values that
are compatible with the theoretical predictions. In particular the $^3PF_2$ is seen to be necessarily quite small,
of the same order as the theoretical results. The proton pairing gap is deduced to be indeed restricted to
relatively lower baryon density, mainly outside the inner core. For the neutron $^1S_0$ the constraints are less
stringent, but anyhow compatible with most of the theoretical predictions.\par The possible onset of exotic matter
components, like hyperons and quarks, complicates noticeably the analysis. These components can be also
superfluid, but unfortunately their pairing gaps are even more uncertain than in the nucleonic case.
From the observational data on glitches one can extract indications on the pinning energy of vortices, and
therefore indirectly on the pairing gap. However this phenomenon needs still a complete dynamical theory, before
any conclusion on both the pairing gap and the pinning energy can be deduced.
\par
The possible link of these results with the pairing phenomenon in nuclei is discussed in section \ref{gapnu}.


\section{Connection with Nuclear Structure}
One of the basic questions that was posed since the first developments of Nuclear Physics is to what extent the
properties of the bulk nuclear medium can be transferred to finite nuclei or to what extent they influence the
general trends that are observed in Nuclear Structure studies. Otherwise stated, is there a link between the
nuclear medium properties and the structure of finite nuclei ? The simplest scheme where this link is exploited is
the Liquid Drop Model, introduced in section \ref{mass}. As explained, in this semi-classical model the assumed
constant saturation energy per particle is corrected by the surface and Coulomb energy to explain the binding
energy of finite nuclei. In more sophisticated variants of the model other terms are introduced, but in any case a
set of parameters are introduced which cannot be derived from the theory. Furthermore, quantal characteristics
like shell effects are not included. We will briefly describe some of the methods that have been used and
developed to understand more deeply this possible link and, at the same time, to devise practical and general
theoretical scheme for Nuclear Structure.
\subsection{The Thomas-Fermi approximation and implementations \label{TF}}
Historically, the first method to relate the properties of finite quantal systems to the corresponding homogeneous
system was the semi-classical Thomas-Fermi scheme. It was devised to calculate the ground state properties of
large atoms and molecules in the case of independent particle limit. In fact, even in this limit, the quantal
calculations can become quite complex. The original form of the scheme is equivalent to take the zero order term
in the expansion in $\hbar$ of the density matrix for the independent particle wave function, but it can be easily
derived by assuming that the system is locally equivalent to a free Fermi gas at the local density and local
potential. We remind here some features of the approximation that are useful for the development of the
presentation. In its simplest version it can be formulated within the Density Functional method, i.e. assuming
that the energy of the system can be written as a functional of the density $\rho(\mathbf{r})$
\beq
E_{TF} \,=\, T_F(\{\rho\}) \,+\, V_c(\{\rho\}) \,+\, V_{pp}(\{\rho\})
\label{eq:funTF}\eeq
\noindent where $T_F$ is the kinetic energy contribution, $V_c$ the external potential and $V_{pp}$ the
particle-particle interaction. In case of atoms, $V_c$ is the energy due to the central Coulomb potential and
$V_{pp}$ is the electron-electron Coulomb interaction energy
\beq
\begin{array} {rl}
\!\!\!\!\!\!\!\!\!\!\!\! T_F(\{\rho\}) &=\, {3 \over 5}\int d^3r E_F(\rho(\mathbf{r}))\rho(\mathbf{r}) \ \ \ ; \ \
\ V_c(\{\rho\}) \,=\, -
Z\int d^3r' v(r)\rho(\mathbf{r'}) \nonumber \\
\   &\  \nonumber\\
\!\!\!\!\!\!\!\!\!\!\!\! V_{pp}(\{\rho\}) &=\, \int d^3r \int d^3r' v(|\mathbf{r} - \mathbf{r'}|)
\rho(\mathbf{r})\rho(\mathbf{r'})
\end{array}
\label{eq:termsTF}\eeq
\noindent with $Z$ the charge of the nucleus, $v = 4\pi e^2/r$ the Coulomb potential between two electrons (of
charge $e$) and $E_F$ the Fermi energy for a free gas at the density $\rho(\mathbf{r})$. The Euler-Lagrange
equation corresponding to the minimization of this functional with the constraint of a fixed number of particles
is
\beq
E_F(\rho(\mathbf{r})) \,-\, Z v(r) \,+\, \int d^3r' v(|\mathbf{r} - \mathbf{r'}|) \rho(\mathbf{r'}) \,=\, \mu
\label{eq:TF}\eeq
\noindent where $\mu$ is the Lagrange multiplier, that has the meaning of chemical potential. Notice that $\mu$ is
a constant, independent of the position. The solution of this equation gives the density and then the energy of
the ground state. A way of solving this equation is to apply the Laplacian differential operator and use the
Poisson equation $\Delta V_C \,=\, - 4\pi e^2 \rho(\mathbf{r})$, where $V_C$ is the last term at the left hand
side of equation (\ref{eq:TF}), that is the potential produced by all electrons at $\mathbf{r}$. This gives the
familiar differential equation of the Thomas-Fermi scheme in its simplest form \cite{March}. We will not discuss
the many refinements of the approximation that have been developed, to include e.g. the exchange interaction, but
rather we consider the Thomas-Fermi approximation in the nuclear case. The physical situation is rather different.
There is no central interaction, and the binding must come from the NN interaction. The latter is not long range,
like the Coulomb potential, but it is short range, even zero range if we adopt a Skyrme effective interaction.
Then equation (\ref{eq:TF}) has no solution, except in the trivial case of an homogeneous system. To get any
sensible result for a nucleus one has to go to the next order in the expansion in $\hbar$ of the kinetic energy
term in the functional. This introduces gradient terms \cite{RS}. To order $\hbar$ only one term contributes,
proportional to $(\nabla \rho)^2$. With the inclusion of this term, a surface can develop and the nuclear
Thomas-Fermi approximation can describe the density profile of a nucleus. \par This discussion makes clear that in
the nuclear case the kinetic term must be treated at different order in the expansion in $\hbar$ and at a
different level of approximation. The full quantal treatment of the kinetic term must be considered to construct
any accurate nuclear energy functional. \par Despite that, the Thomas-Fermi approximation, or more advanced
expansions in $\hbar$ can be useful. In fact the Thomas Fermi approximation and its implementations are expected
to describe the smooth part e.g. of the density of states, leaving outside of their possibility the description of
the shell effects, that are a pure quantal effect. Indeed the expansion in $\hbar$ of many quantities is actually
asymptotic, and the quantal effects depend non-analytically on $\hbar$. This property of the $\hbar$ expansions
can be used to evaluate the shell effects in the so-called microscopic-macroscopic approach, introduced in section
\ref{mass}. The difference between the exact quantal calculations and the results of the expansion is taken as an
estimate of the shell effects. Implicitly, it is assumed that shell effects are essentially the same in the
independent particle model as in the fully correlated system. The smooth part, obtained from the liquid drop or
droplet models, are expected to include in an effective way the correlation contribution. Along these lines,
recently in reference \cite{MarioPeter} the Kirkwood $\hbar$ expansion to fourth order has been used to estimate
the shell effects by comparing the results with the quantal calculations. This method, as similar ones based on
$\hbar$ expansions, are methods alternative to the Strutinski smoothing method \cite{Strut}.\par A step further
along these lines is the construction of general effective Energy Density functionals that are devised to include
all the correlations in an effective way and without any $\hbar$ expansion. Although they are necessarily in part
phenomenological, i.e. they contain a certain number of parameters, the ambition is to relate their
characteristics in terms of the properties of the nuclear medium. On the other hand, if they are treated at purely
phenomenological level, they turn out to be extremely accurate. These items will be discussed in the next section.

\subsection{The Density Functional Method \label{EDF}}
A fully quantal microscopic method to connect the nuclear medium properties and the structure of finite nuclei is
to introduce an effective NN force, whose parameters are fitted, on one hand, to reproduce the nuclear matter EoS,
as derived phenomenologically and microscopically, and on the other hand to reproduce binding energy and radii of
a representative set of nuclei. This is the scheme of the Skyrme forces. These effective forces have been
developed along the years with increasing success and have been widely used in nuclear structure and spectroscopic
studies. It is surely not possible to review the enormous literature on the subject, and we limit here to sketch
the main features of the method. The simplest Skyrme force can be written, for symmetric system
\beq
V = t_0\rho(\v{r})\delta(\v{r} - \v{r}') \,+\, t_3\rho(\v{r})^2 \delta(\v{r} - \v{r}') \delta(\v{r}' - \v{r}'')
\eeq
where $\rho$ is the nucleon number density at the point $\mathbf{r}$ and the parameters $t_0$ (negative) and $t_3$
(positive) are adjusted to reproduce the saturation point of Nuclear Matter, where the force depends only on the
constant total density. They correspond to effective two-body and three-body interactions, respectively. Besides
these parameters, an effective mass $m^*$ is also usually introduced in the kinetic part, so that the total
Hamiltonian can be written
\beq
 H \,=\, T \,+\, V \ \ \ \ ,\ \ \ \  T \,=\, - {\hbar^2 \over 2 m^*} {\mathbf \nabla}\delta(\v{r} - \v{r}')
  {\mathbf \nabla}
\eeq
Then the force is used to calculate the ground state of nuclei in the Hartree-Fock (mean field) approximation,
thus establishing an indirect link between nuclear matter and finite nuclei. Much more elaborated forces have been
developed, in which density gradient terms and asymmetry dependent terms are introduced. Then the number of
parameters increase, but the precision of the fit on the overall mass table can be really impressive, see e.g.
references \cite{Chamel1,Chamel2}. More elaborated terms involving higher degrees of the density derivatives can
be still introduced, and the accuracy of the fit can be astonishingly good \cite{Doba1,Doba2}. However, along this
line, the connection with NN bare forces and nuclear matter EoS is gradually lost, since the additional terms have
a form loosely connected with the NN interaction and are all vanishing in uniform matter.\par Another method that
tries to keep more closely the connection with the properties of the bulk nuclear medium is based on the Kohn and
Sham (KS) \cite{KS,jones,eschrig,Mat02,Per01,Per03,Per05} approach, first devised for atomic, molecular and solid
systems and developed also for nuclear system. Let us consider the microscopic bulk EoS, as reported in figure
(\ref{fig:EOS}), extended to asymmetric nuclear matter. This can be taken as the bulk contribution to the Energy
Density Functionals (EDF). For numerical applications it can be written in a polynomial form. The functional must
be then implemented mainly by three additional contributions. The first one is the Coulomb energy, that can be
calculated with different degrees of sophistication, e.g. by including the exchange and short range parts. The
second one must take into account the presence of the surface, that in nuclei is sharply localized, within a
length of the order of 1 fm, and therefore cannot be described only by the bulk part. The additional contribution
has to be localized at the surface, and the simplest way to do so is to introduce density gradient terms or
non-local short range convolution terms. The surface terms are connected to the surface tension of nuclear matter,
because, in the macroscopic limit, they modify the surface energy of the system. Finally, it is mandatory to add a
spin-orbit term, since it strongly affects the single particle level scheme and it is mandatory in order to
reproduce the shell sequence (i.e. the "magic numbers"). The spin-orbit interaction is roughly proportional to the
gradient of the single particle potential, and therefore it is also localized at the surface. The strength of this
term is severely constrained by phenomenology, but it has still some degree of uncertainty. It is desirable to
keep the number of surface and spin-orbit terms to a minimum, since they introduce additional phenomenological
parameters. It is one of the ambition of the Energy Density Functional method to get these parameters from
microscopic many-body theory, but in order to get a high precision fit to the wide set of nuclear binding and
radii throughout the mass table, they must be finely tuned beyond the possibility in accuracy of any microscopic
theory. The possibility still remains to have a guidance to their values within a reasonable accuracy and to get a
deeper understanding of their microscopic origin. This program has still to be developed.\par
Following the above considerations, the EDF can be written
\begin{equation}
E = T_0 + E^{s.o.} + E_{int}^{\infty} + E_{int}^{FR} + E_C. \label{eq:edf}
\end{equation}
For the surface term, following reference \cite{BCP1}, one can take a simple finite range term
\begin{eqnarray}
E_{int}^{FR}[\rho_n,\rho_p ] &=& \frac{1}{2}\sum_{t,t'}\int \int d^3r d^3r'\rho_{t}({\bf r}) v_{t,t'}({\bf r}-{\bf
r'})\rho_{t'}({\bf r'} )
\nonumber \\
&-& \frac{1}{2}\sum_{t,t'} \gamma_{t,t'}\int d^3r {\rho_{t}({\bf r})} \rho_{t'}({\bf r}) \label{eq:surf}
\end{eqnarray} with $t=$ proton/neutron and $\gamma_{t,t'}$ the volume integral of $v_{t,t'}(r)$. The
substraction in (\ref{eq:surf}) is made in order not to contaminate the bulk part, determined from the microscopic
infinite matter calculation $E_{int}^{\infty}$. Finite range terms have been used in e.g. \cite{BKN,Umar,Fay00},
generalizing usual Skyrme functionals. In references \cite{BCP1,BCP2}, for the finite range form factor
$v_{t,t'}(r)$ a simple Gaussian ansatz: $v_{t,t'}(r)=V_{t,t'}e^{-r^2/{r_0}^2}$ was taken, so that a minimum of
three open parameters was introduced : $V_{p,p}=V_{n,n}=V_L, V_{n,p}=V_{p,n}=V_U$, and $r_0$.
\par
\noindent In equation (\ref{eq:edf}), $ E_{int}^{FR} $ and $ E_C $ are the spin-orbit and Coulomb parts,
respectively. More detail on their determination can be found in references \cite{BCP1,BCP2}. The first piece
$T_0$ in equation (\ref{eq:edf}) corresponds to the uncorrelated part of the kinetic energy and within the KS
method it is written as
\begin{equation}
 T_0=\frac{{\hbar}^2}{2m} \sum_{i,s,t} \int
d^3r|\nabla \psi_i( {\bf r},s,t ) |^2. \label{eq:kin} \end{equation} \noindent
\noindent where the functions $\psi_i({\bf r} )$ form an auxiliary set of $A$ orthonormal single particle wave
functions , being $A$ the number of particles, and the density is assumed to be given by
\begin{equation}
 \rho( {\bf r} ) \, =\, \Sigma_{i,s,t} | \psi_i( {\bf r},s,t ) |^2
\label{e:eq1} \end{equation} \noindent where $s$ and $t$ stand for spin and iso-spin indices. At each point ${\bf
r}$ the bulk term equals the nuclear matter EoS at the corresponding local density (and asymmetry).
 Then, upon variation to minimize the EDF, one gets a closed
set of $A$ Hartree-like equations with an effective potential, the functional derivative of the interaction part
with respect to the local density $\rho({\bf r})$. Since the latter depends on the density, and therefore on the
 $\psi_i$'s, a self-consistent procedure is
necessary. The equations are exact if the exact EDF would be known. The existence of the latter is proved by the
Hohenberg-Kohn (HK) theorem
 \cite{HK}, which states that for a Fermi system, with a non-degenerate
ground state, the total energy can be expressed as a functional of the density $\rho({\bf r})$ only. Such a
functional reaches its variational minimum when evaluated with the exact ground state density. In practice of
course a reliable approximation must be found for the otherwise unknown density functional, taking inspiration
from physical considerations and microscopic input, as discussed above. It has to be stressed that in the KS
formalism the exact ground state wave function is actually not known, the density being the basic quantity. It has
also to be noticed that in the standard KS scheme, the kinetic energy term includes the bare nucleon mass, but
variants with an effective mass are possible (to incorporate the correlated part of the kinetic energy).
\par
\begin{figure}
 \begin{center}
\vskip 0.9 cm
\includegraphics[bb= 140 0 300 790,angle=-90,scale=0.45]{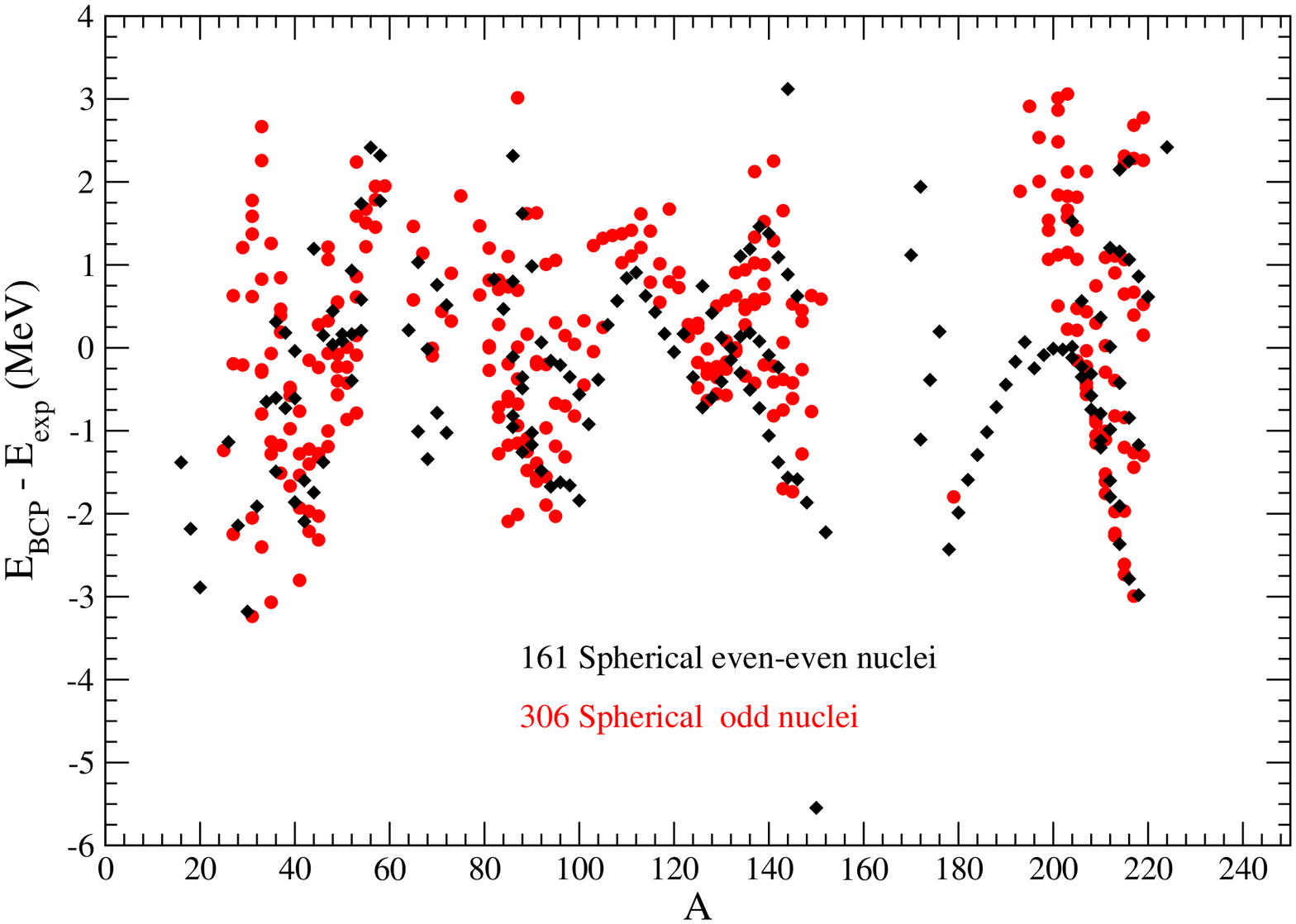}
\end{center}
\vspace{3.8 cm} \caption{(Color online) Energy differences as a function of mass number for a set of 161 spherical
nuclei (diamonds, black points) and 306 spherical odd nuclei (full circles, red points). }
    \label{fig:DE}
\end{figure}

As an illustration of the method, in figure (\ref{fig:DE}), taken from reference \cite{BCP2}, is reported the
difference in total binding energy between the calculated and the experimental values for a wide set of spherical
nuclei.
The parameters of the functional have been fixed by fitting a set of deformed nuclei, both normal and super-heavy
(functional BCP) . The choice of fitting first deformed nuclei is suggested by the consideration that these nuclei
should be better described by mean field, while spherical nuclei need corrections due to zero-point motion in the
ground state, e.g. RPA correlations in the ground state. The mean error in binding energy in the fit of deformed
nuclei is about $\sigma_E \,=\, 0.52$ MeV, while the discrepancy for the spherical nuclei, where the functional
was fixed and no fit was any more performed, is about $\sigma_E \,=\, 1.2$ MeV.
\begin{figure}
 \begin{center}
\vskip 0.9 cm
\includegraphics[bb= 140 0 300 790,angle=-90,scale=0.45]{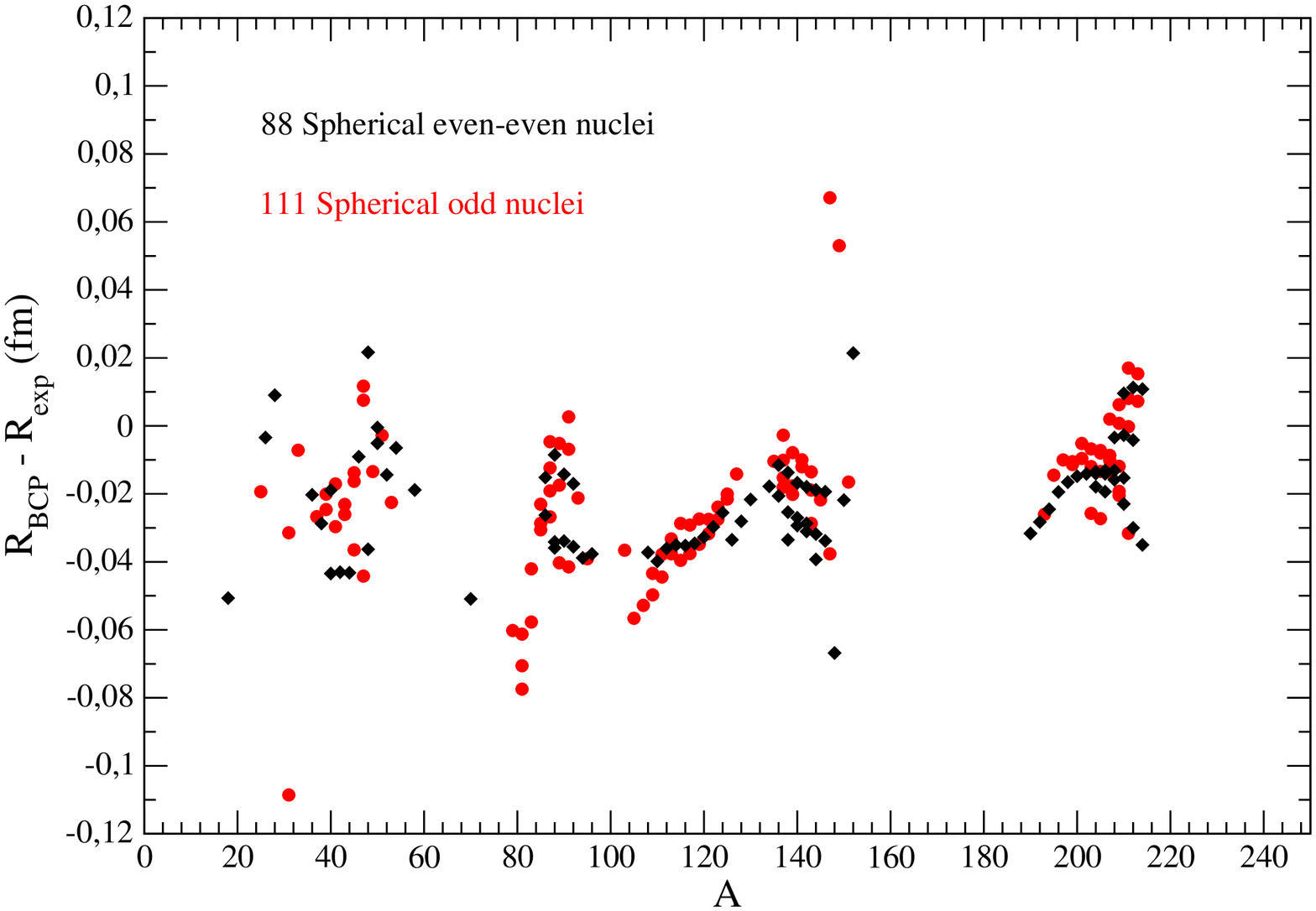}
\end{center}
\vspace{3.8 cm} \caption{(Color online) Differences of radii are shown as a function of mass number for a set of
88 spherical nuclei (diamonds, black points) and 111 spherical odd nuclei (full circles, red points)}
    \label{fig:DR}
\end{figure}
In figure (\ref{fig:DR}) the corresponding deviations for the root mean square radii are reported for the
spherical nuclei.
\par The performance is comparable with the best Skyrme forces, like the Gogny D1S \cite{D1S}, but of lower
quality than the one from HFB calculations of e.g. references \cite{Chamel1,Chamel2}. It must be stressed again
that the functional of equation (\ref{eq:edf}) has been developed by introducing a minimal set of finite size
terms, i.e. surface and spin-orbit terms, in addition to the bulk part fixed once for ever from microscopic
nuclear matter EoS. In this way one can clearly separate the bulk and surface contribution to the binding and the
radius in finite nuclei and establish a link between the properties of the nuclear medium (EoS and the bulk
symmetry energy) and the structure of finite nuclei. It would be interesting to further analyze this link on the
basis of an extended set of results.\par Finally it is worth mentioning that an additional term in the functional
should be included, the so called "Wigner term" \cite{Chamel1}, that introduces an additional binding for
symmetric nuclei, but rapidly vanishes as the nucleus becomes asymmetric. In a purely phenomenological approach it
has a parametrical form of the type \cite{Chamel1}
\beq E_W \,=\, V_W \exp \left[ - \lambda \left({N-Z\over A}\right)^2 \right] \,+\, V_W' |N - Z| \exp \left[ -
\left({A\over A_0}\right)^2 \right]
\eeq
\noindent and it can be ascribed to neutron-proton correlation/pairing, that indeed tends to disappear as soon as
neutrons and protons occupy different shells. This is expected to produce a strong improvement in the overall
quality of the fit. Again, it is an ambition of the microscopic approach to calculate, at least approximately, the
values of the parameters appearing in this expression, following their physical interpretation
\cite{Chamel1,Chamel2}.
\subsection{Pairing in nuclei \label{gapnu}}
Soon after the formulation of BCS theory \cite{BCS}, it was shown \cite{BMP} that pairing correlation is present
in nuclei. Since then, pairing correlation has played a major role in the development of nuclear structure.
Despite the enormous development in this field, the main physical question that remains still unanswered is the
origin of the attractive pairing interaction in nuclei in terms of the NN interaction. In most applications the
pairing interaction is treated as a phenomenological force. This approach is quite successful, particularly with
the Gogny force \cite{D1S}, but still the physical processes at the basis of these forces is not yet clarified. We
will focus here only to pairing between like-particles, because the neutron-proton pairing is restricted mainly to
symmetric or almost symmetric nuclei, where isospin symmetry is valid. We have seen that the bare NN potential is
able to produce in nuclear matter a substantial pairing gap mainly at sub-saturation density. It is not trivial to
relate these results to pairing in nuclei, since the bulk density is at saturation and the nuclear surface is too
sharp to justify a local density approximation. Furthermore in nuclei surface modes can play a role in the
physical processes responsible of the effective pairing interaction \cite{milan1,milan2,milan3}. \par
Besides the many-body aspects of the problem, at least two other features of the nuclear pairing have to be
mentioned. One is related to the fact that the pairing phenomenon occurs close to the Fermi surface, while the
bare nucleon-nucleon (NN) potential necessarily involves also scattering to high energy (or momentum) due to its
strong hard core component, which is one of the main characteristics of the nuclear interaction. It looks
therefore natural to develop a procedure which removes the high energy states and "renormalize" the interaction
into a region close to the Fermi energy. This can be done in different ways, among which the most commonly used
seems to be the Renormalization Group (RG) Method \cite{Schwenk}. A second feature is the relevance of the single
particle spectrum, not only because the density of states at the Fermi surface plays of course a major role, but
also because the whole single particle spectrum has influence on the effective pairing interaction.
\par
 In the last
few years relevant progress has been made in the microscopic calculations of pairing gap in nuclei
\cite{milan1,milan2,milan3,Pankr1,Pankr2,Pankr3,Dug1,Dug2}. The main established result is that the bare NN
interaction, renormalized by projecting out the high momenta, is a reasonable starting point that is able to
produce a pairing gap which shows a discrepancy with respect to the experimental value not larger than a factor 2.
In view of the great sensitivity of the gap to the effective interaction this result does not appear obvious. The
effective pairing interaction constructed within this renormalization scheme \cite{Baldo_1998} explains
qualitatively also the surface relevance. Indeed, this interaction at the surface can exceed the value inside by
one order of magnitude. Direct effect of the surface enhancement of the gap was presented in
\cite{Baldo_1999,Farine_1999} for semi-infinite nuclear matter and in \cite{Baldo_2003} for a nuclear slab.
Besides the renormalization of the bare interaction of the high momentum components, other physical effects should
be included in a microscopic approach, like the ones related to the effective mass, or more generally, the single
particle spectrum, and the many-body renormalization of the pairing interaction. \par
Let us first consider the problem  of reducing the interaction to an effective one close to the Fermi energy. In
general language this can be seen as a typical case that can be approached by an "Effective Theory", where the low
energy phenomena are decoupled from the high energy components. In this procedure the resulting low energy
effective interaction is expected to be independent of the particular form of the high energy components. However
the procedure is not unique. The RG method has been particularly developed for the reduction of the general NN
interaction  keeping the deuteron properties and NN phase shifts up to the energy where they are well established.
In this way a potential, phase equivalent to a known realistic NN potential, can be obtained, which contains only
momenta up to a certain cut-off. The extension of this procedure to the many-body problem appears in general to
require the introduction of strong three-body forces. It has been applied also to the pairing problem. To
illustrate the difficulty of decoupling low and high momentum components for the pairing problem, we consider the
simple case of nuclear matter in the BCS approximation discussed in section \ref{super}, see equation
(\ref{eq:gap}). Taking for the pairing interaction the bare NN interaction $V(k,k')$,
it is possible to project out the momenta larger than a cutoff $k_c$ by introducing the interaction $\Vef(k,k')$,
which is restricted to momenta $k < k_c$. It satisfies the integral equation \be
 \Vef(k,k') = V(k,k')
  - \!\!\sum_{k'' > k_c}
 \!\! \frac { V(k,k'') \Vef(k'',k')} {2E_{k''} }
 \:,
\label{eq:vl1} \ee
 \noindent The gap equation, restricted to momenta $k < k_c$ and with the original interaction $V$ replaced by $\Vef$,
is exactly equivalent to the original gap equation. The relevance of this equation is that for a not too small
cut-off the gap $\Delta(k)$ can be neglected in $E(k)$ to a very good approximation and the effective interaction
$\Vef$ depends only on the normal single particle spectrum above the cutoff $k_c$. As such $\Vef$ contains
necessarily some dependence on the in-medium single particle spectrum at high momenta.
In the RG approach, if the procedure of constructing the low momenta interaction V$_{low}(k,k')$, discussed in
section \ref{Vlow}, is carried out in vacuum, it implicitly assumes a free spectrum at high momenta. In the medium
it faces the same uncertainty. \par
This procedure of projecting out the high momenta can be extended to finite nuclei, and the same uncertainty
persists.
\begin{table}
\caption{\label{tab:delta}  the Argonne v$_{18}$ potential.}
\bigskip
\begin{indented}
\item[]\begin{tabular}{|c|c|c|c|c|}
\hline
 $\lambda $ &  SLy4 & Sly4-1 & Sly4-2 &Sly4-3 \\
\hline
3$s_{1/2}$   &1.23 &1.10 &0.83& 0.56\\
2$d_{5/2}$   &1.32 &1.18 &0.89& 0.61\\
2$d_{3/2}$   &1.34 &1.20 &0.92& 0.63\\
1$g_{7/2}$   &1.48 &1.31 &0.96& 0.64\\
2$h_{11/2}$  &1.27 &1.13 &0.85& 0.57\\
\hline
$\Delta_{\rm F}$&1.34&1.19 &0.89& 0.60\\

\hline
\end{tabular}
\end{indented}
\end{table}
To illustrate this point in table \ref{tab:delta}, taken from reference \cite{JPG_dob}, are reported the values of
the pairing gap for $^{120}$Sn. The diagonal pairing gap matrix elements for the levels around the Fermi energy
and the corresponding average values are compared for different types of calculations. For the mean field the Sly4
Skyrme force has been used. In the procedure of projecting out the high momenta the effective mass has been put
equal to the bare one above a certain cut-off $\Lambda$. This can be done within the so-called Local Potential
Approximation (LPA) \cite{Rep,ST}, that has been proved to be a quite accurate approximation. In the first column
the effective mass has been taken equal (and constant) to the original Sly4 value only inside the model space,
i.e. within the states where the effective interaction is calculated after the projection of the high momentum
components, according to equation (\ref{eq:vl1}). The model space corresponds to single particle energies
$\epsilon_\lambda < 40$ MeV.  For the second column $\Lambda = 3$fm$^{-1}$, for the third $\Lambda = 4$fm$^{-1}$
and for the fourth $\Lambda = 6.2$fm$^{-1}$. The original interaction is the Argonne $v_{18}$ NN potential. In all
cases the original density dependence of the effective mass was kept within the LPA scheme. The strong sensitivity
of the pairing gap to the value of $\Lambda$ indicates that the separation between small and high momenta
components can be problematical.
\par
We have seen in section \ref{super} that in nuclear matter one of the main open questions is the role of the
medium in screening the effective pairing interaction. It turns out \cite{milan1,milan2,milan3} that the same
processes of exchange of collective excitations between quasi-particles produce an overall attractive interaction,
usually indicated as "induced interaction", and could contribute substantially to the strength of the effective
pairing interaction. This striking difference could be due to the strong collectivity of the low lying surface
modes in nuclei and to the corresponding small strength of the spin modes. The above described calculations that
include only the bare NN interaction, on comparison with the experimental data, can put limits on the relevance of
the induced interaction. Due to the mentioned uncertainties, it is not yet possible to draw any quantitative
conclusions. In any case this is an active field of research and some answers, at least partial, could come form
future works.\par
Finally it is important to mention one of the main issues that is involved when one tries to relate pairing in
nuclear matter and finite nuclei. The size $\xi$ of the Cooper pairs in nuclear matter can be estimated in the
weak coupling limit
\beq
\xi \,\approx\, {\hbar v_F \over \sqrt{8}\Delta_F }
\label{size}\eeq
\noindent In nuclear matter for the neutron pairing gap obtained with the free NN interaction one gets that $\xi
> 5$ fm.
\begin{figure}
 \begin{center}
\vskip -11.5 cm
\includegraphics[bb= 140 0 300 790,angle=0,scale=0.65]{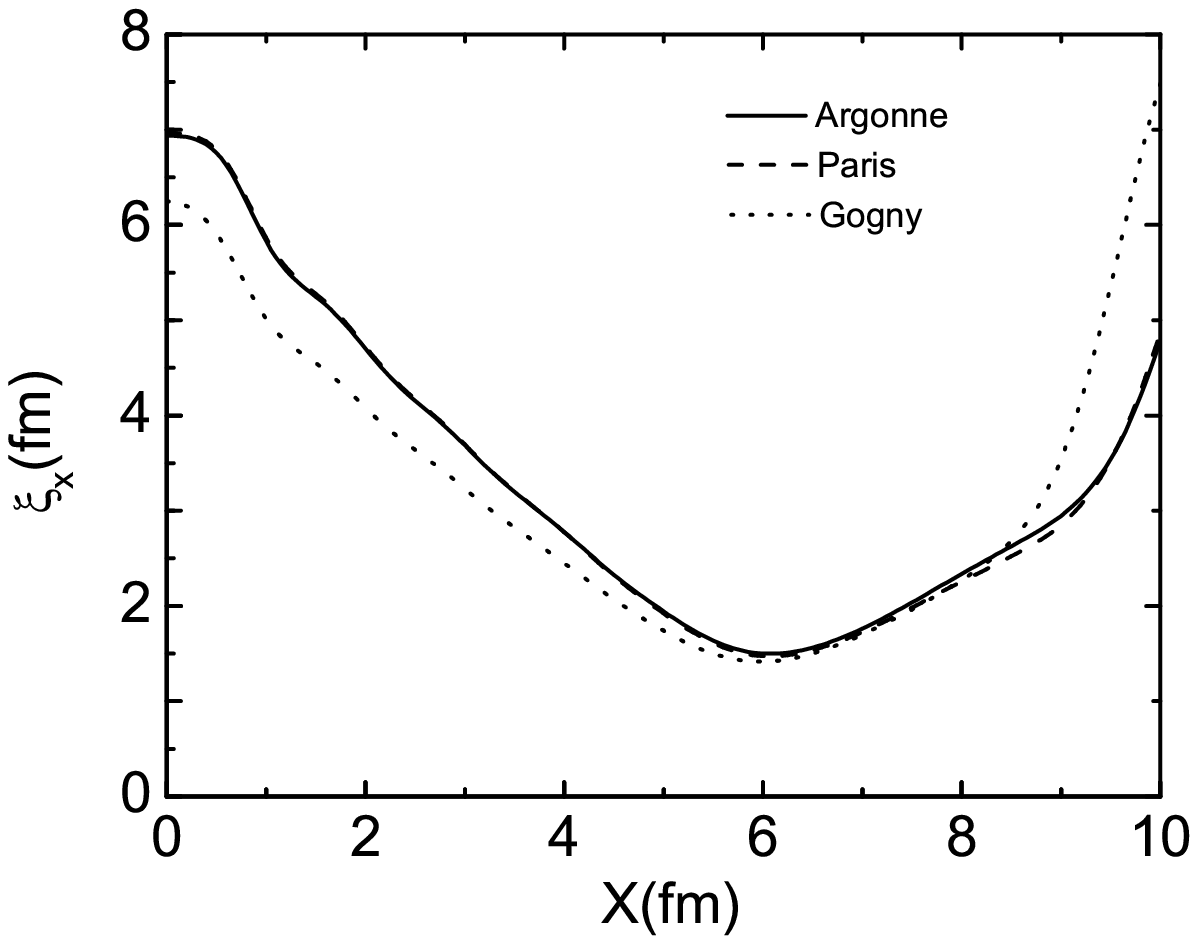}
\end{center}
\vspace{-0.6 cm} \caption{Size of the Cooper pair in the direction perpendicular to a slab of nuclear matter. The
surface of the slab is located at about X = 6 fm, very close to the position of the minimum.}
    \label{fig:size}
\end{figure}
If one extrapolates this expression to finite nuclei and takes a typical value for the gap of 1-2 MeV, then the
value of $\xi$ can exceed the size of the nucleus. It looks that we cannot describe pairing correlation in nuclei
with a simple spacial picture. Also in this case the role of surface is essential. It has been found that the
pairing correlation length $\xi$ depends on the center of mass of the pair, and it shrinks just at the surface of
the nucleus. In figure (\ref{fig:size}), taken from reference \cite{Spat_cor}, is reported the case of a slab of
nuclear matter. Here the Cooper pair size $\xi_x$ in the direction $x$ perpendicular to the slab is reported as a
function of the center of mass position $X$. The minimum of $\xi_x$ falls exactly at the surface of the slab,
where $\xi_x \approx 2$ fm, independently of the pairing interaction used. This is in line with the results
presented in references \cite{Pillet} for realistic cases, where the Cooper pair size was also found to be minimal
at the surface and around 2 fm. References to previous works on this subject can be found in this paper. Even if
this shrinking of the pair size at the surface could be a general effect, not necessarily connected with the
pairing interaction but rather just with the available phase space, it looks that pairing correlation has room
mainly at the surface. This is in agreement with the already mentioned enhancement of the local pairing gap at the
surface. Loosely speaking, one can imagine that the Cooper pairs are mainly formed around the surface, where the
pairing correlation can develop.

\subsection{Neutron and Proton Radii}
It is not easy to establish at a formal level a direct connection of the properties of nuclei and nuclear matter.
It has been then pursued a different strategy, that can be considered a semi-empirical theoretical method. One
considers a wide set of Skyrme forces with different characteristics and one looks for a possible correlation
between a specific nuclear matter property and a physical parameter in nuclei as the force is varied within the
given set.
\begin{figure}
 \begin{center}
\vskip 0.8 cm
\includegraphics[bb= 140 0 300 790,angle=-90,scale=0.45]{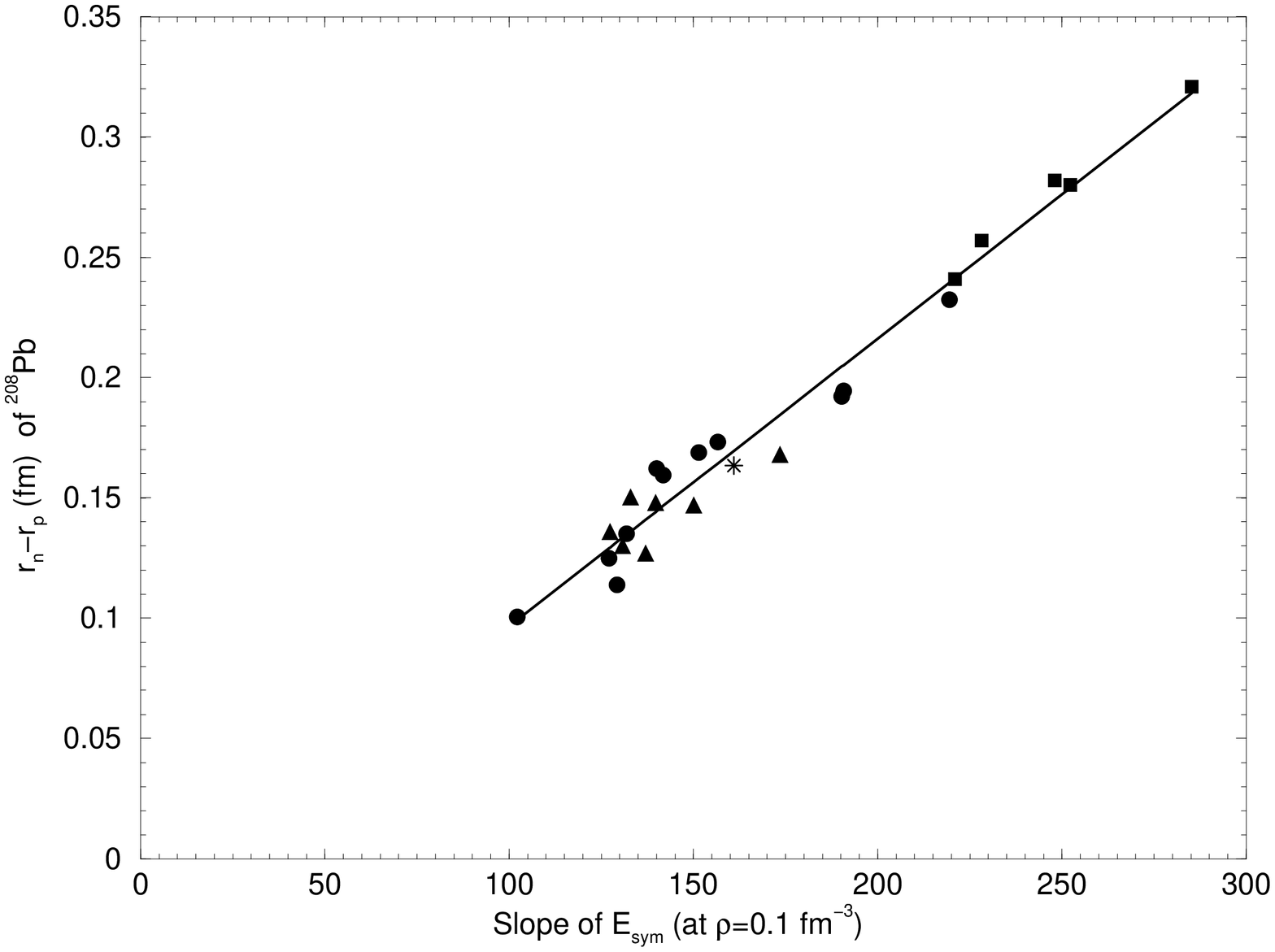}
\end{center}
\vspace{3.6 cm} \caption{Linear correlations between the neutron and proton radii and the slope ( Mev fm$^3$ ) of
the symmetry energy at the density $\rho = 0.1 $fm$^{-3}$. Squares from the top correspond to$NL1$ \cite{NL1},
$NL3$ \cite{NL3}, $G1$, $G2$ \cite{Furns} and $Z271$ \cite{Pic}. Triangles to Gogny forces $D280$, $D300$, $D250$,
$D260$ $D1$ and $D1S$ \cite{Blaizot2}. Circles correspond to the Skyrme forces $SV$ \cite{Giai}, $SIV$
\cite{Giai}, $SkM$ , $SkM^*$ \cite{skm}, $SLy4$, $SLy5$ \cite{sly4}, $T6$ \cite{t6}, $SGII$ \cite{sg}, $SI$
\cite{VB}, $SII$ \cite{VB},$SIII$ and $SVI$ \cite{Giai}. The result for the BCP functional is labeled by a star. }
    \label{fig:rn-rpL}
\end{figure}
This approach was followed in reference \cite{ABrown}, where it was shown that the difference between the neutron
and proton root mean square radii ("neutron skin") of Pb is linearly correlated to the slope of the neutron matter
EoS at the density of 0.1 fm$^{-3}$. This linear correlation plot is shown in figure (\ref{fig:rn-rpL}), taken
from reference \cite{BCP-NPA}.
The linear correlation was shown to hold also for other nuclei and also if one considers, instead of the neutron
matter EoS, the slope of the symmetry energy as a function of density in symmetric nuclear matter at approximately
the same density \cite{Xavier2}. Even if the approach is not formally microscopic, it is quite fruitful because it
links in a direct way the nuclear EoS to the structure of nuclei. In other words, measuring a physical quantity in
nuclei would give direct information on the nuclear EoS. Unfortunately it is not easy to measure the neutron
radius, since the hadron probes are strongly interacting, at variance of electrons that provide the charge radius.
There is a large expectation on the parity violation electron scattering experiment PREX that is going on at JLAB
\cite{PREX}, because these measurements will give directly the difference between neutron and proton radii in
$^{208}Pb$. It has to be seen if the accuracy of the data will be enough to distinguish among different
functionals. Partial justifications of some of these correlations have been presented in the literature
\cite{dieper}, but further microscopic investigations are surely needed to clarify the field. As an illustration,
in figure (\ref{fig:rn-rpI}) is reported the results for the neutron skin value from one of the version of the
functional BCP throughout the mass table. The set of results looks in overall agreement with the phenomenological
data. It would be desirable to identify the main properties of the forces or functionals that determine the value
of the radii difference \cite{Naz,Xavier2} (or other physical quantity), but this goal has still to be reached.
This would be indeed a real progress in our understanding of the fundamental properties of the nuclear medium,
from nuclear matter to finite nuclei.
\begin{figure}
 \begin{center}
\vskip 1.3 cm
\includegraphics[bb= 140 0 300 790,angle=-90,scale=0.45]{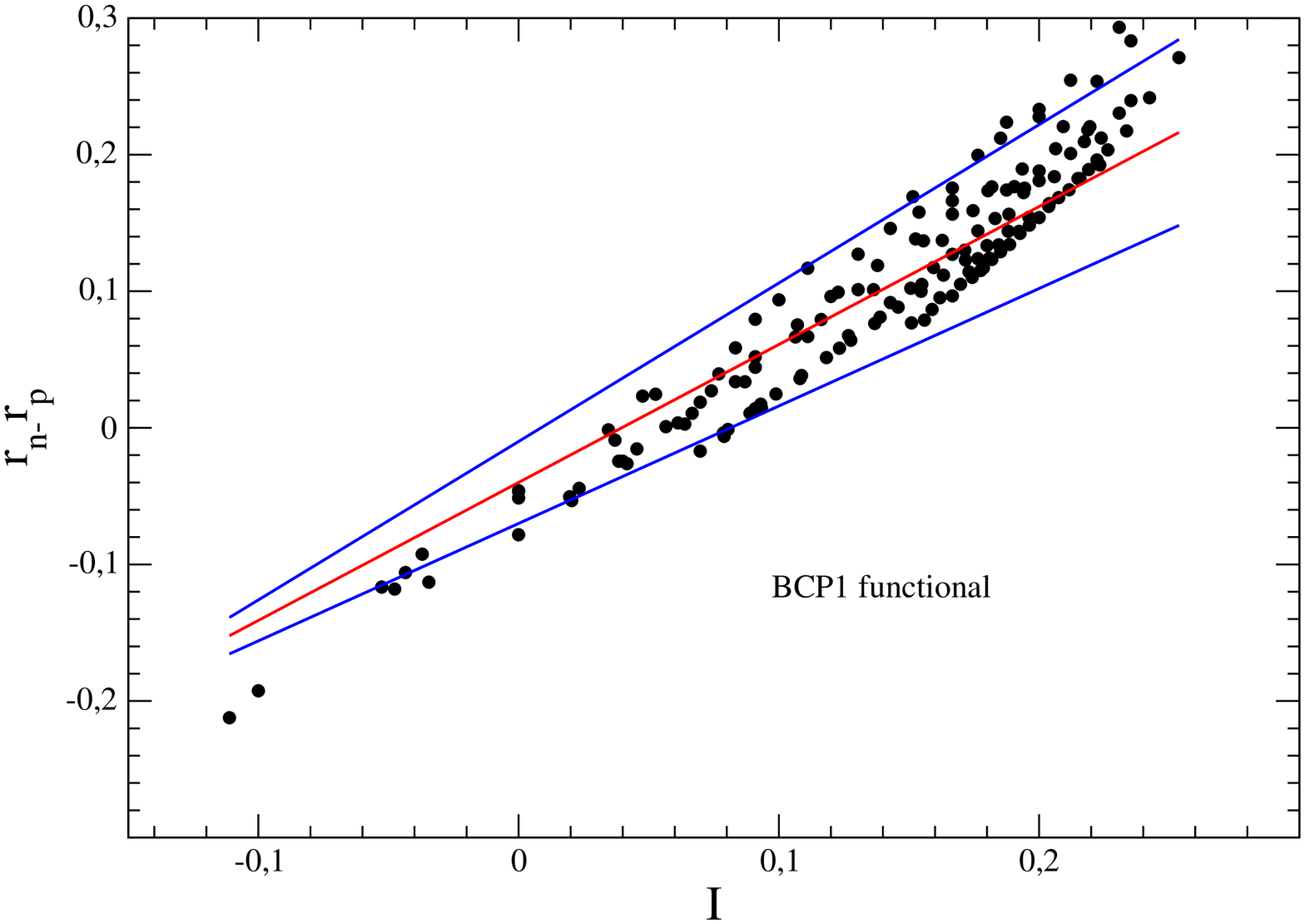}
\end{center}
\vspace{3.9 cm} \caption{Neutron skin calculated with the BCP functional throughout the mass table. The uppermost
line and lowermost line are the phenomenological boundaries obtained in reference \cite{Trzc}. The middle line is
just the average value of the phenomenological boundaries.}
    \label{fig:rn-rpI}
\end{figure}
\subsection{Exotic Nuclei}
The physics of exotic nuclei, i.e. nuclei far away from the stability valley that are not present on the Earth, is
a field of rapid development where a large effort is concentrated in many laboratories throughout the world. Long
term projects with large collaboration networks have been established in different countries or Continents (e.g.
EURISOL, FIRB) to produce radioactive beams. They will allow to study systematically the properties of exotic
nuclei that are not yet available, extending nuclear structure considerably along the asymmetry axis of the mass
table. This is a challenge for the existing most sophisticated EDF, whose predictions for the forthcoming exotic
nuclei often disagree. This line of research will provide invaluable information about the behavior of nuclei and
the nuclear medium at increasing asymmetry. New phenomena are expected to occur at large enough asymmetry, like
the onset of new nuclear shells. However, these phenomena depend on delicate nuclear structure features, and lie
outside the scope of the present review.

\subsection{The Neutron Star Crust}
\begin{figure} [hb]
 \begin{center}
\includegraphics[bb= 200 0 360 790,angle=0,scale=0.5]{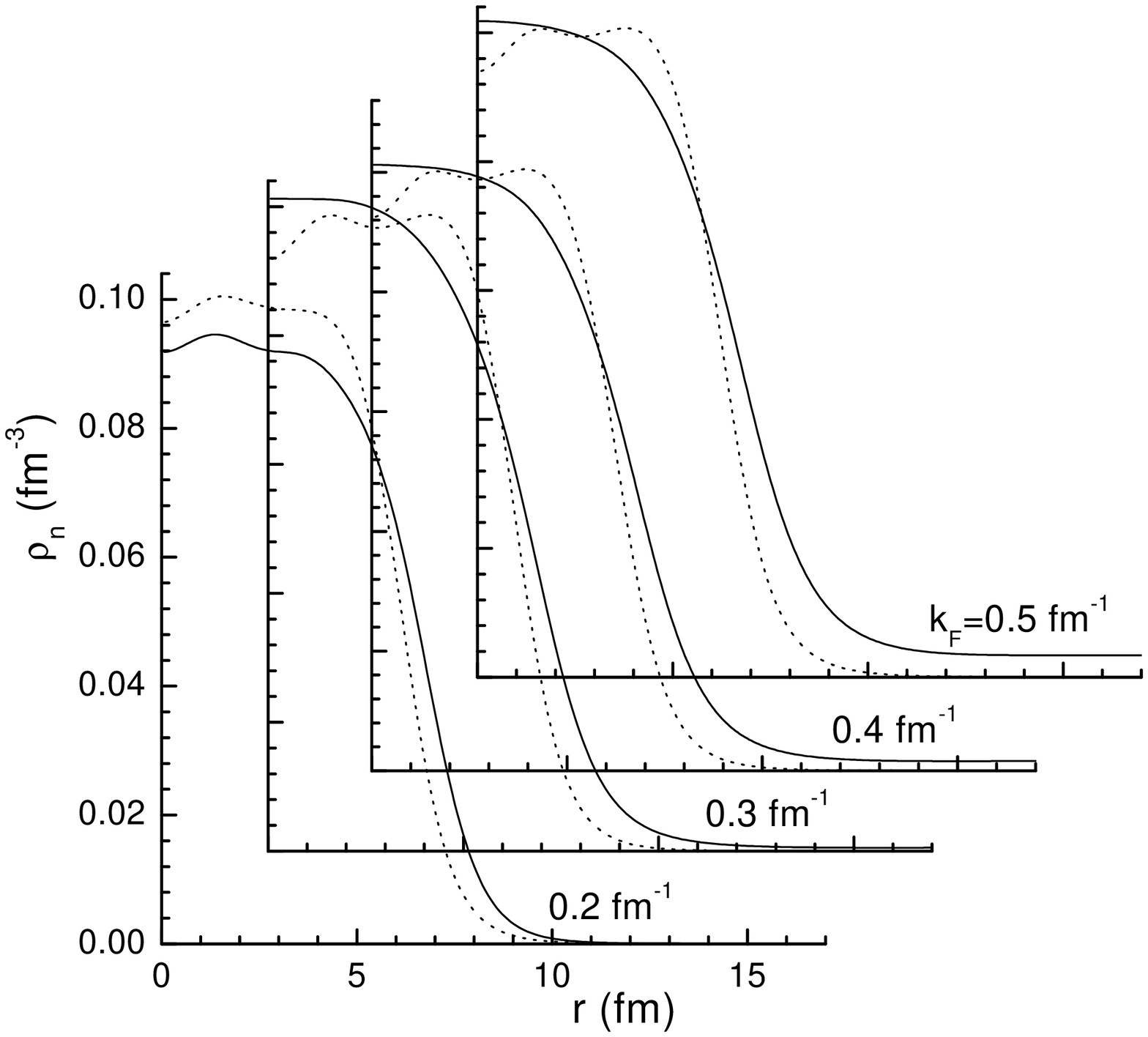}
\end{center}
\vspace{-6.2 cm} \caption{Neutron density profile of the nuclei at the upper edge of the inner crust of a neutron
star. The dotted lines indicate the corresponding neutron density profile of a nucleus with the same atomic and
mass numbers of the "cluster" of matter at the center of the Wigner-Seitz cell.}
    \label{fig:nuclei_crust}
\end{figure}

We have seen that the crust of Neutron Stars is the place in the Universe where the most asymmetric nuclei are
present in a stable manner. The reason is of course the existence of the electron gas that is surrounding the
nuclei and prevents their beta decay due to the Pauli blocking. In the inner crust also a neutron gas exists, but
at not too high density it is possible to separate out, at least approximately, the nucleus at the center of the
lattice cell. In this way one can picture the inner crust as a lattice of nuclei surrounded by a neutron gas
(besides the electron gas). These nuclei are therefore unstable even with respect to the strong interaction
(neutrons are "dripping" from them), but they are in equilibrium with the surrounding neutron gas. To illustrate
this point, in figure (\ref{fig:nuclei_crust}) are reported the results of the calculations in reference
\cite{DRIP}, where the neutron density profile in the WS cell is compared, at different density, with the
corresponding profile of a nucleus with the same atomic number and a mass number equal to the number of nucleons
inside the radius of the "blob" (approximately estimated). They are evaluated with the same functional, and are
stable with respect to strong interaction. One observes the formation of the neutron gas outside the "nucleus" at
the center of the WS cell as the density increases. Of course these nuclei would be strongly unstable with respect
to weak interaction.
Beyond the reported maximum density, it is not possible any more to find finite nuclei corresponding to the "blob"
at the center of the WS cell, because they are unstable even with respect to strong interaction. Then it is not
any more possible to distinguish the "blob" as a nucleus separated from the surrounding neutron gas.
\par
\begin{figure}
 \begin{center}
\vskip 0.5 cm
\includegraphics[bb= 200 0 360 790,angle=0,scale=0.45]{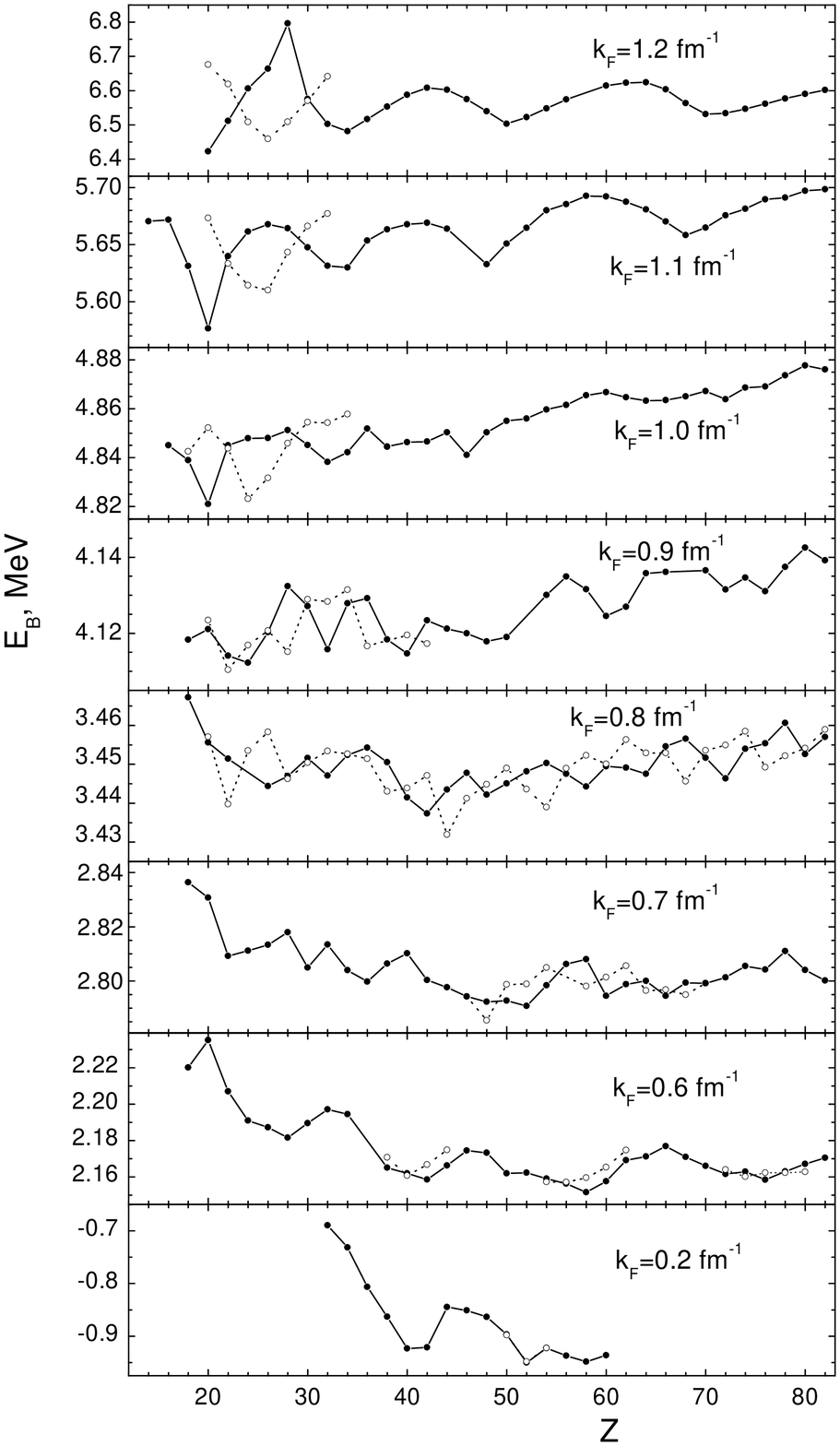}
\end{center}
\vspace{-0.5 cm} \caption{Energy per particle as a function of the atomic number in the Wigner-Seitz cell. The
minimum corresponds to the actual configuration of the crust at the given density. The dotted lines indicate the
results for a different choice of the boundary conditions at the edge of the cell, see reference \cite{REAL}. The
discrepancy between the full and dotted lines is a measure of the uncertainty in the Wigner-Seitz approximation.}
    \label{fig:EAZ}
\end{figure}

The structure of all these WS cells can be studied solely on the basis of a strong extrapolation of the
theoretical methods, in particular of the various EDF, developed and checked along the available mass table.
Unfortunately, despite all EDF must agree, within a certain accuracy, for the nuclei that can be produced in
Laboratory, their predictions on the nuclei of such a large asymmetry are often diverging. This indicates that we
are still far from having under control the microscopic theory of the asymmetry dependence of the nuclear medium
properties. This uncertainty reflects into the uncertainty of the Neutron Star crust, and the discrepancy extends
to the region of higher density, where it is not possible to separate any more a definite nucleus at the center of
cell, until the transition to homogeneous matter occurs. As illustration in table (\ref{tab:Z}) are reported the
values of the atomic numbers of nuclei in the inner crust calculated in the classical paper by Negele and
Vautherin \cite{NV} and with a different functional \cite{REAL}. The latter includes also the pairing correlations
with three different strengths (P1, P2 and P3). The reason of these type of discrepancies remain to be clarified,
but it has to be stressed that the position of the minimum in the energy as a function of the atomic number is
quite delicate, because it can often happen that different local minima are competing among each others. In the
table are reported also the values of the Wigner-Seitz cell radius, that turns out to be a less sensitive
quantity.
\begin{table}
 \caption{Atomic number of the matter inside the Wigner-Seitz cell throughout the inner crust and
corresponding cell radius. The calculations labeled P1, P2, P3 are for three different pairing strengths, reported
in reference \cite{REAL}, in comparison with the results (NV) of reference \cite{NV}.}
\begin{center}
\begin{tabular}{|c|c|c|c|c|c|c|c|}
\hline
$k_{\rm F},$&\multicolumn{4}{|c|}{$Z$}&\multicolumn{3}{|c|}{$R_{\rm c},\;$fm}\\
\cline{2-8}
 \rule{0pt}{13pt}${\rm fm}^{-1}$ & P1&  P2 & P3 &   N\&V  & P1 &     P2  &  P3\\
\hline
  0.6 & 58 & 56 & 56 &   50 &     37.51 &   36.85 &   36.92\\
\hline
  0.7 & 52 & 46 & 46 &   50 &     32.02 &   30.31 &   30.27\\
\hline
  0.8 & 42 & 40 & 40  &  50  &    26.90  &  25.97  &  25.97\\
\hline
  0.9 & 24 & 20 & 20  &  40  &    20.26 &   18.34  &  18.39\\
\hline
  1.0 & 20 & 20 & 20 &   40   &   16.69 &   16.56  &  16.56\\
\hline
  1.1 & 20 & 20 & 20 &   40   &   14.99 &   15.05  &  15.05\\
\hline
  1.2 & 20 & 20 & 20 &   40   &   13.68 &   13.73  &  13.74\\
\hline
\end{tabular}
\end{center}
\label{tab:Z}
\end{table}
It must be pointed out that these discrepancies persist at lower density, down to the drip point and slightly
below, as systematically explored in reference \cite{ruster} for a wide set of Skyrme forces and relativistic mean
field functionals. To illustrate the difficulty and uncertainty in these calculations, in figure (\ref{fig:EAZ})
are reported typical energy curves for the WS cell as a function of the atomic number \cite{REAL}.
\noindent Besides the apparent competition among different energy minima, it has to be noticed the tiny variation
of the binding energy. The accuracy needed to predict the absolute minimum is of the order of 5-10 keV per
particle, which is at the limit of the performance of the best EDF, and surely the extrapolation to so asymmetric
matter is not yet under control.
 In any case, since the structure of these nuclei cannot be studied in laboratory, one has to
look for observational data on Neutron Stars that are sensitive to their properties. The main physical parameters
of the crust is the values of the atomic number of the nuclei in the lattice as a function of density and the
lattice spacing. Other quantities, like the shear modulus or the incompressibility are functions of these
parameters on the basis of well known properties of Coulomb lattices. In fact, even in the inner crust, the effect
of the neutron gas on these quantities is negligible. The observation of NS oscillations during flares of X-ray or
gamma ray emission in accreting processes or in magnetar quakes can be a possibility of studying the structure of
the crust. These observations are now numerous \cite{israel,Alta1,Alta2,Watts}.
\begin{figure}
 \begin{center}
\vskip -1.5 cm
\includegraphics[bb= 200 0 360 790,angle=0,scale=0.6]{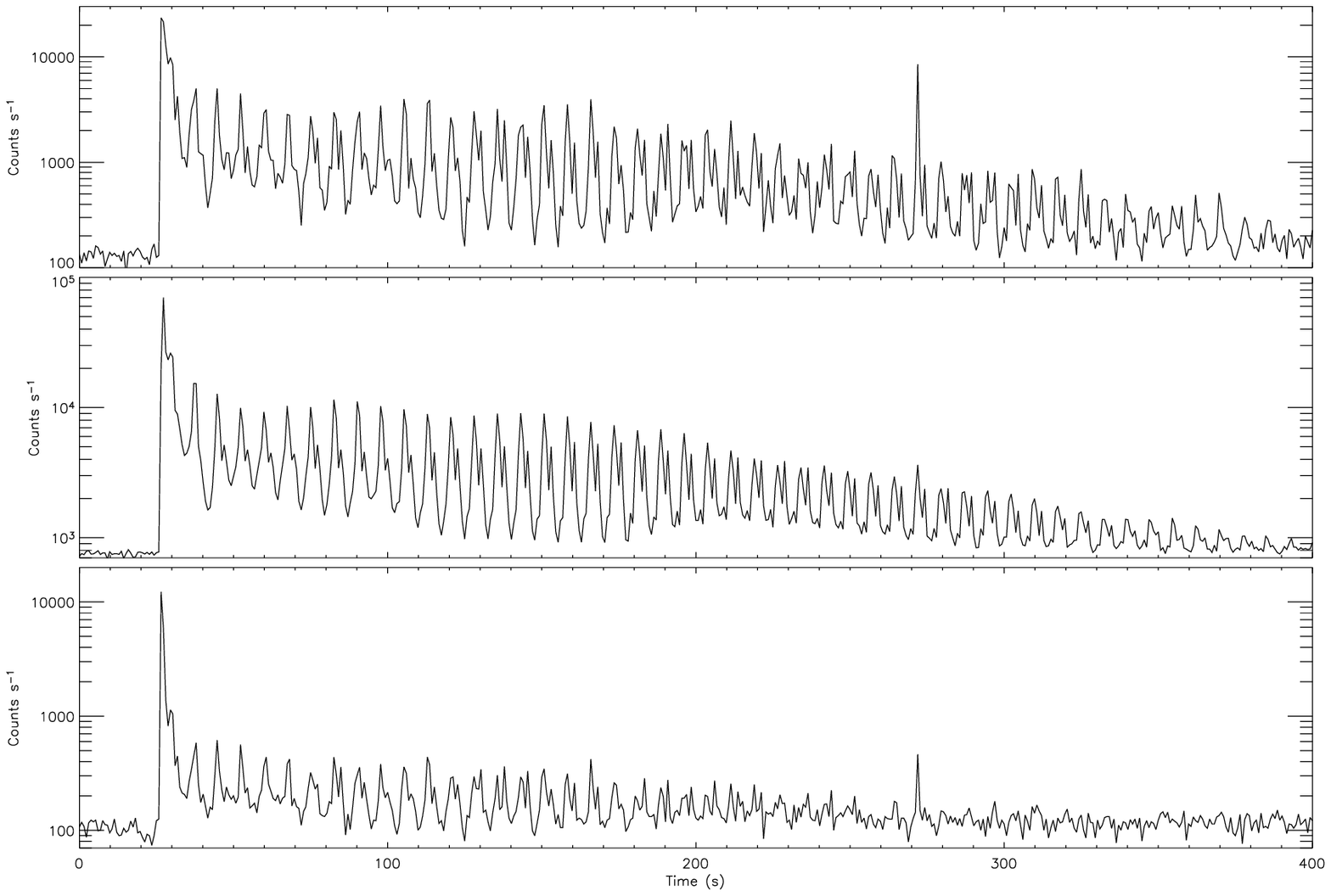}
\end{center}
\vspace{-8. cm} \caption{Light curves in the tail of the hyperflare from the SGR 1806-20. Observations and
analysis of reference \cite{Watts}.}
    \label{fig:osci}
\end{figure}
If in the oscillations the crust is decoupled from the core, the spectral analysis of these data can be used to
constrain the shear modulus and, consequently, the density dependence of the symmetry energy \cite{O_Steiner}.
This types of analysis are partly model dependent, but they open a window on the possibility of studying, even if
in a very indirect way, the nuclear medium for very high asymmetry at and below saturation density. As
illustration we report in figure (\ref{fig:osci}) the luminosity oscillations discovered in reference \cite{Watts}
in the tail of an hyper-flare of SGR 1806-20. The three panels correspond to three different bands of the
radiation frequencies \cite{Watts}.
\section{Liquid-gas phase transition \label{finiteT}}

In the latest stage of the supernovae collapse the EOS of asymmetric nuclear matter at finite temperature plays a
major role in determining the final evolution. Microscopic calculations of the nuclear EOS at finite temperature
are quite few. The variational calculation by Friedman and Pandharipande \cite{FriedPand} was one of the first few
semi-microscopic investigation of the finite temperature EOS. The results appear fairly close to the predictions
based on Skyrme force models: symmetric nuclear matter undergoes a liquid-gas phase transition, with a critical
temperature  $T_c = 18-20$ MeV. This is a fundamental property of the nuclear medium. Different types of Skyrme
forces give different critical temperatures, but they lie all close to this range of values. Later, Brueckner-like
calculations at finite temperature \cite{LG} confirmed these findings with very similar values of $T_c$. The full
finite temperature formalism by Bloch and De Dominicis (BD) \cite{BDD} was followed. In this section we sketch the
main results and the method followed in the finite temperature case. We then discuss the connection with
Laboratory experiments and data.

\subsection{The critical temperature}

\begin{figure}
 \begin{center}
\vskip -3. cm
\includegraphics[bb= 140 0 280 700,angle=0,scale=0.7]{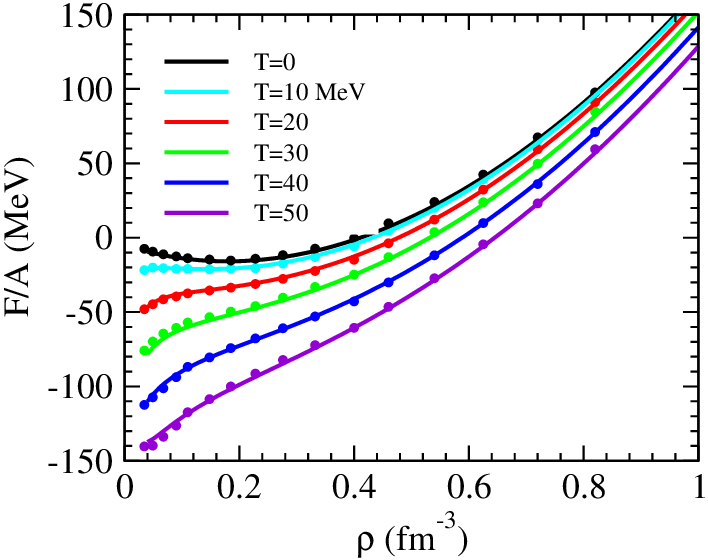}
\end{center}
\vspace{-7.7 cm} \caption{(Color on line) Free energy of symmetric nuclear matter as a function of density at
different temperatures. The lines are fits to the calculated points. The curves decrease systematically as the
temperature increases.}
    \label{fig:free}
\end{figure}

\begin{figure}
 \begin{center}
\vskip -3. cm
\includegraphics[bb= 140 0 280 700,angle=0,scale=0.7]{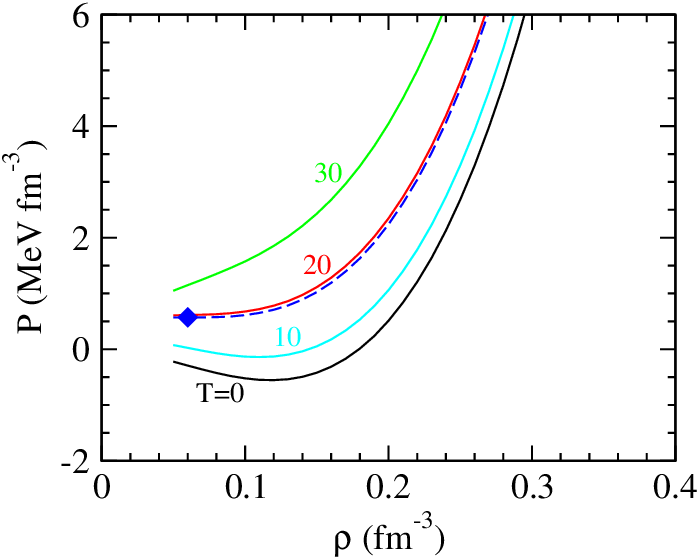}
\end{center}
\vspace{-7.7 cm} \caption{(Color on line) The isotherms of symmetric nuclear matter. The full diamond marks the
critical point of the liquid-gas phase transition.}
    \label{fig:vdw}
\end{figure}

The starting point in the many-body theory of finite temperature EoS is the calculations of the grand-canonical
potential $\Omega$. In the BD formalism,in line with the Brueckner scheme, a self-consistent single particle
potential $U(\mathbf{k})$ is introduced and finally the grand-canonical potential is given by
\begin{equation}
\Omega = \Omega'_0 + \Delta \Omega
\label{eq:delom}
\end{equation}
\noindent where
\begin{equation}
\Omega'_0  = - {2 V\over \pi^2} \int_0^{+\infty} k^2 dk [\,\, {1\over \beta}
 \log (1 + e^{-\beta(e_k \mu)})
+ U(k) n(k)\,\, ]  \label{eq:omep}
\end{equation}
\noindent is the grand canonical potential for independent particles with hamiltonian
\beq
  H'_0 = \sum_k e_k a^\dagger_k a_k =
 \sum_k ( {\hbar^2 k^2\over 2m} + U(k)) a^\dagger_k a_k
\le{hamil} \eeq
\noindent and $\mu$ is the chemical potential. The interaction part $\Delta \Omega$ of the grand canonical
potential is given by
\begin{equation}
\Delta \Omega \,=\, {1\over 2} e^{2\beta\mu}\int_{-\infty}^\infty {e^{-\beta\omega} \over 2\pi }d\omega Tr_2
\left[ \arctan \left(\kk(\omega)\pi \delta (H'_0 -\omega)\right) \right].
\label{eq:arctan}
\end{equation}
\cap The trace in the previous equation $Tr_2$ is taken in the space of antisymmetrized two-body states and the
two-body scattering matrix ${\cal K}$ is defined by,
\begin{equation}
\bar{rl}
  \bra k_1\!\!\! &k_2 \vert \kk(\omega) \vert k_3 k_4 \ket
  = \\
 &                    \\
  &(n_{>}(k_1)n_{>}(k_2)n_{>}(k_3)n_{>}(k_4))^{1\over 2}
   \bra k_1 k_2 \vert K(\omega) \vert k_3 k_4 \ket
\ear \le{kmatt}
\end{equation}
\cap where the scattering matrix $K$ satisfies the integral equation,
\begin{equation}
\bar{rl}
 \bra\!\!\! &k_1 k_2 \vert K(\omega) \vert k_3 k_4 \ket\!\!=\!\!  \bra
 k_1 k_2 \vert v \vert k_3 k_4 \ket +\   \\
  &                 \\
 &\sum_{k_3' k_4'}
\bra k_1 k_2 \vert v \vert k_3' k_4' \ket\ {n_{>}(k_3')n_{>}(k_4') \over \omega - e} \
 \bra k_3' k_4' \vert K(\omega) \vert k_3 k_4 \ket.
\ear \label{eq:kkk}
\end{equation}
\cap In these equations $n_>(k) \,=\, 1 \,-\, n(k)$, with $n(k)$ the Fermi distribution function at a given
temperature and for the single particle spectrum $e(k)$. Then equation (\re{kkk}) coincides with the Brueckner
equation for the $G$ matrix in the zero temperature limit. It has to be noticed, that only the principal part has
to be considered in the integration, thus making $K$ a real matrix. The appearance of the $\arctan$ in equation
(\ref{eq:arctan}) looks peculiar, but it comes from a ladder summation similar to the one for the zero temperature
G-matrix. More detail on the derivation and on the numerical treatment of the equations can be found in reference
\cite{LG}. In this approach one calculates the free energy $F \,=\, \Omega \,+\, \mu N$ and then the pressure from
the thermodynamical relationship
\beq
p \,=\, \rho^2{\partial f \over \partial \rho}
\label{eq:press} \eeq
\noindent where $\rho$ is the total number density and $f$ the free energy per particle $F/N$. A typical result is
reported in figure (\ref{fig:free}), where the full lines are interpolations of the calculated points, suitable
for differentiation.
%
The resulting EoS at finite temperature, i.e. pressure vs. density, is reported in figure (\ref{fig:vdw}). One
recognizes the familiar Van der Waals shape, which entails a liquid-gas phase transition, with a definite critical
temperature, i.e. the temperature at which the minimum in the Van der Waals isotherm disappears. This is clearly a
fundamental property of the nuclear medium : it behaves macroscopically at finite temperature in a way similar to
a classical liquid. The critical temperature turns out to be around $T_c = 18 - 20$ MeV.
\par A difficulty in this approach is the lack of thermodynamical consistency. In fact the
thermodynamical relation $p \,=  -f' \,+\, \mu \rho$, usually referred as the "Hughenoltz-Van Hove theorem", is
not satisfied. Here $f'$ is the free energy per unit volume $F/V$. In other words, the pressure calculated from
equation (\ref{eq:press}) does not coincide with the pressure calculated from $p \,=\, - \Omega/V$. This point,
that is not necessarily a too serious drawback, is discussed in the next section.
\subsection{Theoretical uncertainties and challenges}
We have seen that symmetric nuclear matter undergoes a liquid-gas phase transition. This is the result of
calculations with microscopic calculations, but also with effective forces, e.g. Skyrme forces. The values of the
critical temperature, however, depends on the theoretical scheme, as well as on the particular effective force
adopted. In particular, it turns out that the critical temperature within Dirac-Brueckner scheme is definitely
smaller \cite{Malf,Weber} than in the non-relativistic scheme, about 10 MeV against 18-20 MeV. This cannot be
ascribed to relativistic effects, since the critical density is well below, about $1/3$, the saturation density.
Probably this is due to a different behavior of the Dirac-Brueckner EoS at low density. This point still needs
clarification.\par As anticipated in the previous sub-section, another uncertainty stems from the violation of the
Hughenoltz-Van Hove (HVH) theorem within the extension to finite temperature of the non-relativistic Brueckner
scheme, as implemented by Bloch and De Dominicis \cite{BDD}. In the applications, the pressure is calculated from
the derivative of the free energy per particle and the theorem is actually automatically satisfied. The difficulty
is then that the chemical potential determined by fixing the density in the Fermi distribution is not strictly the
one extracted from the derivative of the free energy per unit volume, as it should be. In any case the procedure
looks the most reliable within the Brueckner scheme, since the HVH theorem, as a thermodynamic relationship, is
satisfied.\par A general approximation scheme that strictly satisfies the theorem have been devised by G. Baym
\cite{GBca}. It is based on the self-consistent Green' s function method, where the single particle self-energy is
approximated by a functional of the Green' s function itself, which is then calculated in a self-consistent
manner. Numerical calculations have indeed \cite{ARios1} shown that the HVH is satisfied. The results at the
two-body correlation level, at least when only two-body forces are used, in some cases are similar to the
Brueckner ones, in some others they differ appreciably, according to the forces used \cite{ARios1,ARios2}. The
main difference with the Brueckner scheme is the introduction in the ladder summation also of the hole-hole
propagation. The expansion scheme is therefore at variance with the hole-line expansion, and actually it is not
clear how to proceed to improve the approximation or if convergence has been reached. Therefore, on one hand we
have the hole-line expansion at zero temperature that has some definite sign of convergence already at the two
hole-line (Brueckner) level, on the other hand at finite temperature we have a different truncation scheme to
satisfy the HVH theorem, that however is not proved to be a satisfactory approximation and does not reduce to the
Brueckner scheme at zero temperature. It would be quite desirable to have a scheme that is able to have both
requirements satisfied, i.e. to have a good degree of convergence and the fulfillment of the HVH theorem (at zero
and finite temperature). Although this sort of dilemma is a challenge that requires further studies, the gross
properties of nuclear matter at finite temperature appear well established.
\subsection{Isospin dependence}
If the nuclear matter is asymmetric, the existence of two components, neutrons and protons, complicate quite a bit
the phase transition picture. The spinodal region is still well defined, the bulk incompressibility at a given
asymmetry is negative in specific portions of the various possible thermodynamical planes. The coexistence line
presents a new feature, the "distillation" phenomenon. The name is suggested by the analogy with the process used
in the distillation of liquors. The chemical equilibrium between liquid and vapor requires the equality of the
proton and and neutron chemical potentials. They are however different if matter is asymmetric, and therefore, in
view of the density dependence of the symmetry energy, the fraction of neutrons and protons are different in the
liquid and in the vapor. In general, it turns out that the vapor is expected to be more neutron rich. The effect
is not dramatic, but it has been claimed that in heavy ion reactions, where the phase transition similar to the
liquid-gas phase transition is expected to occur, this effect should be seen \cite{maria}. The distillation has
also the effect that the Maxwell construction for asymmetric matter is modified with respect to the symmetric
case. The horizontal line that characterizes the construction in the pressure vs. density plane, within the
coexistence region, is replaced by a non-horizontal line, see figure (\ref{fig:NPA771_fig12}), taken from
reference \cite{camilled}.
\begin{figure}
 \begin{center}
\vskip -0.5 cm
\includegraphics[bb= 200 0 320 700,angle=0,scale=0.35]{figures/NPA771_fig12.eps}
\end{center}
\vspace{-0.5 cm} \caption{Isotherms of asymmetric nuclear matter at the indicated protons/nucleons ratio Z/A.
Upper panel, neutron chemical potential. Lower panel, pressure. The points A and B indicate the endpoints of the
coexistent region.}
    \label{fig:NPA771_fig12}
\end{figure}
In fact, if the fractions of vapor and liquid are changed, while they have different compositions the overall
asymmetry must be kept constant. This can be achieved only by changing the equilibrium pressure in order to shift
properly the neutron and proton chemical potentials.
\par
\begin{figure}
\vskip -7. cm
 \begin{center}
\includegraphics[bb= 240 0 380 700,angle=0,scale=0.5]{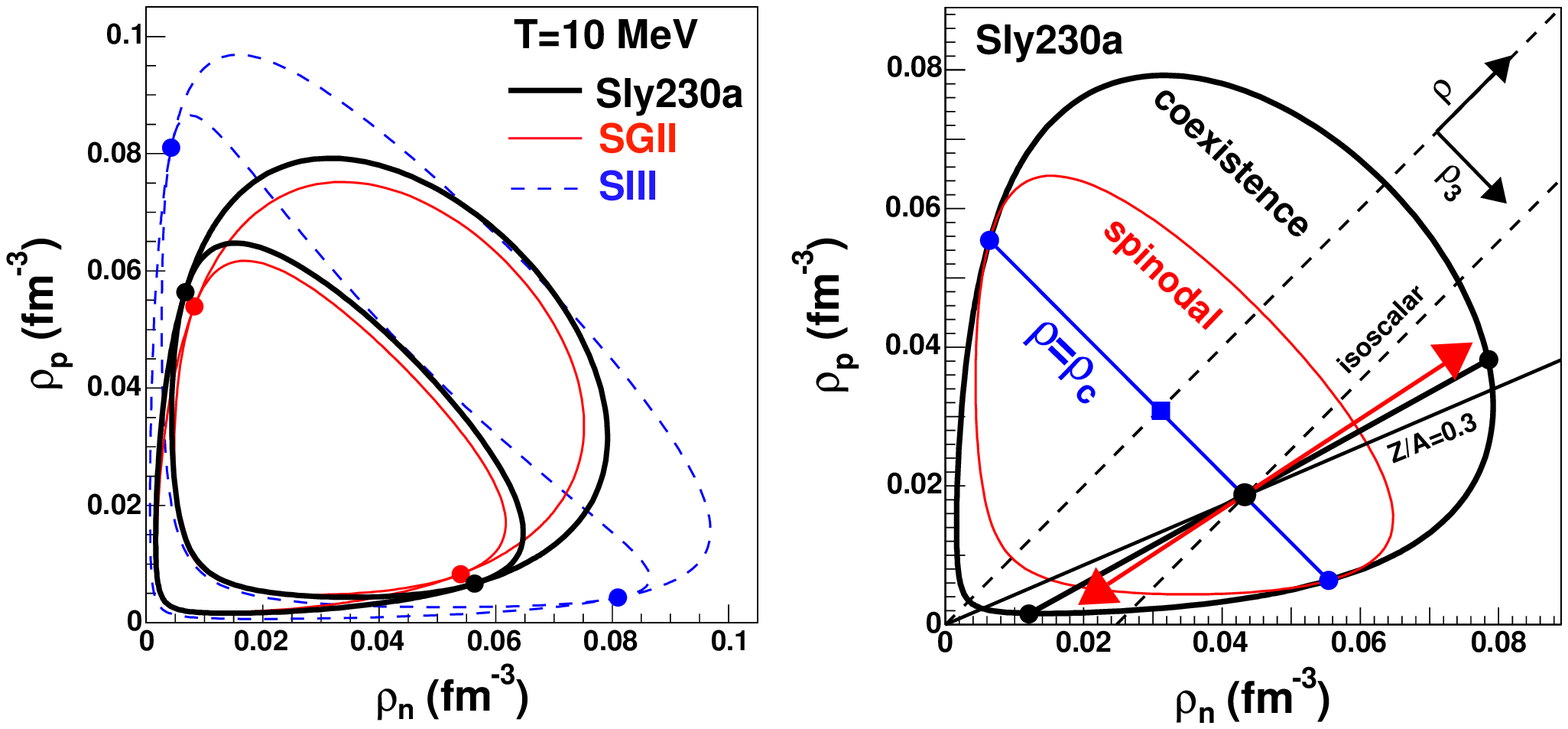}
\end{center}
\vspace{-0.3 cm} \caption{(Color on line) Left panel. Coexistence (outer region) and spinodal (inner region)
boundaries for three different Skyrme forces. In the right panel the arrows indicate the direction of minimal
curvature of the free energy.}
    \label{fig:NPA781_fig2}
\end{figure}

 Both the spinodal decomposition \cite{camilles} and the distillation phenomenon \cite{camilled} in asymmetric
matter at finite temperature have been studied within the Skyrme functional scheme. The line that marks the onset
of the spinodal instability is now characterized not only by the values of the total density and temperature (at a
given overall asymmetry) but also by the direction along which the instability can develop, i.e. the fractions of
protons and neutrons. According to this direction, the curvature of the free energy in the plane of proton vs.
neutron chemical potentials can vary. The direction where the curvature is minimal should indicate the most
unstable direction and therefore the most probable composition of the liquid clusters that are produced due to the
instability. This is illustrated in figure (\ref{fig:NPA781_fig2}), taken form reference \cite{camilles}.
\par This can have direct relevance in astrophysics for the formation process of the NS crust, where however it is
necessary to introduce the electron component and the Coulomb interaction.

\subsection{Phenomenology : the limiting temperature}
We have seen that symmetric nuclear matter undergoes a liquid-gas phase transition. However, if this phase
transition exists, it does not possess a direct correspondence in finite nuclei, due to the Coulomb interaction
and finite size effects. In particular, the Coulomb force is long range and strong enough to modify the nature
of the phase transition. However some authors \cite{LevitBonche,SongSu} have pointed out that the nuclear EoS
can be linked to the maximal temperature a nucleus can sustain before reaching mechanical instability. This
``limiting temperature" $T_{lim}$ is mainly the maximal temperature at which a nucleus can be observed.
\par
It has to be stressed that the reaction dynamics can prevent the formation of a true compound nucleus. The onset
of incomplete fusion reactions can mask completely the possible presence of fusion or quasi-fusion processes. At
higher energies, the heavy ion reaction can be fast enough that no (nearly) thermodynamical equilibrium can be
reached, as demanded in a genuine standard fusion-evaporation reaction. However, combined theoretical and
experimental analysis \cite{natprc} indicate that a nearly equilibrium condition is reached in properly selected
multi-fragmentation heavy ion reactions at intermediate energy. The main experimental observation is the
presence of a plateau in the so-called ''caloric curve", i.e. in the plot of temperature vs. total excitation
energy \cite{poch,wada,cibor,cussol}. This behavior was qualitatively predicted by the Copenhagen statistical
model \cite{bondorf} of nuclear multi-fragmentation. The relation between multi-fragmentation processes and the
nuclear EoS was extensively studied by several authors within the statistical approach to heavy ion reaction at
intermediate energy \cite{bondorf2,bondorf3,botvina,gross,friedman,csernai,bondorf4}.
\par
In different experiments, various methods were used to extract from the data the values of the temperature of
the source which produces the observed fragments, but a careful analysis of the data \cite{natprc} seems to
indicate a satisfactory consistency of the results. In refs. \cite{natprc,natprl} an extensive set of
experimental data was analyzed and it was shown that the temperature at which the plateau starts is decreasing
with increasing mass of the residual nucleus which is supposed to undergo fragmentation. Both the values and the
decreasing trend of this temperature turn out to be consistent with its interpretation as limiting temperature
$T_{lim}$. According to this interpretation, at increasing excitation energy the point where the temperature
plot deviates from Fermi gas behavior and the starting point of the plateau mark the critical point for
mechanical instability and the onset of the multi-fragmentation regime. The corresponding value of the critical
temperature can be calculated within the droplet model, and indeed many estimates based on Skyrme forces are in
fairly good agreements with the values extracted from phenomenology \cite{natprc,SongSu}. Moreover, the relation
between nuclear matter critical temperature $T_c$ and $T_{lim}$ appears to be quite stable and independent on
the particular EoS and method used, which allows \cite{natprl} to estimate $T_c$ from the set of values of
$T_{lim}$.
\par
In general, one can expect that $T_{lim}$ is substantially smaller than the critical one, $T_c$. In fact, both the
Coulomb repulsion and the lowering of the surface tension with increasing temperature tend to destabilize the
nucleus. Since the surface tension goes to zero at the critical temperature, $T_{lim}$ is reached much before
$T_{c}$. These predictions were checked in the seminal paper of reference \cite{LevitBonche}, as well as in
further studies based on macroscopic Skyrme forces \cite{SongSu}, for which a simple relationship was established
between $T_{lim}$ and $T_c$. If microscopic EoS are used, then the relationship between $T_{lim}$ and $T_c$ is not
so simple and the ratio $T_{lim}/T_c$ depends on the detail of the EoS \cite{Tlim}. In principle the comparison
with the phenomenological data can discriminate among different EoS. Indeed, most of the microscopic EoS reproduce
the empirical saturation point, but their behavior at finite temperature can be quite different. This is mainly
because the critical temperature, and therefore the limiting temperature, is very sensitive to the details of the
EoS. In fact $T_c$ is determined by the behavior of the derivative of the pressure with respect to the density,
which in turn is the second derivative of the free energy. If the pressure is extracted directly from the grand
canonical potential as a function of the chemical potential, still a strong sensitivity to the low density and
high temperature properties of the EoS remains.
\begin{figure}
 \begin{center}
\vskip 1.3 cm
\includegraphics[bb= 140 0 280 700,angle=-90,scale=0.4]{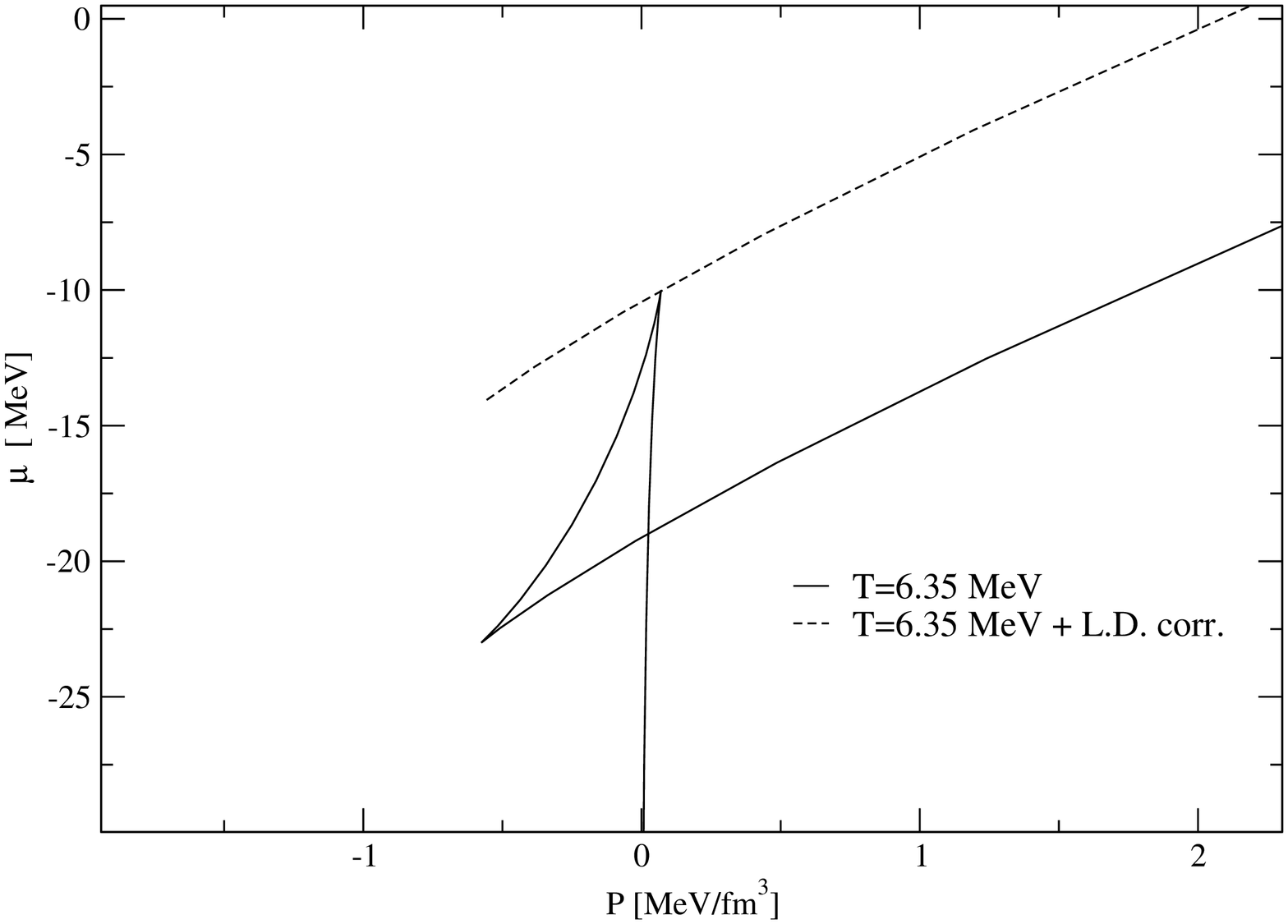}
\end{center}
\vspace{3.3 cm} \caption{Chemical potential vs. pressure for symmetric nuclear matter (full line). The line
starting from the lower cusp is the liquid branch, the one (almost vertical) starting from the upper cusp is the
vapor branch. The intersection point is the coexistence point at that temperature. The line joining the two cusps
is the unstable branch. The dashed line corresponds to a finite nucleus. It is obtained within a simplified Liquid
Drop model. The figure corresponds to the case of a temperature equal to the limiting temperature.}
    \label{fig:pmu}
\end{figure}
In figure (\ref{fig:pmu}) is reported the pressure as a function of the chemical potential in symmetric nuclear
matter, where one can recognize the liquid branch (the branch starting from the lowest cusp) and the vapor branch
(the almost vertical one that starts from the upper cusp) \cite{Tlim}. Their intersection gives the coexistence
point, while the smooth branch joining the two cusps is the unstable part of the EoS, corresponding to the Maxwell
construction. The upward shifted liquid branch (dashed line) is the branch of a finite nucleus that takes into
account the Coulomb interaction and finite size effects. Since the vapor branch is assumed almost unchanged, the
shifted branch gives the shifted coexistent point between the nucleus and the vapor (assumed uncharged). If, as in
the figure, the liquid branch touches the upper cusp, then the corresponding temperature is the searched limiting
temperature \cite{Tlim}, since at increasing temperature the cusp is lowered and no coexistence point exists
anymore.
\begin{figure} [ht]
 \begin{center}
\vskip 0.7 cm
\includegraphics[bb= 140 0 280 700,angle=-90,scale=0.4]{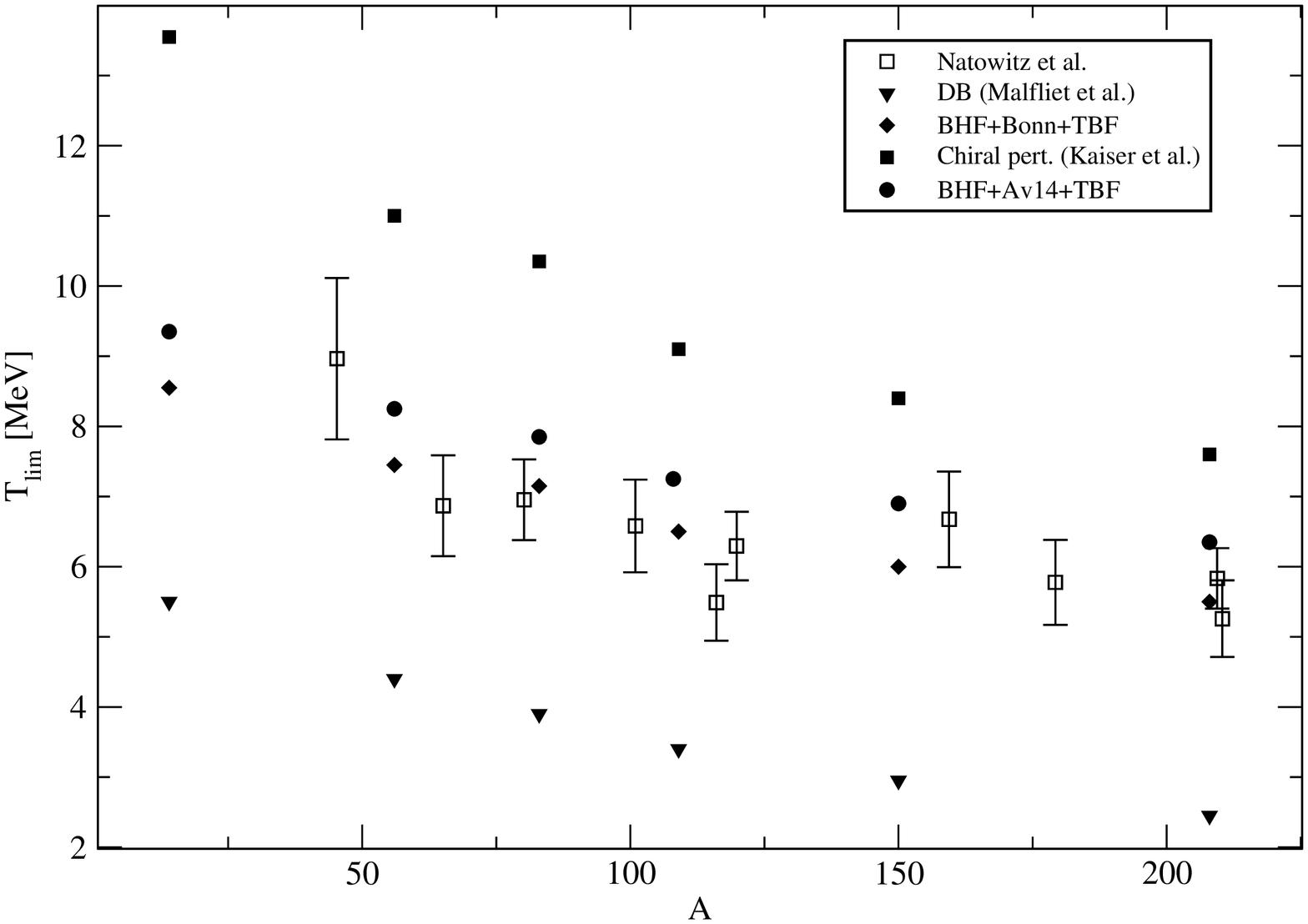}
\end{center}
\vspace{3.3 cm} \caption{Limiting temperature calculated with different EoS in comparison wit the experimental
data (squares with error bars) From the bottom, full triangles corresponds to the Dirac-Brueckner EoS \cite{Malf},
full diamonds to EoS calculated within the finite temperature BHF and with the Bonn + TBF interaction, full
circles to the same method but with the Av$_{14}$ + TBF interaction \cite{Tlim}, and finally full squares to the
EoS from the chiral perturbation theory of reference \cite{Kaiser_T}.}
    \label{fig:Tlim}
\end{figure}
 In figure (\ref{fig:Tlim}), taken from reference
\cite{Tlim}, are reported the results of a systematic calculation of $T_{lim}$ for different nuclei and for
different EoS in comparison with the values extracted from experimental data \cite{natprc,natprl}. One notices the
strong dependence on the EoS derived from different microscopic schemes, and at the same time the relative small
sensitivity to the NN force within the same scheme (BHF). It seems that the non-relativistic EoS based on the BHF
approximation is the favored one.

\subsection{The astrophysical link}

The outer part of cold Neutron Stars is formed by a solid crystal of nuclei, while in a supernova, during the
after bounce stage, the temperature is so high that nuclear matter is in the homogeneous fluid phase. During the
cooling process medium-heavy nuclei are formed from the homogeneous matter, and this transition is quite similar
to a liquid-vapor phase transition, where liquid droplets are formed in the mixed phase. The droplet formation is
directly related to the so called "spinodal" instability, i.e. the region in the phase diagram where the
incompressibility is negative.  However, the presence of Coulomb interaction changes the nature of the transition
\cite{Camille_an}. It turns out that, despite it is of first order, the thermodynamical potentials do not display
singularity as a function of the thermodynamical variables. In any case, the transition is accompanied by the
formation of nuclear clusters. In fact one can expect that at sufficiently low temperature, inside the spinodal
region, nuclear matter is composed of nuclei of different sizes, light fragments (tritons, helium-3 and alpha
particles) and nucleons. This phase has been described on the basis of the liquid drop model \cite{Latt-Swe} and
the relativistic Thomas-Fermi scheme \cite{Shen}. In the low density limit a virial expansion has been also
applied \cite{Horo}. The detailed theoretical description of this phase is quite relevant mainly for supernova
simulations. A microscopic theory of the corresponding EoS, at the same level of accuracy as for the homogeneous
matter case, is still missing.\par Finally, it has to be mentioned that at even lower temperature the clusters
should undergo a liquid-solid phase transition, with the formation of a Coulomb crystal. This transition is even
less known and no consensus exists on its general properties like the solidification energy or latent heath. A
discussion on this subject is outside the scope of the present review.

\par

\section{Conclusions}
\par
The excursus we tried to perform on the properties of the peculiar matter, that can be called "nuclear medium",
surely does not make justice of the impressive progress that has taken place both at phenomenological and
theoretical level. The continuous and vivid interest on the subject is due to the extremely wide realm of
phenomena and physical systems, on the Earth and in the Universe, where the nuclear medium plays a central role.
\par
The continuous interplay between theoretical developments on one hand and laboratory experiments and astrophysical
observations on the other hand is the main driving force that makes possible progress in this exciting field. As
we tried to illustrate, our knowledge on the properties of the nuclear medium has widened and sharpened in many
respects. Many physical parameters
%
%
start to be known with a certain accuracy not only at saturation density but also at lower and higher density,
and, in general, the Equation of State has been severely constrained. Taking the risk of being too schematic, let
us try to summarize tentatively our knowledge of those properties of the nuclear medium that we touched in this
review.
\par\noindent
1. {\it Equation of State.} The constraints that has been obtained from the data on heavy ion collisions and from
the analysis of astrophysical observations on Neutron Stars are complementary between each others, since the
nuclear matter asymmetry that is involved is quite different in the two cases. However, it is a challenge for the
theory to be able to describe the nuclear medium in both physical conditions. If one takes together the two sets
of data, one can get a severe test for the different microscopic theories. Although these constraints have to be
confirmed and further analyzed, one can expect that in the near future new data and new results will put under
serious exam all existing microscopic theories.
\par\noindent
2. {\it Incompressibility.} The value of the symmetric matter incompressibility near saturation seems now to be
constrained to an interval approximately between 210 and 250 MeV. Microscopic theories are in fair agreements with
these values. More difficult is to put constraints on the incompressibility at higher density, since its value is
determined by the details of the density dependence of the pressure. Astrophysical observations should test in
principle the incompressibility of very asymmetric nuclear matter, but the uncertainty on the structure of neutron
stars inner core hinders the progress in this direction.
\par\noindent
3. {\it Symmetry energy.} The symmetry energy at saturation can be considered constrained in a narrow interval,
essentially 30 $\pm$ 1 MeV. All microscopic theories agree with these values. Indirect hints about the density
dependence of the symmetry energy come from both heavy ion collision data and astrophysical observation on neutron
star phenomenology. The data analysis does not allow to get any sound conclusion, and discrepancies among
different groups on the interpretation of the data are present. Microscopic theories agree at subsaturation
density and up to 2-3 times saturation density, but disagreements appear at higher density. Correlations between
symmetry energy at sub-saturation density and nuclear structure properties have been found. These can be of great
help in clarifying many open questions in the field.
\par\noindent
4. {\it Transport coefficients.} Several macroscopic phenomena that occur in neutron stars are determined by
transport properties of the nuclear medium. The prediction of the damping of some of the possible overall
oscillations of neutron stars requires the calculation of the shear and bulk viscosity. The study of cooling
evolution of neutron stars needs the knowledge of the thermal properties of the nuclear medium in different
physical conditions. In principle these transport coefficients require only standard techniques based on the Fermi
Liquid theory. What is needed presently is a consistent scheme that is able to describe both the mechanical
properties, like the EoS, of the nuclear medium and its transport properties. Progresses in this direction have
been done and one can be confident that in the near future a coherent picture of the nuclear medium will be
possible on the basis of microscopic many-body theory, at least for not too high density.
\par\noindent
5. {\it Low energy excitations.} The elementary excitations of the nuclear medium are relevant for its thermal
properties and for many phenomena occurring in neutron stars, like emission and propagation of neutrinos. They are
also a guidance to the low energy excitations in nuclei. They can be studied within the Landau theory of Fermi
liquids. The fundamental physical parameters that are needed are the so called "Landau parameters" that fix the
effective nucleon-nucleon interaction in the vicinity of the Fermi surface. The microscopic determination of these
parameters is a difficult theoretical problem, that has been approached with a variety of techniques. Both in
symmetric and pure neutron matter their values are not firmly established. The phenomenology on nuclei excitations
give only indications on some of them, since the physical conditions are quite different and finite size effects,
like the presence of a surface, are essential.
\par\noindent
6. {\it Nuclei and finite size effects.} The connection between the nuclear medium properties and the structure of
nuclei is of course not simple. However, since a long time, the Energy Density Functional method has been
developed also to describe the ground state of nuclei and their elementary excitations. This method is the most
suited to establish a link between nuclear structure and the nuclear medium properties. Indeed the main assumption
that can be taken within this method is to divide the functional to be minimized into a bulk and a surface part
and identify the bulk part with the EoS of nuclear matter. Of course, besides these two building blocks, the
spin-orbit and Coulomb interaction must be included. This is a theory that is semi-microscopic in character, since
all finite size effects are assumed to be included in an effective way in surface terms, to be determined
phenomenologically. However, the physical interpretation of the different terms can provide hints on the values of
the parameters that fix the non-bulk parts of the functional. If the functionals is treated completely
phenomenological and the number and complexity of the surface terms (i.e. containing density gradients) are
increased, the accuracy of these functionals can be excellent throughout the mass table. The semi-microscopic and
the purely phenomenological approach are complementary, and substantial refinements and increase of accuracy are
expected to occur in the future.
\par\noindent
7. {\it Pairing and superfluidity.} The pairing correlation is of paramount relevance both in the physics of
neutron stars and in nuclear structure. Substantial progress has been made in the theoretical determination of the
effective pairing interaction in neutron matter and in neutron star matter for the different superfluid channels.
Due to the extreme sensitivity of the pairing gap to the strength of the effective interaction, the various
pairing gaps are still uncertain, but their overall trends as a function of density can be considered established,
at least for not too high density. In nuclei it is not yet established to what extent the bare nucleon-nucleon
interaction, taken as pairing interaction, is able to reproduce the experimental pairing gaps. However, a
substantial fraction can surely come from the bare interaction, while the remaining part is a matter of debates.
\par\noindent
8. {\it Transition to quark matter.} Neutron stars could be the only place in the Universe where macroscopic
portion of quark matter is present in stable conditions. Massive neutron stars are indeed expected to have so high
density baryonic matter in their interior that a transition to the deconfined phase could be possible. This is of
course not firmly established, but progress in this field seems to be rapid. The maximum mass of observed neutron
stars, with a fairly good degree of confidence, has increased in the last few years, up to a value close to 2
$M_\odot$. This result, if confirmed, would put definite constraints not only on the nuclear medium EoS but also
on the possible quark matter EoS. This shows the interplay between the "traditional" nuclear physics and the
theory of the QCD matter in the physics of neutron stars. Further exciting and fundamental progresses are expected
in the near future in this field.
\par
This schematic list surely does not exhaust all the facets of the properties of the nuclear medium, but we hope to
have given at least some ideas of the state of the art in the field, of what is going on and of the trends that
can be expected to develop in the future.

\ack We would like the thanks all our collaborators along the years, who appear as co-authors of the papers quoted
in this review. To them we would like to express our gratitude and friendship. One of the author (M.B.) would like
to acknowledge the illuminating discussions with Prof. P. Schuck on the Thomas-Fermi approximation and Prof. X.
Vinas on the Density Functional method.\par
We acknowledge the support of COMPSTAR, a research and training program of the European Science Foundation.

\end{document}